\documentclass[aps,pra,twocolumn,showpacs]{revtex4}%
\usepackage{graphics}
\usepackage{amsmath}
\usepackage{graphicx}
\usepackage{amsfonts}
\usepackage{amssymb}
\newtheorem{theorem}{Theorem}

\newtheorem{corollary}[theorem]{Corollary}

\begin{document}
\title{Quantum Reading of Digital Memories}
\author{Stefano Pirandola}
\affiliation{Department of Computer Science, University of York, York YO10 5GH, United Kingdom}
\date{\today}

\begin{abstract}
We consider a basic model of digital memory where each cell is composed of a
reflecting medium with two possible reflectivities. By fixing the mean number
of photons irradiated over each memory cell, we show that a non-classical
source of light can retrieve more information than any classical source. This
improvement is shown in the regime of few photons and high reflectivities,
where the gain of information can be surprising. As a result, the use of
quantum light can have non-trivial applications in the technology of digital
memories, such as optical disks and barcodes.

\end{abstract}

\maketitle

In recent years, non-classical states of radiation have been
exploited to achieve marvellous results in quantum information and
computation \cite{NielsenBook,BraREV,BraREV2,GaussSTATES}. In the
language of quantum optics, the bosonic states of the
electromagnetic field are called \textquotedblleft
classical\textquotedblright\ when they can be expressed as
probabilistic mixtures of coherent states. Classical states
describe practically all the radiation sources which are used in
today's technological applications. By contrast, a bosonic state
is called \textquotedblleft non-classical\textquotedblright\ when
its decomposition in coherent states is non-positive
\cite{CLASSstate,Prepres}. One of the key properties which makes a
state non-classical is quantum entanglement. In the bosonic
framework, this is usually present under the form of
Einstein-Podolsky-Rosen (EPR) correlations \cite{EPRpaper},
meaning that the position and momentum quadrature operators of two
bosonic modes are so correlated as to beat the standard quantum
limit \cite{EPRcorrelations}. This is a well-known feature of the
two-mode squeezed vacuum (TMSV) state, one of the most important
states routinely produced in today's quantum optics labs.

In this Letter, we show how the use of non-classical light possessing
EPR\ correlations can \textit{widely} improve the readout of information from
digital memories. To our knowledge, this is the first study which proves and
quantifies the advantages of using non-classical light for this fundamental
task, being absolutely non-trivial to identify the physical conditions that
can effectively disclose these advantages (as an example, see the recent no-go
theorems of Ref.~\cite{Nair} applied to quantum illumination \cite{QiLL}). Our
model of digital memory is simple but can potentially be extended to realistic
optical disks, like CDs and DVDs, or other kinds of memories such as barcodes.
In fact, we consider a memory where each cell is composed of a reflecting
medium with two possible reflectivities, $r_{0}$ and $r_{1}$, used to store a
bit of information. This memory is irradiated by a source of light which is
able to resolve every single cell. The light focussed on, and reflected from,
a single\ cell is then measured by a detector, whose outcome provides the
value of the bit stored in that cell. Besides the \textquotedblleft
signal\textquotedblright\ modes irradiating the target cell, we also consider
the possible presence of ancillary \textquotedblleft idler\textquotedblright%
\ modes which are directly sent to the detector. The general aim of these
modes is to improve the performance of the output measurement by exploiting
possible correlations with the signals. Adopting this model and fixing the
mean number of photons irradiated over each memory cell, we show that a
\textit{non-classical} source of light with EPR correlations between signals
and idlers can retrieve more information than any classical source of light.
In particular, this is proven for high reflectivities (typical of optical
disks) and few photons irradiated. In this regime the difference of
information can be surprising, up to 1 bit per cell (corresponding to the
extreme situation where only quantum light can retrieve information).
As we will discuss in the conclusion, the chance of reading
information using few photons can have remarkable consequences in
the technology of digital memories, e.g., in terms of
data-transfer rates and storage capacities.\begin{figure}[ptbh]
\vspace{-0.4cm}
\par
\begin{center}
\includegraphics[width=0.5\textwidth] {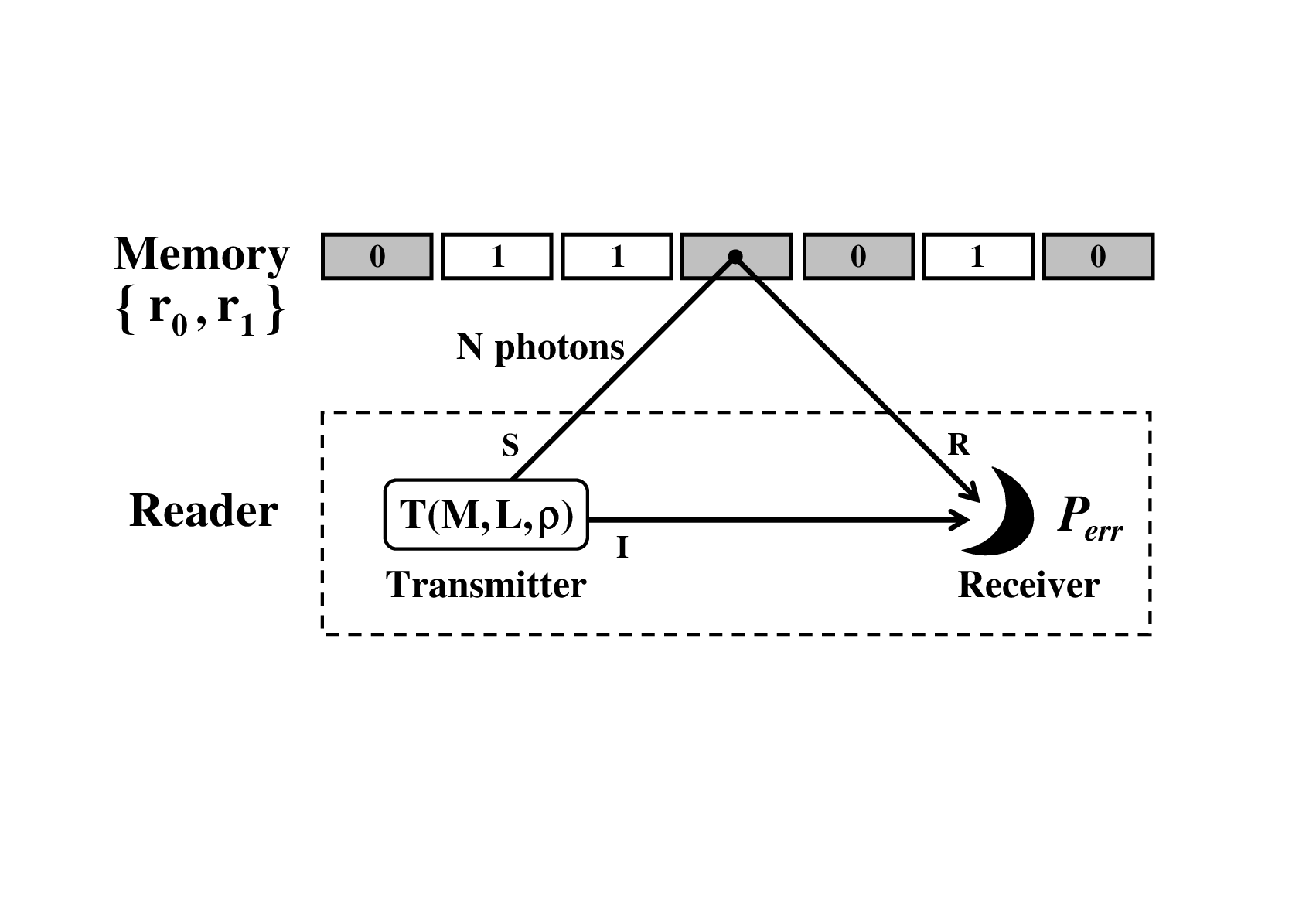}
\end{center}
\par
\vspace{-1.4cm}\caption{\textbf{Basic model of memory}. Digital
information is stored in a memory whose cells have different
reflectivities: $r=r_{0}$ encoding bit-value $u=0$, and $r=r_{1}$
encoding bit-value $u=1$. \textbf{Readout of the memory}. In
general, a digital reader consists of transmitter and receiver.
The transmitter $T(M,L,\rho)$ is a bipartite bosonic system,
composed by a signal system $S$ (with $M$ modes) and an idler
system $I$ (with $L$ modes), which is given in some global state
$\rho$. The signal
$S$ emitted by this source has \textquotedblleft bandwidth\textquotedblright%
\ $M$\ and \textquotedblleft energy\textquotedblright\ $N$\ (mean number of
photons). The signal is directly shined over the cell, and its reflection $R$
is detected together with the idler $I$ at the output receiver, where a
suitable measurement retrieves the value of the bit up to an error probability
$P_{err}$.}%
\label{QreadPIC}%
\end{figure}

Let us consider a digital memory where each cell can have two possible
reflectivities, $r_{0}$ or $r_{1}$, encoding the two values of a logical bit
$u$ (see\ Fig.~\ref{QreadPIC}). Close to the memory, we have a digital reader,
made up of transmitter and receiver, whose goal is to retrieve the value of
the bit stored in a target cell. In general, we call the \textquotedblleft
transmitter\textquotedblright\ a bipartite bosonic system, composed by a
signal system $S$ with $M$ modes and an idler system $I$ with $L$ modes, and
globally given in some state $\rho$. This source can be completely specified
by the notation $T(M,L,\rho)$. By definition, we say that the transmitter $T$
is \textquotedblleft classical\textquotedblright\ (\textquotedblleft
non-classical\textquotedblright) when the corresponding state $\rho$ is
classical (non-classical), i.e., $T_{c}=T(M,L,\rho_{c})$\ and $T_{nc}%
=T(M,L,\rho_{nc})$. The signal $S$ emitted by the transmitter is associated
with two basic parameters: the number of modes $M$, that we call the
\textquotedblleft bandwidth\textquotedblright\ of the signal, and the mean
number of photons $N$, that we call the \textquotedblleft
energy\textquotedblright\ of the signal \cite{NOTEonN}. The signal $S$ is
shined directly on the target cell, and its reflection $R$ is detected
together with the idler $I$ at the output receiver. Here a suitable
measurement yields the value of the bit up to an error probability $P_{err}$.
Repeating the process for each cell of the memory, the reader retrieves an
average of $1-H(P_{err})$\ bits per cell, where $H(\cdot)$ is the binary
Shannon entropy.

The basic mechanism in our model of digital readout is quantum channel
discrimination. In fact, encoding a logical bit $u\in\{0,1\}$\ in a pair of
reflectivities $\{r_{0},r_{1}\}$ is equivalent to encoding $u$ in a pair of
attenuator channels $\{\mathcal{E}(r_{0}),\mathcal{E}(r_{1})\}$, with linear
losses $\{r_{0},r_{1}\}$ acting on the signal modes. The readout of the bit
consists in the statistical discrimination between $r_{0}$\ and $r_{1}$, which
is formally equivalent to the channel discrimination between $\mathcal{E}%
(r_{0})$ and $\mathcal{E}(r_{1})$. The error probability affecting the
discrimination $\mathcal{E}(r_{0})\neq\mathcal{E}(r_{1})$\ depends on both
transmitter and receiver. For a \textit{fixed} transmitter $T(M,L,\rho)$, the
pair $\{\mathcal{E}(r_{0}),\mathcal{E}(r_{1})\}$\ generates two possible
output states at the receiver, $\sigma_{0}(T)$ and $\sigma_{1}(T)$. These are
expressed by $\sigma_{u}(T)=[\mathcal{E}(r_{u})^{\otimes M}\otimes
\mathcal{I}^{\otimes L}](\rho)$, where $\mathcal{E}(r_{u})$ acts on the
signals and the identity $\mathcal{I}$ on the idlers. By optimizing over the
output measurements, the minimum error probability which is achievable by the
transmitter $T$ in the channel discrimination $\mathcal{E}(r_{0}%
)\neq\mathcal{E}(r_{1})$\ is equal to $P_{err}(T)=(1-D)/2$, where $D$ is the
trace distance between $\sigma_{0}(T)$ and $\sigma_{1}(T)$. Now the crucial
point is the minimization of $P_{err}(T)$ over the transmitters $T$. Clearly,
this optimization must be constrained by fixing basic parameters of the
signal. Here we consider the most general situation where only the signal
energy $N$ is fixed. Under this energy constraint the optimal transmitter $T$
which minimizes $P_{err}(T)$ is unknown. For this reason, it is non-trivial to
ask the following question: does a non-classical transmitter which outperforms
any classical one exist? In other words: given two reflectivities
$\{r_{0},r_{1}\}$, i.e., two attenuator channels $\{\mathcal{E}(r_{0}%
),\mathcal{E}(r_{1})\}$, and a fixed value $N$\ of the signal energy, can we
find any $T_{nc}$ such that $P_{err}(T_{nc})<P_{err}(T_{c})$ for every $T_{c}%
$? In the following we reply to this basic question, characterizing the
regimes where the answer is positive. The first step in our derivation is
providing a bound which is valid for every classical transmitter (see Appendix
for the proof).

\begin{theorem}
[\textbf{classical discrimination bound}]\label{THEOmain}\textit{Let us
consider the discrimination of two reflectivities }$\{r_{0},r_{1}%
\}$\textit{\ using a classical transmitter }$T_{c}$\textit{\ which signals
}$N$\textit{\ photons. The corresponding error probability satisfies}%
\begin{equation}
P_{err}(T_{c})\geq\mathcal{C}(N,r_{0},r_{1}):=\frac{1-\sqrt{1-e^{-N(\sqrt
{r_{1}}-\sqrt{r_{0}})^{2}}}}{2}~. \label{CB_cread}%
\end{equation}

\end{theorem}

\noindent According to this theorem, all the classical transmitters $T_{c}$
irradiating $N$ photons on a memory with reflectivities $\{r_{0},r_{1}\}$
cannot beat the classical discrimination bound $\mathcal{C}(N,r_{0},r_{1})$,
i.e., they cannot retrieve more than $1-H(\mathcal{C})$ bits per cell.
Clearly, the next step is constructing a non-classical transmitter which can
violate this bound. A possible design is the \textquotedblleft
EPR\ transmitter\textquotedblright, composed by $M$ signals and $M$\ idlers,
that are entangled pairwise via two-mode squeezing. This transmitter has the
form $T_{epr}=T(M,M,\left\vert \xi\right\rangle \left\langle \xi\right\vert
^{\otimes M})$, where $\left\vert \xi\right\rangle \left\langle \xi\right\vert
$ is a TMSV state entangling signal mode $s\in S$ with idler mode $i\in I$. In
the number-ket representation $\left\vert \xi\right\rangle =(\cosh\xi
)^{-1}\sum_{n=0}^{\infty}(\tanh\xi)^{n}\left\vert n\right\rangle
_{s}\left\vert n\right\rangle _{i}$, where the squeezing parameter $\xi$
quantifies the signal-idler entanglement. An arbitrary EPR\ transmitter,
composed by $M$ copies of $\left\vert \xi\right\rangle \left\langle
\xi\right\vert $, irradiates a signal with bandwidth $M$ and energy
$N=M\sinh^{2}\xi$. As a result, this transmitter can be completely
characterized by the basic parameters of the emitted signal, i.e., we can set
$T_{epr}=T_{M,N}$. Then, let us consider the discrimination of two
reflectivities $\{r_{0},r_{1}\}$\ using an EPR transmitter $T_{M,N}$ which
signals $N$ photons. The corresponding error probability is upper-bounded by
the quantum Chernoff bound \cite{QCbound,MinkoPRA}
\begin{equation}
P_{err}(T_{M,N})\leq\mathcal{Q}(M,N,r_{0},r_{1}):=\frac{1}{2}\left[
\inf_{t\in(0,1)}\mathrm{Tr}(\theta_{0}^{t}\theta_{1}^{1-t})\right]  ^{M},
\label{QCB_qread}%
\end{equation}
where $\theta_{u}:=[\mathcal{E}(r_{u})\otimes\mathcal{I}](\left\vert
\xi\right\rangle \left\langle \xi\right\vert )$. In other words, at least
$1-H(\mathcal{Q})$ bits per cell can be retrieved from the memory. Exploiting
Eqs.~(\ref{CB_cread}) and~(\ref{QCB_qread}), our main question simplifies to
finding $\bar{M}$\ such that $\mathcal{Q}(\bar{M},N,r_{0},r_{1})<\mathcal{C}%
(N,r_{0},r_{1})$. In fact, this implies $P_{err}(T_{\bar{M},N})<\mathcal{C}%
(N,r_{0},r_{1})$, i.e., the existence of an EPR\ transmitter $T_{\bar{M},N}$
able to outperform any classical transmitter $T_{c}$. This is the result of
the following theorem (see Appendix for the proof).

\begin{theorem}
[threshold energy]\textit{For every pair of reflectivities }$\{r_{0},r_{1}%
\}$\textit{\ with }$r_{0}\neq r_{1}$\textit{, and signal energy }%
\begin{equation}
N>N_{th}(r_{0},r_{1}):=\frac{2\ln2}{2-r_{0}-r_{1}-2\sqrt{(1-r_{0})(1-r_{1})}%
}~,
\end{equation}
\textit{there is an }$\bar{M}$\textit{\ such that }$P_{err}(T_{\bar{M}%
,N})<\mathcal{C}(N,r_{0},r_{1})$.
\end{theorem}

\begin{figure}[ptbh]
\vspace{-0.14cm}
\par
\begin{center}
\includegraphics[width=0.47\textwidth] {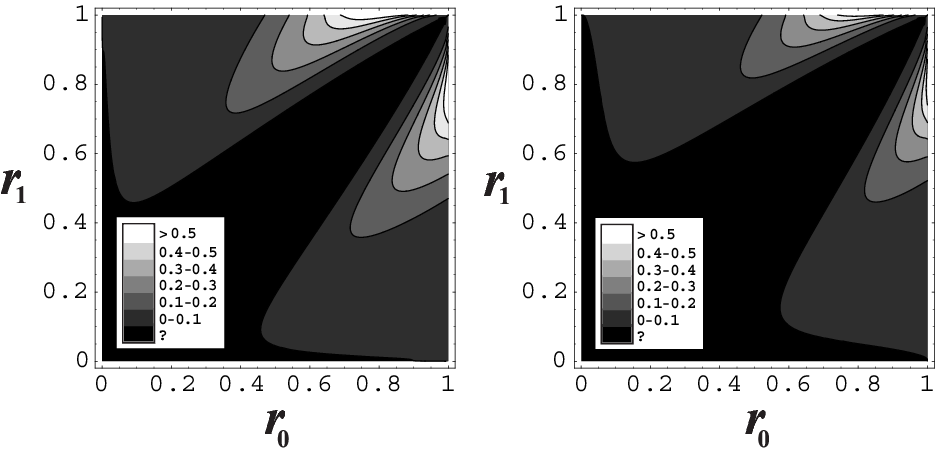}
\end{center}
\par
\vspace{-0.59cm}\caption{\textbf{Left.} Minimum information gain $G$ over the
memory plane $\{r_{0},r_{1}\}$. For a few-photon signal ($N=30$), we compare a
narrowband EPR transmitter ($M=30$) with all the classical transmitters.
Inside the black region ($r_{0}\approx r_{1}$) our investigation is
inconclusive. Outside the black region, we have $G>0$. \textbf{Right.} $G$
plotted over the plane $\{r_{0},r_{1}\}$ in the presence of decoherence
($\varepsilon=\bar{n}=10^{-5}$). For a few-photon signal ($N=30$), we compare
a narrowband EPR transmitter ($M=30$) with all the classical transmitters
$T(M,L,\rho_{c})$\ having $M\leq M^{\ast}=5\times10^{6}$.}%
\label{colorbar}%
\end{figure}

\noindent Thus we get the central result of the paper: for every memory and
above a threshold energy, there is an EPR transmitter which outperforms any
classical transmitter. Remarkably, the threshold energy $N_{th}$ turns out to
be low ($<10^{2}$) for most of the memories $\{r_{0},r_{1}\}$ outside the
region $r_{0}\approx r_{1}$. This means that we can have an enhancement in the
regime of few photons ($N<10^{2}$). Furthermore, for low energy $N$, the
critical bandwidth $\bar{M}$\ can be low too. In other words, in the regime of
few photons, narrowband EPR transmitters are generally sufficient to overcome
every classical transmitter. To confirm and quantify this analysis, we
introduce the \textquotedblleft minimum information gain\textquotedblright%
\ $G(M,N,r_{0},r_{1}):=1-H(\mathcal{Q})-[1-H(\mathcal{C})]$. For given memory
$\{r_{0},r_{1}\}$ and signal energy $N$, this quantity lowerbounds the number
of bits per cell which are gained by an EPR\ transmitter $T_{M,N}$ over any
classical transmitter $T_{c}$ \cite{G}. Numerical investigations (see
Fig.~\ref{colorbar}) show that narrowband EPR transmitters are able to give
$G>0$ in the regime of few photons and high reflectivities, corresponding to
having $r_{0}$ or $r_{1}$ sufficiently close to\ $1$ (as typical of optical
disks). In this regime, part of the memories display remarkable gains ($G>0.5$).

Thus the enhancement provided by quantum light can be dramatic in the regime
of few photons and high reflectivities. To investigate more closely this
regime, we consider the case of ideal memories, defined by $r_{0}<r_{1}=1$. As
an analytical result, we have the following (see Appendix for the proof).

\begin{theorem}
[ideal memory]\textit{For every }$r_{0}<r_{1}=1$\textit{\ and }$N\geq
N_{th}:=1/2$\textit{, there is a minimum bandwidth }$\bar{M}$\textit{\ such
that }$P_{err}(T_{M,N})<\mathcal{C}(N,r_{0},r_{1})$\textit{\ for every
}$M>\bar{M}$.
\end{theorem}

\begin{figure}[ptbh]
\vspace{-0.13cm}
\par
\begin{center}
\includegraphics[width=0.47\textwidth] {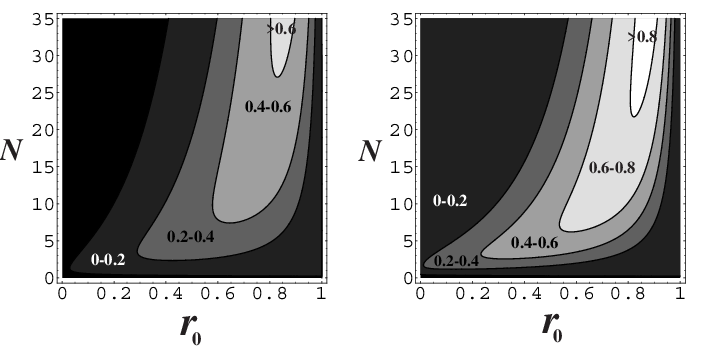}
\end{center}
\par
\vspace{-0.55cm}\caption{Minimum information gain $G$\ versus $r_{0}$ and $N$.
Left picture refers to $M=1$, right picture to $M\rightarrow\infty$. (For
arbitrary $M$ the scenario is intermediate.) Outside the inconclusive black
region we have $G>0$. For $M\rightarrow\infty$ the black region is completely
collapsed below $N_{th}=1/2$.}%
\label{idealPIC}%
\end{figure}

\noindent Thus, for ideal memories and signals above $N_{th}=1/2$ photon,
there are infinitely many EPR transmitters able to outperform every classical
transmitter. For these memories, the threshold energy is so low that the
regime of few photons can be fully explored. The gain $G$\ increases with the
bandwidth, so that optimal performances are reached by broadband
EPR\ transmitters ($M\rightarrow\infty$). However, narrowband EPR transmitters
are sufficient to give remarkable advantages, even for $M=1$ (i.e., using a
single TMSV state). This is shown in Fig.~\ref{idealPIC}, where $G$ is plotted
in terms of $r_{0}$ and $N$, considering the two extreme cases $M=1$ and
$M\rightarrow\infty$. According to Fig.~\ref{idealPIC}, the value of $G$ can
approach $1$ for ideal memories and few photons even if we consider narrowband
EPR transmitters. 

\textit{Presence of decoherence}.~Note that the previous analysis does not
consider the presence of thermal noise. Actually this is a good approximation
in the optical range, where the number of thermal background photons is around
$10^{-26}$ at about $1~\mu$m and $300~\mathrm{K}$. However, to complete the
analysis, we now show that the quantum effect exists even in the presence of
stray photons hitting the upper side of the memory and decoherence within the
reader. The scattering is modelled as white thermal noise with $\bar{n}$
photons per mode entering each memory cell. Numerically we consider $\bar
{n}=10^{-5}$ corresponding to non-trivial diffusion. This scenario may occur
when the light, transmitted through the cells, is not readily absorbed by the
drive (e.g., using a bucket detector just above the memory) but travels for a
while diffusing photons which hit neighboring cells. Assuming the presence of
one photon per mode travelling the \textquotedblleft optimistic
distance\textquotedblright\ of one meter and undergoing Rayleigh scattering,
we get roughly $\bar{n}\simeq10^{-5}$ \cite{Rayleigh}. The internal
decoherence is modelled as a thermal channel\ $\mathcal{N}(\varepsilon)$
adding Gaussian noise of variance $\varepsilon$ to each signal/reflected mode,
and $2\varepsilon$ to the each idler mode (numerically we consider the
non-trivial value $\varepsilon=\bar{n}=10^{-5}$). Now, distinguishing between
two reflectivities $\{r_{0},r_{1}\}$ corresponds to discriminating between two
Gaussian channels $\mathcal{S}_{u}\otimes\mathcal{N}(2\varepsilon)$ for
$u\in\{0,1\}$. Here $\mathcal{S}_{u}:=\mathcal{N}(\varepsilon)\circ
\mathcal{E}(r_{u},\bar{n})\circ\mathcal{N}(\varepsilon)$ acts on each signal
mode, and contains the attenuator channel $\mathcal{E}(r_{u},\bar{n})$ with
conditional loss $r_{u}$\ and thermal noise $\bar{n}$. To solve this scenario
we use Theorem~\ref{THEOmain} with the proviso of generalizing the classical
discrimination bound. In general, we have $\mathcal{C}=(1-\sqrt{1-F^{M}})/2$,
where $F$ is the fidelity between $\mathcal{S}_{0}(|\sqrt{n_{S}}\rangle
\langle\sqrt{n_{S}}|)$ and $\mathcal{S}_{1}(|\sqrt{n_{S}}\rangle\langle
\sqrt{n_{S}}|)$, the two outputs of a single-mode coherent state $|\sqrt
{n_{S}}\rangle$ with $n_{S}:=N/M$ mean photons. Here the expression for
$\mathcal{C}$ depends also on the bandwidth $M$\ of the classical transmitter
$T_{c}=T(M,L,\rho_{c})$. Since $\mathcal{C}$ decreases to zero for
$M\rightarrow\infty$, our quantum-classical comparison is now restricted to
classical transmitters $T(M,L,\rho_{c})$ with $M$ less than a maximal value
$M^{\ast}<\infty$. Remarkably we find that, in the regime of few photons and
high reflectivities, narrowband EPR\ transmitters are able to outperform all
the classical transmitters up to an extremely large bandwidth $M^{\ast}$. This
is confirmed by the numerical results of Fig.~\ref{colorbar}, proving the
robustness of the quantum effect $G>0$ in the presence of decoherence. Note
that we can neglect classical transmitters with extremely large bandwidths
(i.e., with $M>M^{\ast}$) since they are not meaningful for the model. In
fact, in a practical setting, the signal is an optical pulse with carrier
frequency $\nu$ high enough to completely resolve the target cell. This pulse
has frequency bandwidth $w\ll\nu$ and duration $\tau\simeq w^{-1}$. Assuming
an output detector with response time $\delta t\lesssim\tau$ and
\textquotedblleft reading time\textquotedblright\ $t>\tau$, the number of
modes which are excited is roughly $M=wt$. In other words, the bandwidth\ of
the signal $M$ is the product of its frequency bandwidth $w$ and the reading
time of the detector $t$. Now, the limit $M\rightarrow\infty$ corresponds to
$\delta t\rightarrow0$ (infinite detector resolution) or $t\rightarrow\infty
$\ (infinite reading time). As a result, transmitters with too large an $M$
can be discarded.

\begin{figure}[ptbh]
\par
\begin{center}
\includegraphics[width=0.46\textwidth] {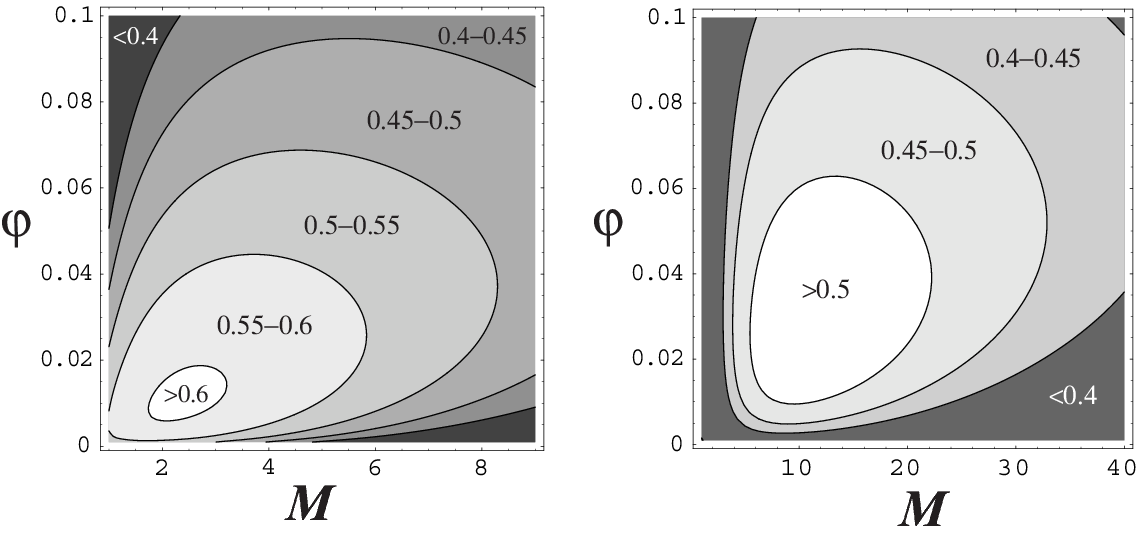}
\end{center}
\par
\vspace{-0.6cm}\caption{\textbf{Left. }$G$ optimized over $M$ and $\varphi$.
$G$ can be higher than $0.6$ bit per cell. Results are shown in the absence of
decoherence ($\varepsilon=\bar{n}=0$) considering $r_{0}=0.85$, $r_{1}=1$ and
$N=35$. \textbf{Right. }$G$ optimized over $M$ and $\varphi$. Results are
shown in the presence of decoherence ($\bar{n}=\varepsilon=10^{-5}$)
considering $r_{0}=0.85$, $r_{1}=0.95$, $N=100$ and $M^{\ast}=10^{6}$.}%
\label{BellPIC}%
\end{figure}

\textit{Sub-optimal receiver}.~The former results are valid assuming optimal
output detection. Here we show an explicit receiver design which is (i)\ easy
to construct and (ii)\ able to approximate the optimal results. This
sub-optimal receiver consists of a continuous variable Bell measurement (i.e.,
a balanced beam-splitter followed by two homodyne detectors) whose output is
classically processed by a suitable $\chi^{2}$-test with significance level
$\varphi$ (see Appendix for details). In this case the information gain $G$
can be optimized jointly over the signal bandwidth $M$ (i.e., the number of
input TMSV states)\ and the significance level of the output test $\varphi$.
As shown in Fig.~\ref{BellPIC}, the advantages of quantum reading are fully
preserved. 

\textit{Error correction}.~In our basic model of memory we store one bit of
information per cell. In an alternative model, information is stored in block
of cells by using error correcting codes, so that the readout of data is
practically flawless. In this configuration, we show that the error correction
overhead which is needed by EPR transmitters can be made very small. By
contrast, classical transmitters are useless since they may require more than
100 cells for retrieving a single bit of information in the regime of few
photons (see Appendix for details).

\textit{Conclusion}.~Quantum reading is able to work in the regime of few
photons. What does it imply? Using fewer photons means that we can reduce the
reading time of the cell, thus accessing higher data-transfer rates. This is a
theoretical prediction that can be checked with a pilot experiment (see
Appendix). Alternatively, we can fix the total reading time of the memory
while increasing its storage capacity (see Appendix for details). The chance
of using few photons leads to another interesting application: the safe
readout of photodegradable memories, such as dye-based optical disks or
photo-sensitive organic microfilms (e.g., containing confidential
information.) Here faint quantum light can retrieve the data safely, whereas
classical light could only be destructive.
More fundamentally, our results apply to the binary discrimination of
attenuator channels.

\textit{Acknowledgments}.~This work was partly supported by a Marie Curie
Action of the European Union. The author would like to thank the warm
hospitality of the W. M. Keck center for extreme quantum information
processing (xQIT) at the Massachusetts Institute of Technology. The author
also thanks S. L. Braunstein, S. Lloyd, J. H. Shapiro, A. Aspuru-Guzik, R.
Nair, R. Munoz-Tapia, C. Ottaviani, N. Datta, M. Mosheni, J. Oppenheim, C.
Weedbrook, C. Lupo, S. Mancini, P. Tombesi, G. Adesso, V. P. Belavkin, S.
Weigert, M. Paternostro, and G. Gribakin for comments and discussions.

\bigskip

\begin{center}
{\LARGE Appendix}{\Large  }

{\Large (Supplemental Material)}
\end{center}

In this appendix, we start by providing some introductory notions on bosonic
systems and\ Gaussian channels (Sec.~\ref{APP_INTRO}). Then, we explicitly
connect our memory model with the basic problem of quantum channel
discrimination (Sec.~\ref{APP_connection}). This connection is explicitly
shown in two paradigmatic cases: the basic \textquotedblleft pure-loss
model\textquotedblright, where the memory cell is represented by a
conditional-loss beam-splitter subject to vacuum noise, and the more general
\textquotedblleft thermal-loss model\textquotedblright, where thermal noise is
added in order to include non-trivial effects of decoherence. In the
subsequent sections we consider both these models and we provide the basic
tools for making the quantum-classical comparison. In particular, in
Sec.~\ref{APP_CLASS_DISCR}, we provide the fundamental bound for studying
classical transmitters, i.e., the \textquotedblleft classical discrimination
bound\textquotedblright. Then, in the next Sec.~\ref{APP_EPR_TRA}, we review
the mathematical tools for studying EPR\ transmitters. By using these
elements, we compare EPR\ and classical transmitters in
Sec.~\ref{APP_comparison}, where we explain in detail how to achieve the main
results presented in our Letter. In particular, for the pure-loss model, we
can provide analytical results: the \textquotedblleft threshold
energy\textquotedblright\ and \textquotedblleft ideal memory\textquotedblright%
\ theorems, which are proven in Sec.~\ref{APP_Pure_Loss}. Then, in
Sec.~\ref{APP_Bell}, we show how the advantages of quantum reading persist
when the optimal output detection is replaced by an easy-to-implement
sub-optimal receiver. This receiver consists of a continuous variable Bell
measurement followed by a suitable classical processing (one-tailed $\chi^{2}%
$-test). In Sec.~\ref{SEC_ECcodes}, we show an alternative model of memory
where information is stored in block of cells by means of error correcting
codes. Here, we compare the amount of error correction overhead which is
needed by EPR and classical transmitters in order to provide a
\textquotedblleft flawless\textquotedblright\ readout of logical data. This
alternative approach is useful for a future practical implementation of the
scheme. Sec.~\ref{Hamming_SEC} recalls standard results in classical error
correction. Finally, in Sec.~\ref{APP_discussion}, we discuss the implications
of the few-photon regime, where quantum reading outperforms every classical
strategy. Here we give general long-term predictions but also an estimation of
the current technological facilities in order to realize a pilot experiment.

\section{Introduction to bosonic systems\label{APP_INTRO}}

A bosonic system with $n$ modes is a quantum system described by a
tensor-product Hilbert space $\mathcal{H}^{\otimes n}$ and a vector of
quadrature operators
\begin{equation}
\mathbf{\hat{x}}^{T}:=(\hat{q}_{1},\hat{p}_{1},\ldots,\hat{q}_{n},\hat{p}%
_{n})~,
\end{equation}
satisfying the commutation relations \cite{notationCOMM}%
\begin{equation}
\lbrack\mathbf{\hat{x}},\mathbf{\hat{x}}^{T}]=2i\mathbf{\Omega}~,
\label{CommQUAD}%
\end{equation}
where $\mathbf{\Omega}$\ is a symplectic form in $\mathbb{R}^{2n}$, i.e.,%
\begin{equation}
\mathbf{\Omega}:=\bigoplus\limits_{i=1}^{n}\left(
\begin{array}
[c]{cc}%
0 & 1\\
-1 & 0
\end{array}
\right)  ~. \label{Symplectic_Form}%
\end{equation}
By definition a quantum state $\rho$ of a bosonic system is called
\textquotedblleft Gaussian\textquotedblright\ when its Wigner
phase-space representation is Gaussian
\cite{BraREV,BraREV2,GaussSTATES}. In such a case, the state is
completely described by the first and second statistical moments.
In other words, a Gaussian state $\rho$ of $n$ bosonic modes is
characterized by a displacement vector
\begin{equation}
\mathbf{\bar{x}}:=\mathrm{Tr}(\mathbf{\hat{x}}\rho)~,
\end{equation}
and a covariance matrix (CM)
\begin{equation}
\mathbf{V}:=\tfrac{1}{2}\mathrm{Tr}\left(  \left\{  \mathbf{\hat{x},\hat{x}%
}^{T}\right\}  \rho\right)  -\mathbf{\bar{x}\bar{x}}^{T}~,
\end{equation}
where $\{,\}$ denotes the anticommutator \cite{notationCOMM}. The CM\ is a
$2n\times2n$ real and symmetric matrix which must satisfy the uncertainty
principle \cite{SIMONprinc,Alex}%
\begin{equation}
\mathbf{V}+i\mathbf{\Omega}\geq0~. \label{unc_PRINC}%
\end{equation}
An important example of Gaussian state is the two-mode-squeezed vacuum (TMSV)
state for two bosonic modes $\{s,i\}$ (whose number-ket representation is
given in the Letter). This state has zero mean ($\mathbf{\bar{x}}=0$) and its
CM\ is given by%
\begin{equation}
\mathbf{V}=\left(
\begin{array}
[c]{cc}%
(2n_{S}+1)\mathbf{I} & 2\sqrt{n_{S}(n_{S}+1)}\mathbf{Z}\\
2\sqrt{n_{S}(n_{S}+1)}\mathbf{Z} & (2n_{S}+1)\mathbf{I}%
\end{array}
\right)  ~, \label{CM_TMSV}%
\end{equation}
where $n_{S}\geq0$ and%
\begin{equation}
\mathbf{I}=\left(
\begin{array}
[c]{cc}%
1 & 0\\
0 & 1
\end{array}
\right)  ~,~\mathbf{Z}=\left(
\begin{array}
[c]{cc}%
1 & 0\\
0 & -1
\end{array}
\right)  ~.
\end{equation}
Here $n_{S}$ represents the mean number of thermal photons which are present
in each mode. This number is connected with the \textquotedblleft two-mode
squeezing parameter\textquotedblright\ \cite{QObook}\ by the relation%
\begin{equation}
n_{S}=\sinh^{2}\xi~.
\end{equation}
Parameter $\xi$ completely characterizes the state (therefore denoted by
$\left\vert \xi\right\rangle \left\langle \xi\right\vert $) and quantifies the
entanglement between the two modes $s$ and $i$ (being an entanglement monotone
for this class of states).

Another important example of Gaussian state is the (multimode) coherent state.
For $K$ modes this is given by%
\begin{equation}
\left\vert \boldsymbol{\alpha}\right\rangle \left\langle \boldsymbol{\alpha
}\right\vert =\bigotimes\limits_{i=1}^{K}\left\vert \alpha_{i}\right\rangle
\left\langle \alpha_{i}\right\vert ~,
\end{equation}
where $\boldsymbol{\alpha}:=(\alpha_{1},\cdots,\alpha_{K})$ is a row-vector of
amplitudes $\alpha_{i}=(q_{i}+ip_{i})/2$. This state has CM equal to the
identity and, therefore, is completely characterized by its displacement
vector $\mathbf{\bar{x}}$, which is determined by $\boldsymbol{\alpha}$.
Starting from the coherent states, we can characterize all the possible states
of a bosonic system by introducing the P-representation. In fact, a generic
state $\rho$ of $K$ bosonic modes can be decomposed as%
\begin{equation}
\rho=\int d^{2K}\boldsymbol{\alpha~}\mathcal{P}(\boldsymbol{\alpha
})~\left\vert \boldsymbol{\alpha}\right\rangle \left\langle \boldsymbol{\alpha
}\right\vert ~, \label{PdecomP}%
\end{equation}
where $\mathcal{P}(\boldsymbol{\alpha})$ is a quasi-probability distribution,
i.e., normalized to $1$ but generally non-positive (here we use the compact
notation $\int d^{2K}\boldsymbol{\alpha}:=\int d^{2}\alpha_{1}\cdots\int
d^{2}\alpha_{K}$). By definition, a bosonic state $\rho$ is called
\textquotedblleft classical\textquotedblright\ if $\mathcal{P}%
(\boldsymbol{\alpha})$ is positive, i.e., $\mathcal{P}$ is a proper
probability distribution. By contrast, the state is called \textquotedblleft
non-classical\textquotedblright\ when $\mathcal{P}(\boldsymbol{\alpha})$ is
non-positive. It is clear that a classical state is separable, since
Eq.~(\ref{PdecomP}) with $\mathcal{P}$ positive corresponds to a state
preparation via local operations and classical communications (LOCCs). The
borderline between classical and non-classical states is given by the coherent
states, for which $\mathcal{P}$ is a delta function. Note also that the
classical states are generally non-Gaussian. In fact, one can have a classical
state given by a \textit{finite ensemble} of coherent states (whose
$\mathcal{P}$-representation corresponds to a sum of Dirac-deltas).

In general we use the formalism $T(M,L,\rho)$ to denote a \textquotedblleft
bipartite transmitter\textquotedblright\ or, more simply, a \textquotedblleft
transmitter\textquotedblright. This is a compact notation for characterizing
simultaneously a bipartite bosonic system and its state. More specifically,
$M$ and $L$, represent the number of modes present in two partitions of the
system, called signal (sub)system $S$\ and idler (sub)system $I$. Then, given
this bipartite system, $\rho$ represents the corresponding global state. This
notation is very useful when both system and state are variable. By
definition, a \textquotedblleft transmitter\textquotedblright\ $T(M,L,\rho)$
is classical (non-classical) when its state $\rho$\ is classical
(non-classical), i.e., $T_{c}=T(M,L,\rho_{c})$ and $T_{nc}=T(M,L,\rho_{nc})$.
In the comparison between different transmitters, one has to fix some of the
parameters of the signal (sub)system $S$. The basic parameters of $S$ are the
total number of signal modes $M$ (signal bandwidth) and the mean total number
of photons $N$ (signal energy).

\subsection{Gaussian channels}

A Gaussian channel is a completely positive trace-preserving (CPTP) map
\begin{equation}
\mathcal{E}:\rho\rightarrow\sigma:=\mathcal{E}(\rho)
\end{equation}
which transforms Gaussian states into Gaussian states. In particular, a
one-mode Gaussian channel (i.e., acting on one-mode bosonic states) can be
easily described in terms of the statistical moments $\{\mathbf{\bar{x}%
},\mathbf{V}\}$. This channel corresponds to the transformation
\cite{HolevoCAN,CharacATT,AlexJens}%
\begin{equation}
\mathbf{V}\rightarrow\mathbf{KVK}^{T}+\mathbf{N~,~\bar{x}\rightarrow K\bar
{x}+d~,} \label{Gauss_TRASF}%
\end{equation}
where $\mathbf{d}$ is an $\mathbb{R}^{2}$-vector, while $\mathbf{K}$ and
$\mathbf{N}$ are $2\times2$ real matrices, with $\mathbf{N}^{T}=\mathbf{N}%
\geq0$ and
\begin{equation}
\det\mathbf{N}\geq\left(  \det\mathbf{K}-1\right)  ^{2}~. \label{bona_fide_N}%
\end{equation}
Important examples of one-mode Gaussian channels are the following:

\begin{description}
\item[(i)] The \textquotedblleft attenuator channel\textquotedblright%
\ $\mathcal{E}(r,\bar{n})$ with loss $r\in\lbrack0,1]$ and thermal noise
$\bar{n}\geq0$. This channel can be represented by a beam splitter with
reflectivity $r$ which mixes the input mode with a bath mode prepared in a
thermal state with $\bar{n}$ mean photons. This channel implements the
transformation of Eq.~(\ref{Gauss_TRASF}) with%
\begin{equation}
\mathbf{K}=\sqrt{r}\mathbf{I}~,~\mathbf{N}=(1-r)(2\bar{n}+1)\mathbf{I}~,
\label{KN_1}%
\end{equation}
and $\mathbf{d}=0$. In particular, when the thermal noise $\bar{n}$\ is
negligible, the attenuator channel $\mathcal{E}(r):=\mathcal{E}(r,0)$ is
represented by a beam splitter with reflectivity $r$\ and a vacuum bath mode.

\item[(ii)] The thermal channel $\mathcal{N}(\varepsilon)$\ adding Gaussian
noise with variance $\varepsilon\geq0$. This channel implements the
transformation of Eq.~(\ref{Gauss_TRASF}) with%
\begin{equation}
\mathbf{K}=\mathbf{I}~,~\mathbf{N}=\varepsilon\mathbf{I}~, \label{KN_2}%
\end{equation}
and $\mathbf{d}=0$. This kind of channel is suitable to describe the effects
of decoherence when the loss is negligible. This is a typical situation within
optical apparatuses which are small in size (as is the case of our memory reader).
\end{description}

\noindent It is clear that, for every pair of Gaussian channels, $\mathcal{E}
$ and $\mathcal{E}^{\prime}$, acting on the same state space, their
composition $\mathcal{E}\circ\mathcal{E}^{\prime}$ is a also Gaussian channel.
If $\mathcal{E}$ and $\mathcal{E}^{\prime}$ are Gaussian channels acting on
two different spaces, their tensor product $\mathcal{E}\otimes\mathcal{E}%
^{\prime}$ is also Gaussian. Hereafter we call \textquotedblleft bipartite
Gaussian channel\textquotedblright\ a Gaussian channel which is in the
tensor-product form $\mathcal{E}\otimes\mathcal{E}^{\prime}$. The action of a
bipartite Gaussian channel on a two-mode bosonic state is very simple in terms
of its second statistical moments. In fact, let us consider two bosonic modes,
$A$ and $B$, in a state $\rho_{AB}$ with generic CM%
\begin{equation}
\mathbf{V}=\left(
\begin{array}
[c]{cc}%
\mathbf{A} & \mathbf{C}\\
\mathbf{C}^{T} & \mathbf{B}%
\end{array}
\right)  ~,
\end{equation}
where $\mathbf{A},$ $\mathbf{B}$\ and $\mathbf{C}$\ are $2\times2$ real
matrices. At the output of a bipartite Gaussian channel $\mathcal{E}^{A\otimes
B}:=\mathcal{E}_{A}\otimes\mathcal{E}_{B}$, we have the CM%
\begin{equation}
\mathbf{V}_{out}=\left(
\begin{array}
[c]{cc}%
\mathbf{K}_{A}\mathbf{AK}_{A}^{T}+\mathbf{N}_{A} & \mathbf{K}_{A}%
\mathbf{CK}_{B}^{T}\\
\mathbf{K}_{B}\mathbf{C}^{T}\mathbf{K}_{A}^{T} & \mathbf{K}_{B}\mathbf{BK}%
_{B}^{T}+\mathbf{N}_{B}%
\end{array}
\right)  ~, \label{V_out_bip_CH}%
\end{equation}
where the matrices $(\mathbf{K}_{A},\mathbf{N}_{A})$ refer to $\mathcal{E}%
_{A}$ (acting on the first mode), while $(\mathbf{K}_{B},\mathbf{N}_{B})$
refer to $\mathcal{E}_{B}$ (acting on the second mode).

\bigskip

\textbf{Proof.}~~Both the channels $\mathcal{E}_{A}\otimes\mathcal{E}_{B}$ are
dilated, so that we have%
\begin{align}
\mathcal{E}^{A\otimes B}(\rho_{AB})  &  =\mathrm{Tr}_{A^{\prime}B^{\prime}%
}\left[  (U_{A^{\prime}A}\otimes U_{BB^{\prime}})\times\right. \nonumber\\
&  \left.  (\left\vert 0\right\rangle \left\langle 0\right\vert _{A^{\prime}%
}\otimes\rho_{AB}\otimes\left\vert 0\right\rangle \left\langle 0\right\vert
_{B^{\prime}})(U_{A^{\prime}A}^{\dagger}\otimes U_{BB^{\prime}}^{\dagger
})\right]  ,\nonumber\\
&  \label{double_dil}%
\end{align}
where $A^{\prime}$\ and $B^{\prime}$ are supplementary sets of bosonic modes
prepared in vacua, while $U_{A^{\prime}A}$ and $U_{BB^{\prime}}$ are Gaussian
unitaries. According to Eq.~(\ref{double_dil}), the input CM $\mathbf{V}$ is
subject to three subsequent operations. First, it must be dilated to
$\mathbf{I}_{A^{\prime}}\oplus\mathbf{V}\oplus\mathbf{I}_{B^{\prime}}$, where
$\mathbf{I}_{A^{\prime}}$ and $\mathbf{I}_{B^{\prime}}$ are identity matrices
of suitable dimensions. Then, it is transformed via congruence by
$\mathbf{S}_{A^{\prime}A}\oplus\mathbf{S}_{BB^{\prime}}$, where $\mathbf{S}%
_{A^{\prime}A}$\ and $\mathbf{S}_{B^{\prime}B}$\ are the symplectic matrices
corresponding to $U_{A^{\prime}A}$ and $U_{BB^{\prime}}$, respectively.
Finally, raws and columns corresponding to $A^{\prime}$\ and $B^{\prime}$ are
elided (trace). By setting%
\begin{equation}
\mathbf{S}_{A^{\prime}A}=\left(
\begin{array}
[c]{cc}%
\mathbf{a} & \mathbf{c}\\
\mathbf{d} & \mathbf{b}%
\end{array}
\right)  ~,~\mathbf{S}_{BB^{\prime}}=\left(
\begin{array}
[c]{cc}%
\mathbf{e} & \mathbf{g}\\
\mathbf{h} & \mathbf{f}%
\end{array}
\right)  ~,
\end{equation}
we get%
\begin{equation}
\mathbf{V}_{out}=\left(
\begin{array}
[c]{cc}%
\mathbf{bAb}^{T}+\mathbf{dd}^{T} & \mathbf{bCe}^{T}\\
\mathbf{eC}^{T}\mathbf{b}^{T} & \mathbf{eBe}^{T}+\mathbf{gg}^{T}%
\end{array}
\right)  ~.
\end{equation}
Now, by setting
\begin{equation}
\mathbf{K}_{A}:=\mathbf{b~},~\mathbf{N}_{A}:=\mathbf{dd}^{T}~,
\end{equation}
and%
\begin{equation}
\mathbf{K}_{B}:=\mathbf{e~},~\mathbf{N}_{B}:=\mathbf{gg}^{T}~,
\end{equation}
we get exactly
Eq.~(\ref{V_out_bip_CH}).~$\blacksquare$\begin{figure}[ptbh]
\vspace{-0.0cm}
\par
\begin{center}
\includegraphics[width=0.48\textwidth] {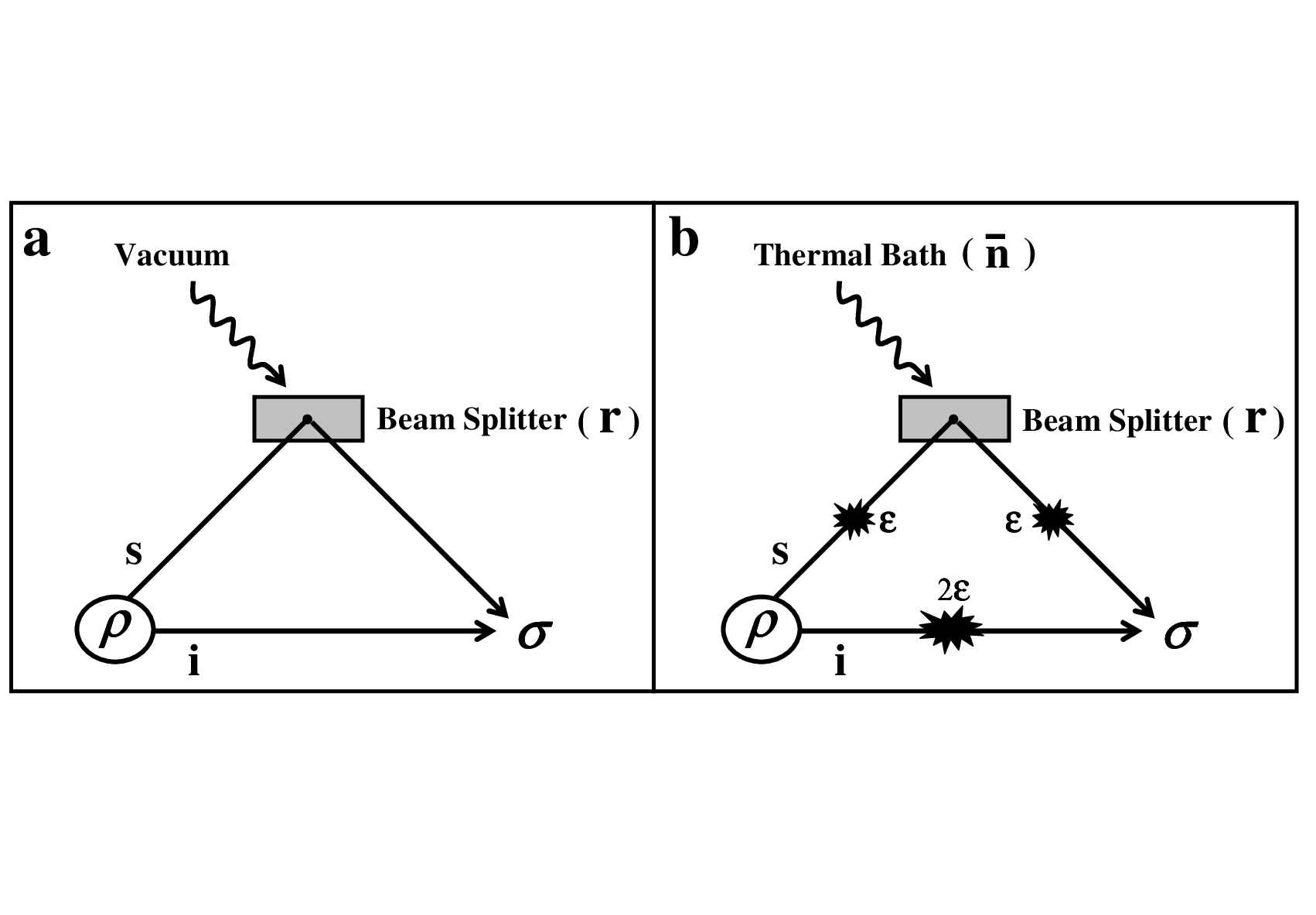}
\end{center}
\par
\vspace{-1.5cm}\caption{\textbf{Inset~a.} Pure-loss model, implementing the
bipartite Gaussian channel $\mathcal{G}=\mathcal{E}\otimes\mathcal{I}$.
\textbf{Inset~b.} Thermal-loss model, implementing the bipartite Gaussian
channel $\mathcal{G}=\mathcal{S}\otimes\mathcal{D}$ (see text).}%
\label{BS}%
\end{figure}

\section{Memory model and Gaussian channel
discrimination\label{APP_connection}}

\begin{figure}[ptbh]
\vspace{-0.4cm}
\par
\begin{center}
\includegraphics[width=0.47\textwidth] {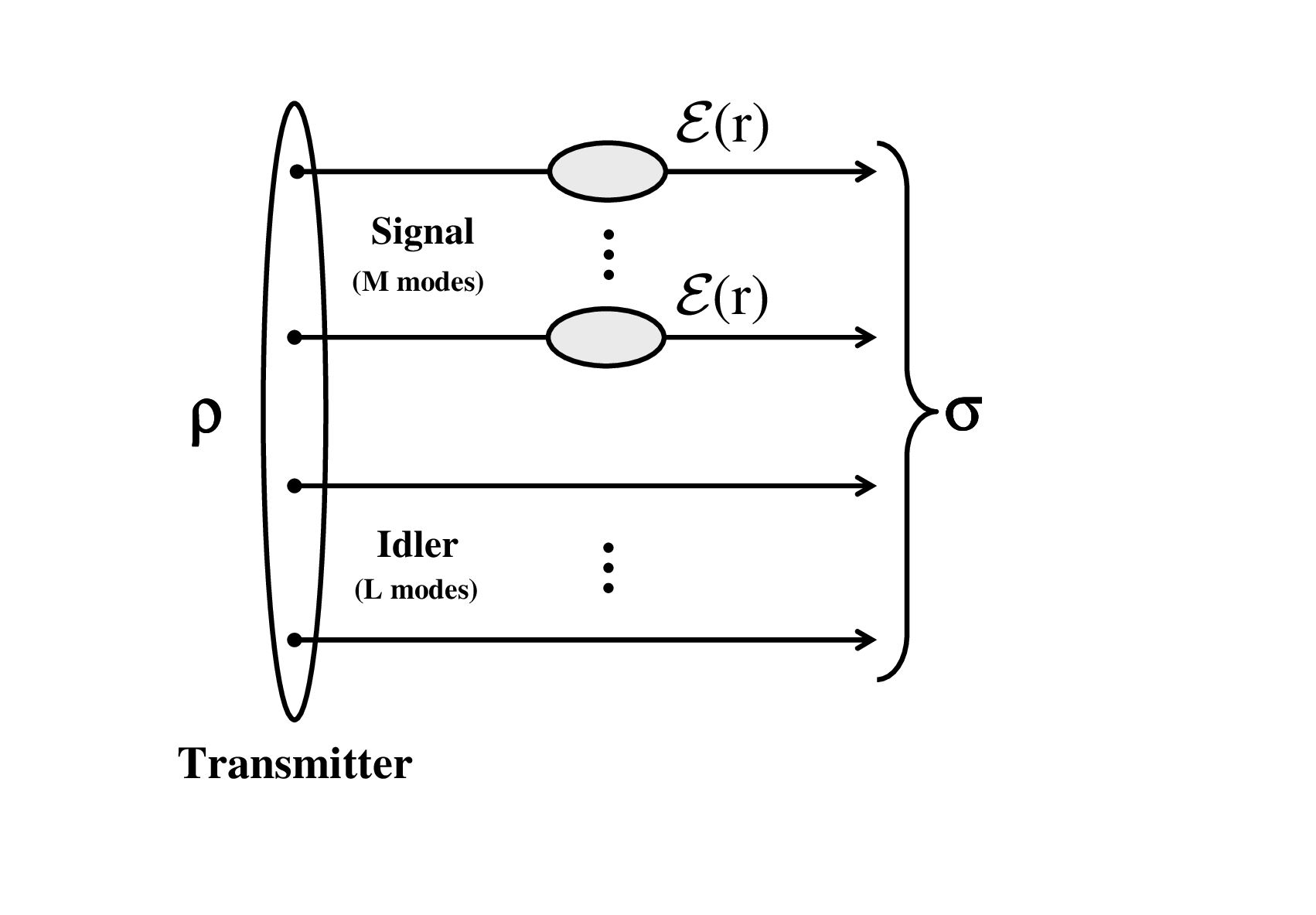}
\end{center}
\par
\vspace{-1.0cm}\caption{Visual representation of Eq.~(\ref{Mless}).}%
\label{Scenario1}%
\end{figure}

Let us consider the beam-splitter scheme of Fig.~\ref{BS}a, where an input
state $\rho$ of two modes (\textquotedblleft signal mode\textquotedblright%
\ $s$\ and \textquotedblleft idler mode\textquotedblright\ $i$) is transformed
into an output state $\sigma$. Hereafter we call this scheme the
\textquotedblleft pure-loss model\textquotedblright. The corresponding
transformation is a bipartite Gaussian channel
\begin{equation}
\mathcal{G}=\mathcal{E}(r)\otimes\mathcal{I},
\end{equation}
where the attenuator channel $\mathcal{E}(r)$ acts on the signal mode, and the
identity channel $\mathcal{I}$ acts on the idler mode. By construction,
$\mathcal{G}$ is a memoryless channel. Thus, if we consider a transmitter
$T(M,L,\rho)$, i.e., $M$\ signal modes $s\in S$ and $L$ idler modes $i\in I$
in a multimode state $\rho$, the input state is transformed into the output
state
\begin{equation}
\sigma=\mathcal{G}^{M,L}(\rho)~,
\end{equation}
where%
\begin{equation}
\mathcal{G}^{M,L}:=\mathcal{E}(r)^{\otimes M}\otimes\mathcal{I}^{\otimes L}~.
\label{Mless}%
\end{equation}
This is schematically depicted in Fig.~\ref{Scenario1}. \begin{figure}[ptbh]
\vspace{-0.0cm}
\par
\begin{center}
\includegraphics[width=0.47\textwidth] {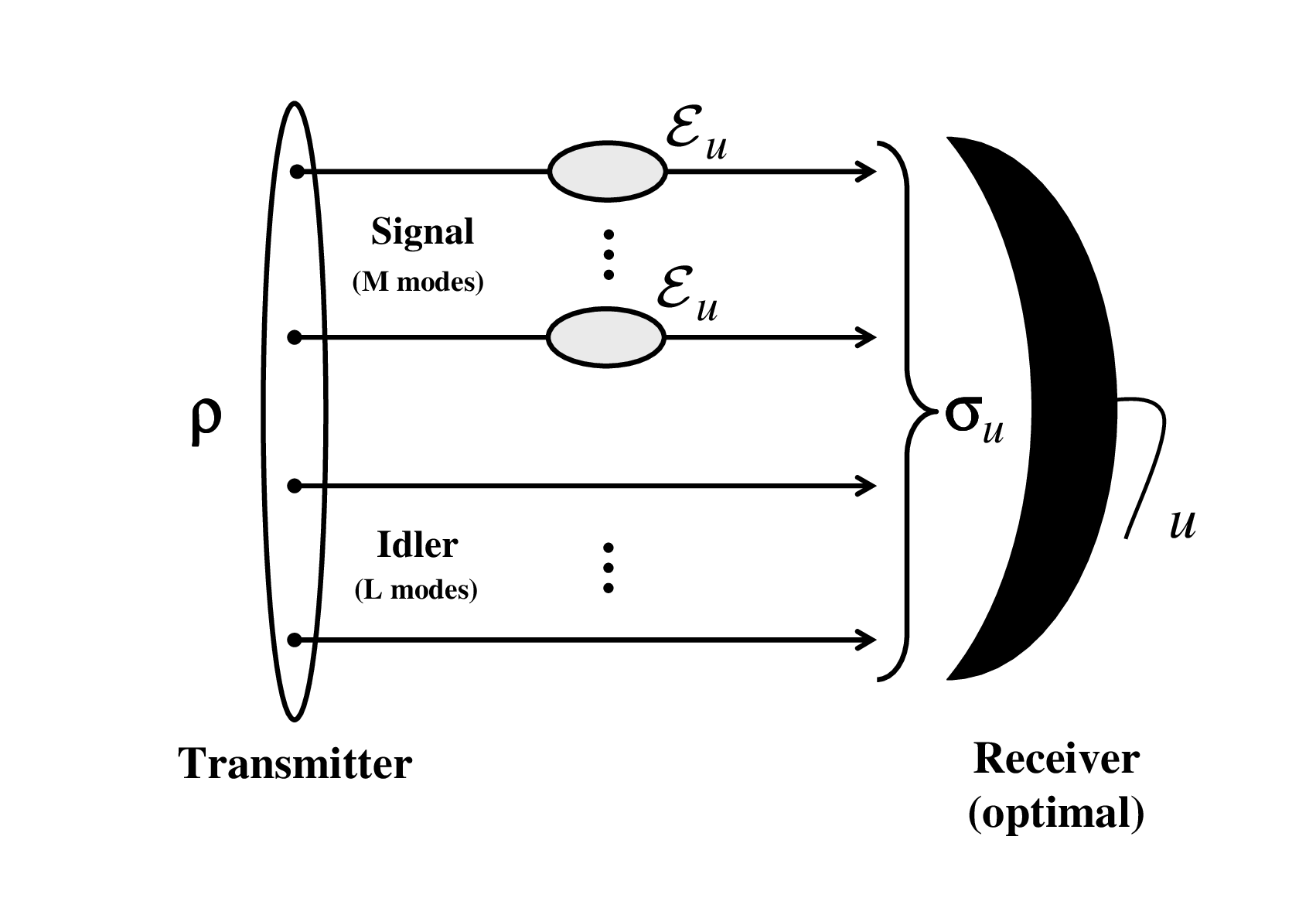}
\end{center}
\par
\vspace{-0.9cm}\caption{Channel discrimination problem (pure-loss model).}%
\label{Scenario2}%
\end{figure}

Now, encoding a logical bit $u=0,1$ in the reflectivity of the beam splitter
$r=r_{u}$ corresponds to encoding the bit into the conditional attenuator
channel $\mathcal{E}_{u}:=\mathcal{E}(r_{u})$. Thus, reading a memory cell
using an input transmitter $T(M,L,\rho)$, and an output receiver, corresponds
to the channel discrimination problem depicted in Fig.~\ref{Scenario2}. The
error probability in the decoding of the logical bit, i.e.,
\begin{equation}
P_{err}=\frac{P(u=0|u=1)+P(u=1|u=0)}{2}%
\end{equation}
corresponds to the error probability of discriminating between the two
equiprobable channels $\mathcal{E}_{0}$ and $\mathcal{E}_{1}$. The
minimization of this error probability involves the optimization of both input
and output. For a \textit{fixed} input $T(M,L,\rho)$, we have two equiprobable
output states, $\sigma_{0}$ and $\sigma_{1}$, with%
\begin{equation}
\sigma_{u}=\mathcal{G}_{u}^{M,L}(\rho)=\left(  \mathcal{E}_{u}^{\otimes
M}\otimes\mathcal{I}^{\otimes L}\right)  (\rho)~.
\end{equation}
The optimal measurement for their discrimination is given by the dichotomic
POVM \cite{Helstrom}%
\begin{equation}
E_{0}=\Pi(\gamma_{+})~,~E_{1}=I-\Pi(\gamma_{+})~, \label{optimal_POVM}%
\end{equation}
where $\Pi(\gamma_{+})$ is the projector onto the positive part $\gamma_{+}$
of the Helstrom matrix $\gamma:=\sigma_{0}-\sigma_{1}$ \cite{Ppart}. Using
this measurement, the two output states, $\sigma_{0}$ and $\sigma_{1}$, are
discriminated with a \textit{minimum} error probability which is provided by
the Helstrom bound, i.e.,
\begin{equation}
P_{err}(\sigma_{0}\neq\sigma_{1})=\frac{1-D(\sigma_{0},\sigma_{1})}{2}~,
\end{equation}
where $D(\sigma_{0},\sigma_{1})$ is the trace distance between $\sigma_{0} $
and $\sigma_{1}$ \cite{Helstrom}. In particular,%
\begin{equation}
D(\sigma_{0},\sigma_{1}):=\frac{1}{2}\left\Vert \sigma_{0}-\sigma
_{1}\right\Vert _{1}~,
\end{equation}
where $\left\Vert \gamma\right\Vert _{1}:=\mathrm{Tr}\sqrt{\gamma^{\dagger
}\gamma}$ is the trace norm. Thus, given two equiprobable channels
$\{\mathcal{E}_{0},\mathcal{E}_{1}\}$ and a \textit{fixed} input transmitter
$T$, we can always assume an optimal output detection. Under this assumption
(optimal detection), the channel discrimination problem $\mathcal{E}_{0}%
\neq\mathcal{E}_{1}$ with fixed transmitter $T=T(M,L,\rho)$ is fully
characterized by the conditional error probability%
\begin{align}
P_{err}(\mathcal{E}_{0}  &  \neq\mathcal{E}_{1}|T):=\left[  \frac
{1-D(\sigma_{0},\sigma_{1})}{2}\right]  _{\sigma_{u}=\left(  \mathcal{E}%
_{u}^{\otimes M}\otimes\mathcal{I}^{\otimes L}\right)  (\rho)}.\nonumber\\
&
\end{align}
Now, the minimal error probability for discriminating $\mathcal{E}_{0}$ and
$\mathcal{E}_{1}$ is given by optimizing the previous quantity over all the
input transmitters, i.e.,%
\begin{equation}
P_{err}(\mathcal{E}_{0}\neq\mathcal{E}_{1})=\min_{T}P_{err}(\mathcal{E}%
_{0}\neq\mathcal{E}_{1}|T)~. \label{OptimizationCH}%
\end{equation}
In general, this error probability tends to zero in the limit of infinite
energy $N\rightarrow+\infty$ (this happens whenever $\mathcal{E}_{0}%
\neq\mathcal{E}_{1}$). For this reason, in order to consider a non-trivial
quantity, we must fix the signal-energy $N$ in the previous minimization. Let
us denote by $T|N$ the class of transmitters signalling $N$ photons. Then, we
can define the conditional error probability%
\begin{equation}
P_{err}(\mathcal{E}_{0}\neq\mathcal{E}_{1}|N)=\min_{T|N}P_{err}(\mathcal{E}%
_{0}\neq\mathcal{E}_{1}|T)~. \label{Optimization2}%
\end{equation}
It is an \textit{open question} to find the optimal transmitter within the
class $T|N$, i.e., realizing the minimization of Eq.~(\ref{Optimization2}).
The central idea of our Letter is a direct consequence of this open question.
In fact, strictly connected with this question, there is another fundamental
problem, whose resolution sensibly narrows the search for optimal
transmitters: within the conditional class $T|N$, can we find a non-classical
transmitter which outperforms any classical transmitter? More exactly, given
two attenuator channels $\{\mathcal{E}_{0},\mathcal{E}_{1}\}$ and fixed
signal-energy $N$, can we find any $T_{nc}$ such that
\begin{equation}
P_{err}(\mathcal{E}_{0}\neq\mathcal{E}_{1}|T_{nc})<P_{err}(\mathcal{E}_{0}%
\neq\mathcal{E}_{1}|T_{c})~, \label{Q1}%
\end{equation}
for every $T_{c}$? In our Letter we solve this problem \cite{Noteshort}. In
particular, we show this is possible for an important physical regime, i.e.,
for channels $\{\mathcal{E}_{0},\mathcal{E}_{1}\}$ corresponding to
high-reflectivities (as typical of optical memories) and signals with few
photons (as typical of entanglement sources).

\subsection{Introducing thermal noise}

\begin{figure}[ptbh]
\vspace{-0.3cm}
\par
\begin{center}
\includegraphics[width=0.47\textwidth] {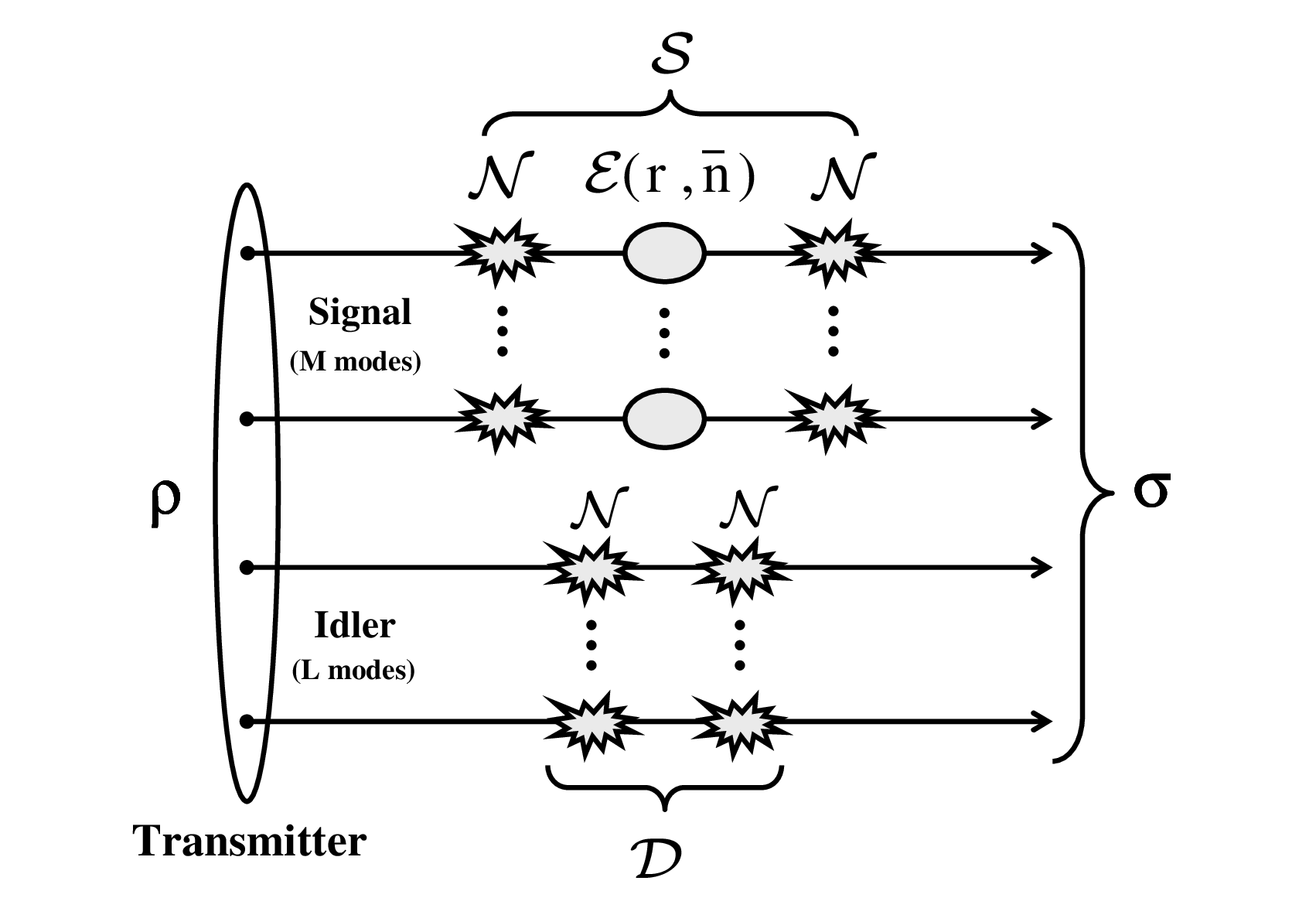}
\end{center}
\par
\vspace{-0.7cm}\caption{Visual representation of Eq.~(\ref{GML2}).}%
\label{Scenario3}%
\end{figure}

\begin{figure}[ptbh]
\vspace{-0.9cm}
\par
\begin{center}
\includegraphics[width=0.47\textwidth] {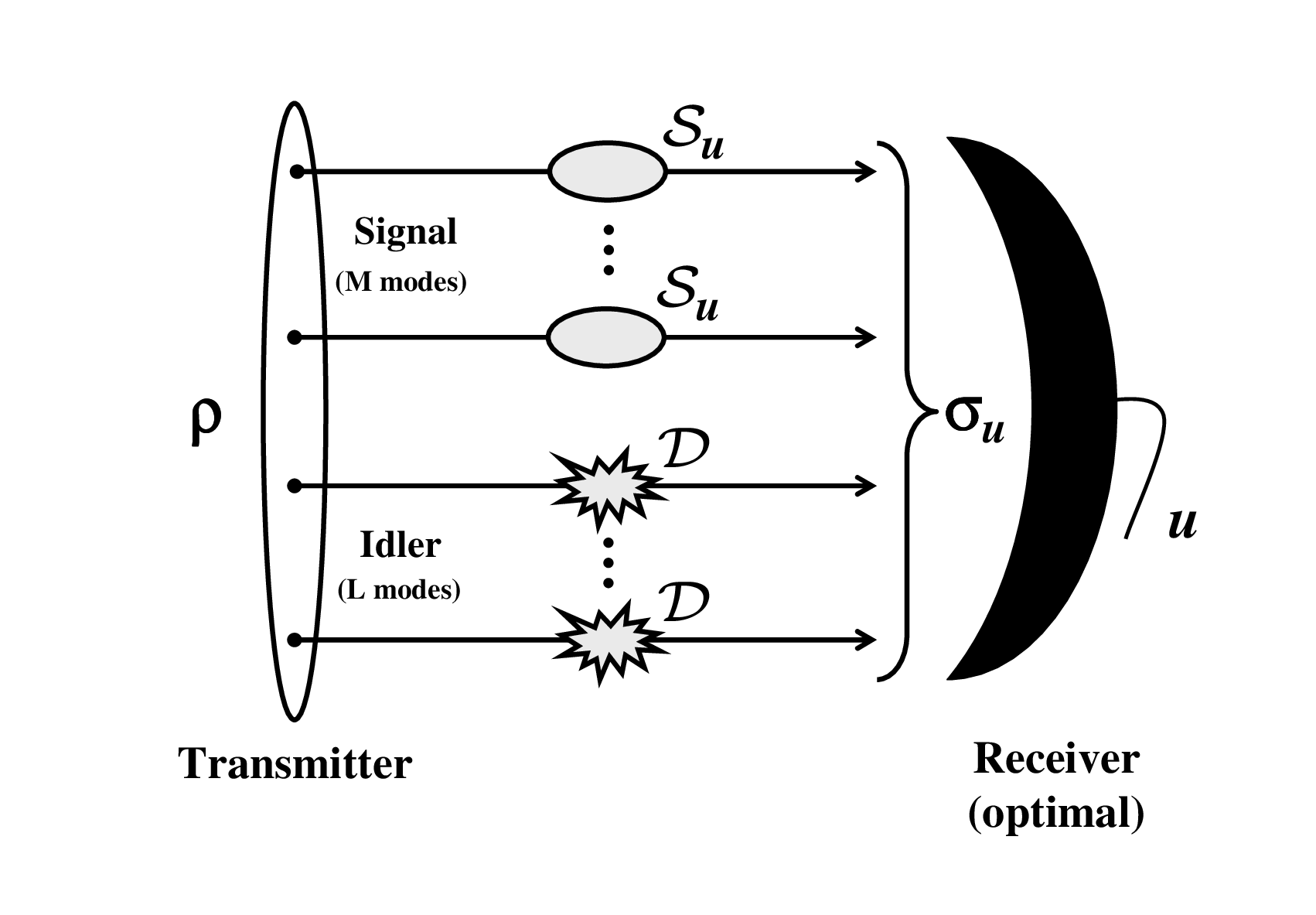}
\end{center}
\par
\vspace{-1.1cm}\caption{Channel discrimination problem (thermal-loss model).}%
\label{Scenario4}%
\end{figure}

In general, the pure-loss model of Fig.~\ref{BS}a represents a very good
description in the optical range if we assume the use of a good reading
apparatus. To complete the analysis and show the robustness of the model with
respect to decoherence, we also consider the presence of thermal noise, as
explicitly stated in the Letter. The noisy scenario is the one depicted in
Fig.~\ref{BS}b, that we call the \textquotedblleft thermal-loss
model\textquotedblright. This corresponds to the bipartite Gaussian channel
\begin{equation}
\mathcal{G}=\mathcal{S}\otimes\mathcal{D}, \label{bip_TL_model}%
\end{equation}
where
\begin{equation}
\mathcal{S}:=\mathcal{N}(\varepsilon)\circ\mathcal{E}(r,\bar{n})\circ
\mathcal{N}(\varepsilon)=\mathcal{S}(r,\bar{n},\varepsilon)
\end{equation}
acts on the signal mode, and
\begin{equation}
\mathcal{D}=\mathcal{N}(\varepsilon)\circ\mathcal{N}(\varepsilon
)=\mathcal{N}(2\varepsilon)
\end{equation}
acts on the idler mode. Exactly as before, $\mathcal{G}$ represents a
memoryless channel. As a result, the state $\rho$ of an input transmitter
$T(M,L,\rho)$ is transformed into an output state $\sigma=\mathcal{G}%
^{M,L}(\rho)$, where%
\begin{equation}
\mathcal{G}^{M,L}:=\mathcal{S}^{\otimes M}\otimes\mathcal{D}^{\otimes L}~.
\label{GML2}%
\end{equation}
This transformation is depicted in Fig.~\ref{Scenario3}. Now, encoding a
logical bit $u=0,1$ in the reflectivity of the beam splitter $r=r_{u}$
corresponds to encoding the bit into the Gaussian channel $\mathcal{S}%
_{u}:=\mathcal{S}(r_{u},\bar{n},\varepsilon)$. It follows that the readout of
our memory cell corresponds to the channel discrimination problem depicted in
Fig.~\ref{Scenario4}. 

It is clear that the present thermal scenario (Fig.~\ref{Scenario4}) can be
formally achieved by the previous non-thermal one (Fig.~\ref{Scenario2}) via
the replacements%
\begin{equation}
\mathcal{E}_{u}\rightarrow\mathcal{S}_{u}~,~\mathcal{I}\rightarrow
\mathcal{D}~.
\end{equation}
In this case, the idlers are subject to a non-trivial decoherence channel
$\mathcal{D}$, which plays an active role in the discrimination problem. In
general, this problem can be formulated as the discrimination of two
\textit{bipartite} channels, $\mathcal{G}_{0}=\mathcal{S}_{0}\otimes
\mathcal{D}$ and $\mathcal{G}_{0}=\mathcal{S}_{1}\otimes\mathcal{D}$, using a
transmitter $T=T(M,L,\rho)$ and an optimal receiver. The corresponding error
probability is defined by%
\begin{align}
P_{err}(\mathcal{G}_{0}  &  \neq\mathcal{G}_{1}|T):=\left[  \frac
{1-D(\sigma_{0},\sigma_{1})}{2}\right]  _{\sigma_{u}=\left(  \mathcal{S}%
_{u}^{\otimes M}\otimes\mathcal{D}^{\otimes L}\right)  (\rho)}.\nonumber\\
&
\end{align}
Clearly this problem is more difficult to study. For this reason, our basic
question becomes the following: given two bipartite channels $\{\mathcal{G}%
_{0},\mathcal{G}_{1}\}$ and fixed signal-energy $N$, can we find any $T_{nc}$
such that
\begin{equation}
P_{err}(\mathcal{G}_{0}\neq\mathcal{G}_{1}|T_{nc})<P_{err}(\mathcal{G}_{0}%
\neq\mathcal{G}_{1}|T_{c})~, \label{Q2}%
\end{equation}
for a \textit{suitable large class} of $T_{c}$? As stated in the Letter (and
explicitly shown afterwards), we can give a positive answer to this question
too. This is possible by excluding broadband classical transmitters which are
not meaningful for the model (see the Letter for a physical discussion).

In the following Secs.~\ref{APP_CLASS_DISCR} and~\ref{APP_EPR_TRA} we
introduce the basic tools that we need to compare classical and non-classical
transmitters, in both the models: pure-loss and thermal-loss. In particular,
Sec.~\ref{APP_CLASS_DISCR} provides one of the central results of the work:
the \textquotedblleft classical discrimination bound\textquotedblright, which
enables us to bound all the classical transmitters. Then,
Sec.~\ref{APP_EPR_TRA} concerns the study of the non-classical EPR
transmitter. At that point we have all the elements to make the comparison,
i.e., replying to the questions in Eqs.~(\ref{Q1}) and~(\ref{Q2}). This
comparison is thoroughly discussed in Sec.~\ref{APP_comparison}. In
particular, for the pure-loss model, we can also derive analytical results, as
shown in Sec.~\ref{APP_Pure_Loss}. These results are the \textquotedblleft
threshold energy\textquotedblright\ theorem and the \textquotedblleft ideal
memory\textquotedblright\ theorem.

\section{Classical discrimination bound\label{APP_CLASS_DISCR}}

\subsection{Thermal-loss model}

Let us consider the discrimination of two bipartite Gaussian channels
\begin{equation}
\mathcal{G}_{0}=\mathcal{S}_{0}\otimes\mathcal{D}=\mathcal{G}(r_{0},\bar
{n},\varepsilon)~,
\end{equation}
and%
\begin{equation}
\mathcal{G}_{1}=\mathcal{S}_{1}\otimes\mathcal{D}=\mathcal{G}(r_{1},\bar
{n},\varepsilon)~,
\end{equation}
by using a bipartite transmitter $T=T(M,L,\rho)$ which signals $N$ photons.
The global input state can be decomposed using the following bipartite
$\mathcal{P}$-representation \cite{Prepres,vectNOT}%

\begin{equation}
\rho=\int d^{2M}\boldsymbol{\alpha}\int d^{2L}\boldsymbol{\beta~}%
\mathcal{P}(\boldsymbol{\alpha},\boldsymbol{\beta})\left\vert
\boldsymbol{\alpha}\right\rangle _{S}\left\langle \boldsymbol{\alpha
}\right\vert \otimes\left\vert \boldsymbol{\beta}\right\rangle _{I}%
\left\langle \boldsymbol{\beta}\right\vert ~, \label{Prepres}%
\end{equation}
where $\boldsymbol{\alpha}:=(\alpha_{1},\cdots,\alpha_{M})$ is a vector of
amplitudes for the signal $M$-mode coherent-state%
\begin{equation}
\left\vert \boldsymbol{\alpha}\right\rangle _{S}\left\langle
\boldsymbol{\alpha}\right\vert =\bigotimes\limits_{k=1}^{M}\left\vert
\alpha_{k}\right\rangle \left\langle \alpha_{k}\right\vert ~,
\end{equation}
and $\boldsymbol{\beta}:=(\beta_{1},\cdots,\beta_{L})$ is a vector of
amplitudes for the idler $L$-mode coherent-state
\begin{equation}
\left\vert \boldsymbol{\beta}\right\rangle _{I}\left\langle \boldsymbol{\beta
}\right\vert =\bigotimes\limits_{k=1}^{L}\left\vert \beta_{k}\right\rangle
\left\langle \beta_{k}\right\vert ~.
\end{equation}
The reduced state for the $M$ signal modes is given by%
\begin{equation}
\rho_{S}=\int d^{2M}\boldsymbol{\alpha}~\mathcal{P}(\boldsymbol{\alpha
})~\left\vert \boldsymbol{\alpha}\right\rangle _{S}\left\langle
\boldsymbol{\alpha}\right\vert ~, \label{Signal_GEN}%
\end{equation}
where%
\begin{equation}
\mathcal{P}(\boldsymbol{\alpha}):=\int d^{2L}\boldsymbol{\beta}~\mathcal{P}%
(\boldsymbol{\alpha},\boldsymbol{\beta})~. \label{Q_alpha_distribution}%
\end{equation}
The mean total number of photons in the $M$-mode signal system can be written
as%
\begin{equation}
N=\int d^{2M}\boldsymbol{\alpha}~\mathcal{P}(\boldsymbol{\alpha})~E~,
\end{equation}
where%
\begin{equation}
E:=\sum_{k=1}^{M}\left\vert \alpha_{k}\right\vert ^{2}~. \label{E_def}%
\end{equation}

Now, conditioned on the value of the bit ($u=0$ or $1$), the global output
state at the receiver can be written as%
\begin{equation}
\sigma_{u}=\int d^{2M}\boldsymbol{\alpha}\int d^{2L}\boldsymbol{\beta
}~\mathcal{P}(\boldsymbol{\alpha},\boldsymbol{\beta})~\sigma_{u}%
(\boldsymbol{\alpha})\otimes\gamma(\boldsymbol{\beta})~, \label{Output_SigmaU}%
\end{equation}
where%
\begin{equation}
\sigma_{u}(\boldsymbol{\alpha}):=\mathcal{S}_{u}^{\otimes M}\left(  \left\vert
\boldsymbol{\alpha}\right\rangle _{S}\left\langle \boldsymbol{\alpha
}\right\vert \right)  =\bigotimes\limits_{k=1}^{M}\mathcal{S}_{u}(\left\vert
\alpha_{k}\right\rangle \left\langle \alpha_{k}\right\vert )~,
\label{SigmaU_Alpha}%
\end{equation}
and%
\begin{equation}
\gamma(\boldsymbol{\beta}):=\mathcal{D}^{\otimes L}\left(  \left\vert
\boldsymbol{\beta}\right\rangle _{I}\left\langle \boldsymbol{\beta}\right\vert
\right)  =\bigotimes\limits_{k=1}^{L}\mathcal{D}(\left\vert \beta
_{k}\right\rangle \left\langle \beta_{k}\right\vert )~.
\end{equation}
Assuming an optimal detection of all the output modes (signals and idlers),
the error probability in the channel discrimination (i.e., bit decoding) is
given by%
\begin{equation}
P_{err}(\mathcal{G}_{0}\neq\mathcal{G}_{1}|T)=\frac{1}{2}\left[
1-D(\sigma_{0},\sigma_{1})\right]  ~,
\end{equation}
where $\sigma_{0}$ and $\sigma_{1}$ are specified by Eq.~(\ref{Output_SigmaU})
for $u=0,1$.

The study of this error probability can be greatly simplified in the case of
classical transmitters. In fact, by assuming a classical transmitter
$T_{c}=T(M,L,\rho_{c})$, the $\mathcal{P}$-representation of
Eq.~(\ref{Prepres}) is positive. As a consequence, the error probability can
be lower-bounded by a quantity which depends on the signal parameters only,
i.e.,%
\begin{equation}
P_{err}(\mathcal{G}_{0}\neq\mathcal{G}_{1}|T_{c})\geq\mathcal{C}(M,N)~.
\end{equation}
In other words, for fixed $\mathcal{G}_{0}$ and $\mathcal{G}_{1}$, we can
write a lower bound which holds for all the classical transmitters emitting
signals with bandwidth $M$ and energy $N$. This is the basic result stated in
the following theorem.

\begin{theorem}
\label{Theo_FORMS} Let us consider the discrimination of two bipartite
Gaussian channels, $\mathcal{G}_{0}=\mathcal{S}_{0}\otimes\mathcal{D}%
=\mathcal{G}(r_{0},\bar{n},\varepsilon)$ and $\mathcal{G}_{1}=\mathcal{S}%
_{1}\otimes\mathcal{D}=\mathcal{G}(r_{1},\bar{n},\varepsilon)$, by using a
classical transmitter $T_{c}=T(M,L,\rho_{c})$ which signals $N$ photons. The
corresponding error probability $P_{err}(\mathcal{G}_{0}\neq\mathcal{G}%
_{1}|T_{c})$ is lower-bounded ($\geq$) by%
\begin{equation}
\mathcal{C}(M,N):=\frac{1-\sqrt{1-F(n_{S})^{M}}}{2}~, \label{Class_Gaussians}%
\end{equation}
where $F(n_{S})$ is the fidelity between $\mathcal{S}_{0}(|\sqrt{n_{S}}%
\rangle\langle\sqrt{n_{S}}|)$ and $\mathcal{S}_{1}(|\sqrt{n_{S}}\rangle
\langle\sqrt{n_{S}}|)$, the two outputs of a single-mode coherent state
$|\sqrt{n_{S}}\rangle$ with $n_{S}:=N/M$ mean photons. In particular, in terms
of all the parameters, we have%
\begin{equation}
F(n_{S})=\omega^{-1}\exp(-\lambda n_{S})~, \label{stronger_F}%
\end{equation}
where $\omega$ and $\lambda$ are defined by%
\begin{equation}
\omega:=\frac{1}{2}\left[  \xi_{0}\xi_{1}+1-\sqrt{(\xi_{0}^{2}-1)(\xi_{1}%
^{2}-1)}\right]  \geq1~, \label{omega_THEO}%
\end{equation}
and
\begin{equation}
\lambda:=\frac{2\left(  \sqrt{r_{0}}-\sqrt{r_{1}}\right)  ^{2}}{\xi_{0}%
+\xi_{1}}\geq0~, \label{lambda_THEO}%
\end{equation}
with%
\begin{equation}
\xi_{u}:=1+2\bar{n}(1-r_{u})+\varepsilon(1+r_{u})\geq1~. \label{CsiU_THEO}%
\end{equation}

\end{theorem}

\bigskip

\textbf{Proof}.~~The two possible states, $\sigma_{0}$ and $\sigma_{1}$,
describing the whole set of output modes (signals and idlers) are given by
Eq.~(\ref{Output_SigmaU}), under the assumption that $\mathcal{P}%
(\boldsymbol{\alpha},\boldsymbol{\beta})$ is positive. In order to lowerbound
the error probability%
\begin{equation}
P_{err}(\mathcal{G}_{0}\neq\mathcal{G}_{1}|T_{c})=\frac{1}{2}\left[
1-D(\sigma_{0},\sigma_{1})\right]  ~, \label{Eq_Pclass_Proof}%
\end{equation}
we upperbound the trace distance $D(\sigma_{0},\sigma_{1})$. Since
$\mathcal{P}(\boldsymbol{\alpha},\boldsymbol{\beta})$ is a proper probability
distribution, we can use the joint convexity of the trace distance
\cite{NielsenBook}. Using this property, together with the stability of the
trace distance under addition of systems \cite{NielsenBook}, we get%
\begin{gather}
D(\sigma_{0},\sigma_{1})\leq\int d^{2M}\boldsymbol{\alpha}\int d^{2L}%
\boldsymbol{\beta~}\mathcal{P}(\boldsymbol{\alpha},\boldsymbol{\beta}%
)\times\nonumber\\
D\left[  \sigma_{0}(\boldsymbol{\alpha})\otimes\gamma(\boldsymbol{\beta
}),\sigma_{1}(\boldsymbol{\alpha})\otimes\gamma(\boldsymbol{\beta})\right]
=\nonumber\\
\int d^{2M}\boldsymbol{\alpha}\int d^{2L}\boldsymbol{\beta}~\mathcal{P}%
(\boldsymbol{\alpha},\boldsymbol{\beta})~D\left[  \sigma_{0}%
(\boldsymbol{\alpha}),\sigma_{1}(\boldsymbol{\alpha})\right]  =\nonumber\\
\int d^{2M}\boldsymbol{\alpha}~\mathcal{P}(\boldsymbol{\alpha})~D\left[
\sigma_{0}(\boldsymbol{\alpha}),\sigma_{1}(\boldsymbol{\alpha})\right]  ~,
\label{Eq_rho0_rho1}%
\end{gather}
where $\mathcal{P}(\boldsymbol{\alpha})$ is the marginal distribution defined
in Eq.~(\ref{Q_alpha_distribution}). In general, it is known that
\cite{Fuchs,FuchsThesis}%
\begin{equation}
D(\rho,\sigma)\leq\sqrt{1-F(\rho,\sigma)}~,
\end{equation}
for every pair of quantum states $\rho$ and $\sigma$. As a consequence, we can
immediately write%
\begin{equation}
D\left[  \sigma_{0}(\boldsymbol{\alpha}),\sigma_{1}(\boldsymbol{\alpha
})\right]  \leq\sqrt{1-F[\sigma_{0}(\boldsymbol{\alpha}),\sigma_{1}%
(\boldsymbol{\alpha})]}~, \label{D1F}%
\end{equation}
where $F[\sigma_{0}(\boldsymbol{\alpha}),\sigma_{1}(\boldsymbol{\alpha})] $ is
the fidelity between the two outputs $\sigma_{0}(\boldsymbol{\alpha})$ and
$\sigma_{1}(\boldsymbol{\alpha})$ defined by Eq.~(\ref{SigmaU_Alpha}). Now we
can exploit the multiplicativity of the fidelity under tensor products of
density operators. In other words, we can decompose%
\begin{gather}
F[\sigma_{0}(\boldsymbol{\alpha}),\sigma_{1}(\boldsymbol{\alpha})]=\nonumber\\
F\left[  \bigotimes_{k=1}^{M}\mathcal{S}_{0}(\left\vert \alpha_{k}%
\right\rangle \left\langle \alpha_{k}\right\vert ),\bigotimes_{k=1}%
^{M}\mathcal{S}_{1}(\left\vert \alpha_{k}\right\rangle \left\langle \alpha
_{k}\right\vert )\right]  =\nonumber\\
\prod_{k=1}^{M}F\left[  \mathcal{S}_{0}(\left\vert \alpha_{k}\right\rangle
\left\langle \alpha_{k}\right\vert ),\mathcal{S}_{1}(\left\vert \alpha
_{k}\right\rangle \left\langle \alpha_{k}\right\vert )\right]  :=\prod
_{k=1}^{M}F_{k}~, \label{Proof_REPL}%
\end{gather}
where $F_{k}$ is the fidelity between the two single-mode Gaussian states
$\mathcal{S}_{0}(\left\vert \alpha_{k}\right\rangle \left\langle \alpha
_{k}\right\vert )$ and $\mathcal{S}_{1}(\left\vert \alpha_{k}\right\rangle
\left\langle \alpha_{k}\right\vert )$. In order to compute $F_{k}$, let us
consider the explicit action of $\mathcal{S}_{u}$ on the coherent state
$\left\vert \alpha_{k}\right\rangle \left\langle \alpha_{k}\right\vert $,
which is a Gaussian state with CM $\mathbf{V}=\mathbf{I}$ and displacement
$\mathbf{\bar{x}}^{T}=(2$\textrm{Re}$(\alpha_{k}),2$\textrm{Im}$(\alpha_{k}%
))$. At the output of $\mathcal{S}_{u}$, we get a Gaussian state
$\mathcal{S}_{u}(\left\vert \alpha_{k}\right\rangle \left\langle \alpha
_{k}\right\vert )$ whose statistical moments are proportional to the input
ones, i.e., $\mathbf{V}_{u}=\xi_{u}\mathbf{I}$ and $\mathbf{\bar{x}}_{u}%
=\sqrt{r_{u}}\mathbf{\bar{x}}$, were $\xi_{u}$ is given in
Eq.~(\ref{CsiU_THEO}). Notice that $\xi_{u}\geq1$ because $\mathbf{V}_{u}$
must be a bona-fide CM \cite{Alex}. Then, by using the formula of
Ref.~\cite{FidFormulas}, we can compute the analytical expression of the
fidelity, which is equal to%
\begin{equation}
F_{k}=\omega^{-1}\exp\left[  -\lambda\left\vert \alpha_{k}\right\vert
^{2}\right]  ~, \label{Fidelity_TwoGaussian}%
\end{equation}
where $\omega$ and $\lambda$ are defined in Eqs.~(\ref{omega_THEO})
and~(\ref{lambda_THEO}), respectively. It is trivial to check that $\omega
\geq1$ and $\lambda\geq0$ using $\xi_{u}\geq1$. \ Now, using
Eq.~(\ref{Fidelity_TwoGaussian}) into Eq.~(\ref{Proof_REPL}), we get%
\begin{equation}
F[\sigma_{0}(\boldsymbol{\alpha}),\sigma_{1}(\boldsymbol{\alpha})]=\omega
^{-M}\exp\left[  -\lambda\sum_{k=1}^{M}\left\vert \alpha_{k}\right\vert
^{2}\right]  =g(E)~, \label{Fe}%
\end{equation}
where%
\begin{equation}
g(E):=\omega^{-M}\exp\left[  -\lambda E\right]  ~,
\end{equation}
and $E$ is defined in Eq.~(\ref{E_def}). Thus, by combining
Eqs.~(\ref{Eq_rho0_rho1}), (\ref{D1F}) and (\ref{Fe}), we get%
\begin{equation}
D(\sigma_{0},\sigma_{1})\leq\mathcal{I}:=\int d^{2M}\boldsymbol{\alpha
}~\mathcal{P}(\boldsymbol{\alpha})~\mathcal{B}(E)~, \label{Integral_I}%
\end{equation}
where we have introduced the $\mathbb{R}^{+}\rightarrow\lbrack0,1]$\ function%
\begin{equation}
\mathcal{B}(E):=\sqrt{1-g(E)}~.
\end{equation}

Since the latter quantity depends only on the real scalar $E$, we can greatly
simplify the integral $\mathcal{I}$\ in Eq.~(\ref{Integral_I}). For this sake,
let us introduce the polar coordinates%
\begin{equation}
\boldsymbol{\alpha}=\sqrt{\mathbf{e}}\exp(i\boldsymbol{\theta})~,
\end{equation}
where $\mathbf{e}:=(e_{1},\cdots,e_{M})$ is a vector with generic element
$e_{k}:=\left\vert \alpha_{k}\right\vert ^{2}$, and $\boldsymbol{\theta
}:=(\theta_{1},\cdots,\theta_{M})$ is a vector of phases (here $d^{2M}%
\boldsymbol{\alpha}=2^{-M}d^{M}\mathbf{e}d^{M}\boldsymbol{\theta}$). Then, we
can write
\begin{equation}
\mathcal{I}=\int_{0}^{+\infty}d^{M}\mathbf{e}~\mathcal{R}(\mathbf{e}%
)~\mathcal{B}(E)~,
\end{equation}
where $\mathcal{R}(\mathbf{e})$ is the radial probability distribution%
\begin{equation}
\mathcal{R}(\mathbf{e}):=\int_{0}^{2\pi}\frac{d^{M}\boldsymbol{\theta}}{2^{M}%
}~\mathcal{P}(\mathbf{e,}\boldsymbol{\theta})~, \label{ProbERRE}%
\end{equation}
and%
\begin{equation}
\mathcal{P}(\mathbf{e,}\boldsymbol{\theta}):=\left[  \mathcal{P}%
(\boldsymbol{\alpha})\right]  _{\boldsymbol{\alpha}=\sqrt{\mathbf{e}}%
\exp(i\boldsymbol{\theta})}~.
\end{equation}
The integral can be further simplified by setting
\begin{equation}
e_{M}=E-\sum_{k=1}^{M-1}e_{k}~,
\end{equation}
which introduces the further change of variables%
\begin{equation}
\mathbf{e}\rightarrow\mathbf{e}^{\prime}:=(e_{1},\cdots,e_{M-1},E)~.
\end{equation}
Then, we get%
\begin{equation}
\mathcal{I}=\int_{0}^{+\infty}dE~\mathcal{R}(E)~\mathcal{B}(E)~,
\label{Integral_Last}%
\end{equation}
where%
\begin{align}
\mathcal{R}(E)  &  :=\int_{0}^{E}de_{1}\int_{0}^{E-e_{1}}de_{2}\times
\nonumber\\
&  \cdots\times\int_{0}^{E-\sum_{k=1}^{M-2}e_{k}}de_{M-1}~\mathcal{R}%
(\mathbf{e}^{\prime})~,
\end{align}
and%
\begin{equation}
\mathcal{R}(\mathbf{e}^{\prime}):=\left[  \mathcal{R}(\mathbf{e})\right]
_{e_{M}=E-\sum_{k=1}^{M-1}e_{k}}~.
\end{equation}
It is clear that the simplification of the integral from Eq.~(\ref{Integral_I}%
) to Eq.~(\ref{Integral_Last}) can be done not just for $\mathcal{B}(E)$, but
for a generic integrable function $f(E)$. As a consequence, we can repeat the
same simplification for $f(E)=E$, which corresponds to write the following
expression for the mean total number of photons
\begin{equation}
N=\int_{0}^{+\infty}dE~\mathcal{R}(E)~E~.
\end{equation}
Analytically, it is easy to check that $\mathcal{B}(E)$ is concave, i.e.,%
\begin{equation}
p\mathcal{B}(E)+(1-p)\mathcal{B}(E^{\prime})\leq\mathcal{B}[pE+(1-p)E^{\prime
}]~,
\end{equation}
for every $E,E^{\prime}\in\mathbb{R}^{+}$ and $p\in\lbrack0,1]$. Then,
applying Jensen's inequality \cite{Jensen}, we get%
\begin{align}
\mathcal{I}  &  \leq\mathcal{B}\left[  \int_{0}^{+\infty}dE~\mathcal{R}%
(E)~E\right]  =\mathcal{B}\left(  N\right) \nonumber\\
&  =\sqrt{1-g(N)}~. \label{I_final}%
\end{align}
Here we can set%
\begin{equation}
g(N)=F(n_{S})^{M}~, \label{g_f}%
\end{equation}
where $n_{S}=N/M$ and $F(n_{S})$ is given in Eq.~(\ref{stronger_F}). According
to Eq.~(\ref{Fidelity_TwoGaussian}), $F(n_{S})$ represents the fidelity
between the two states $\mathcal{S}_{0}(|\sqrt{n_{S}}\rangle\langle\sqrt
{n_{S}}|)$ and $\mathcal{S}_{1}(|\sqrt{n_{S}}\rangle\langle\sqrt{n_{S}}|)$,
i.e., the two possible outputs of the coherent state $|\sqrt{n_{S}}\rangle$.
In conclusion, by combining Eqs.~(\ref{Integral_I}), (\ref{I_final})
and~(\ref{g_f}), we get%
\begin{equation}
D(\sigma_{0},\sigma_{1})\leq\sqrt{1-F(n_{S})^{M}}~.
\end{equation}
Using the latter equation with Eq.~(\ref{Eq_Pclass_Proof}), we obtain the
lower bound of Eq.~(\ref{Class_Gaussians}).~$\blacksquare$

\bigskip

\begin{figure}[ptbh]
\vspace{-0.5cm}
\par
\begin{center}
\includegraphics[width=0.5\textwidth] {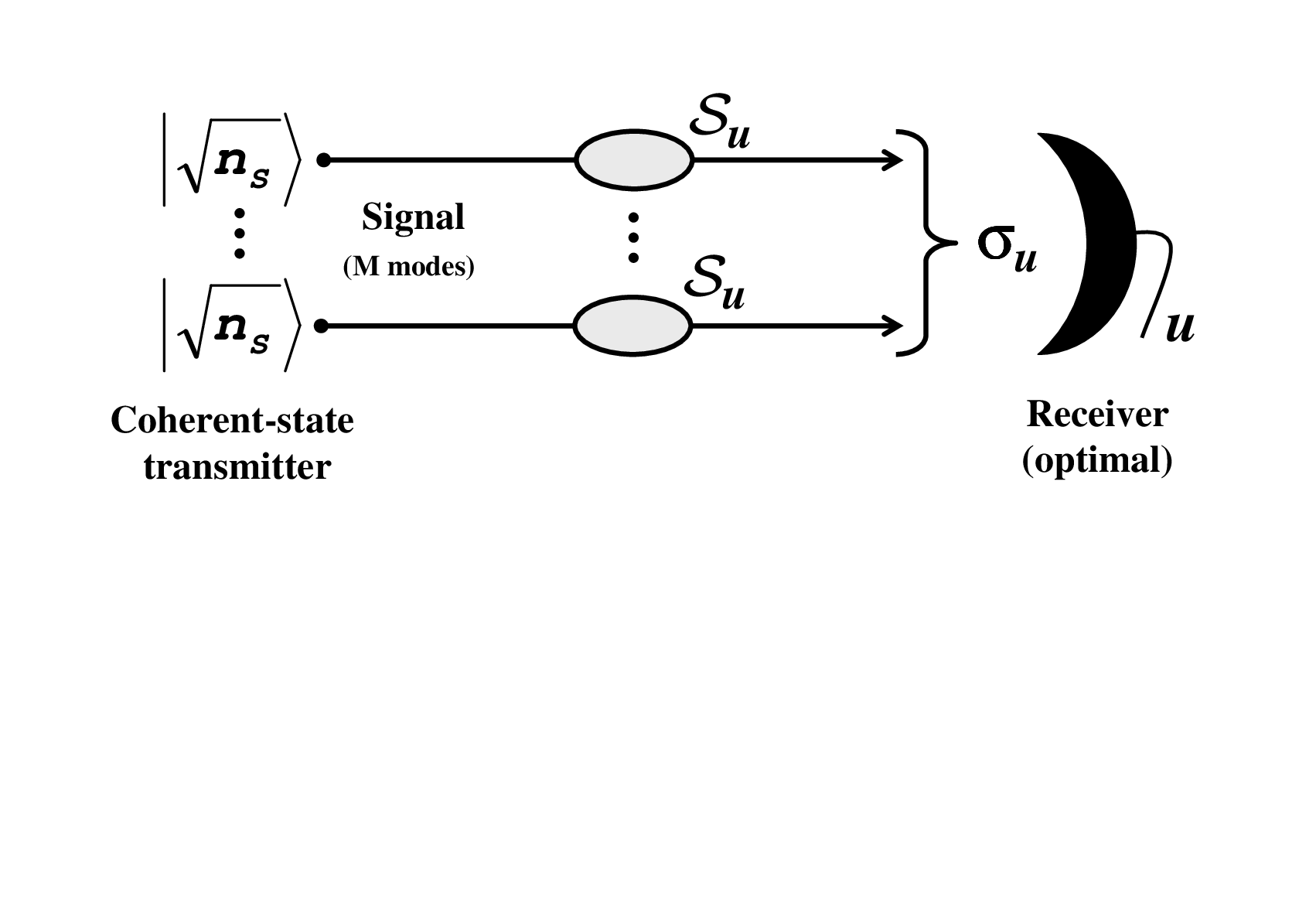}
\end{center}
\par
\vspace{-3.2cm}\caption{Coherent-state transmitter which signals $M$ copies of
a coherent state $|\sqrt{n_{S}}\rangle$ with mean photon number $n_{S}$.}%
\label{COHEpic}%
\end{figure}

According to the previous theorem, the classical discrimination bound
$\mathcal{C}(M,N)$ can be \textit{computed} by assuming a coherent-state
transmitter which signals $M$ identical coherent states with $n_{S}=N/M$ mean
photons each. This transmitter can be denoted by
\begin{equation}
T_{coh}=T(M,0,|\sqrt{n_{S}}\rangle\langle\sqrt{n_{S}}|^{\otimes M})~,
\end{equation}
and is schematically depicted in Fig.~\ref{COHEpic}. Despite the computation
of $\mathcal{C}(M,N)$ can be performed in this simple way, we do not know
which classical transmitter is actually able to reach, or approach, the lower
bound $\mathcal{C}(M,N)$. 

\subsection{Pure-loss model}

For the pure-loss model of Fig.~\ref{Scenario2}, the previous theorem can be
greatly simplified. In particular, we get a classical discrimination bound
which does not depend on the bandwidth $M$, but only on the energy $N$\ of the
signal. This result is stated in the following corollary (which corresponds to
the first theorem in our Letter).

\begin{corollary}
\label{Theo_FORMS_CORO} Let us consider the discrimination of two attenuator
channels, $\mathcal{E}_{0}=\mathcal{E}(r_{0})$ and $\mathcal{E}_{1}%
=\mathcal{E}(r_{1})$, by using a classical transmitter $T_{c}=T(M,L,\rho_{c})$
which signals $N$ photons. Then, we have%
\begin{equation}
P_{err}(\mathcal{E}_{0}\neq\mathcal{E}_{1}|T_{c})\geq\mathcal{C}(N)~,
\label{C_class_bound}%
\end{equation}
where%
\begin{equation}
\mathcal{C}(N):=\frac{1-\sqrt{1-\mathrm{\exp}\left[  -N\left(  \sqrt{r_{1}%
}-\sqrt{r_{0}}\right)  ^{2}\right]  }}{2}~. \label{C_attenuators}%
\end{equation}

\end{corollary}

\bigskip

\textbf{Proof}.~~Let us set $\bar{n}=\varepsilon=0$ in
Theorem~\ref{Theo_FORMS}, so that%
\begin{align}
\mathcal{G}_{u}  &  =\mathcal{S}_{u}\otimes\mathcal{D}=[\mathcal{N}%
(0)\circ\mathcal{E}(r_{u},0)\circ\mathcal{N}(0)]\otimes\lbrack\mathcal{N}%
(0)\circ\mathcal{N}(0)]\nonumber\\
&  =[\mathcal{I}\circ\mathcal{E}(r_{u})\circ\mathcal{I}]\otimes\lbrack
\mathcal{I}\circ\mathcal{I}]=\mathcal{E}_{u}\otimes\mathcal{I}~.
\end{align}
Clearly we can write $P_{err}(\mathcal{G}_{0}\neq\mathcal{G}_{1}%
|T_{c})=P_{err}(\mathcal{E}_{0}\neq\mathcal{E}_{1}|T_{c})$. This quantity is
lower-bounded by using Eq.~(\ref{Class_Gaussians}) where now the fidelity term
$F(n_{S})^{M}$ can be greatly simplified. In fact, because of $\bar
{n}=\varepsilon=0$, here we have $\xi_{u}=1$, which implies%
\begin{equation}
\omega=1~,~\lambda=\left(  \sqrt{r_{0}}-\sqrt{r_{1}}\right)  ^{2}~.
\end{equation}
As a consequence, we have%
\begin{align}
F(n_{S})^{M}  &  =\left\{  \exp\left[  -n_{S}\left(  \sqrt{r_{1}}-\sqrt{r_{0}%
}\right)  ^{2}\right]  \right\}  ^{M}\nonumber\\
&  =\exp\left[  -N\left(  \sqrt{r_{1}}-\sqrt{r_{0}}\right)  ^{2}\right]  ~,
\end{align}
providing the expression of Eq.~(\ref{C_attenuators}).~$\blacksquare$

\section{Quantum transmitter\label{APP_EPR_TRA}}

As explained in the Letter, we consider a particular kind of non-classical
transmitter, that we call \textquotedblleft EPR\ transmitter\textquotedblright%
. This is given by%
\begin{equation}
T_{epr}=T(M,M,\left\vert \xi\right\rangle \left\langle \xi\right\vert
^{\otimes M})~,
\end{equation}
where $\left\vert \xi\right\rangle $ is the TMSV state described in
Sec.~\ref{APP_INTRO}. This transmitter is completely characterized by the
basic parameters of the emitted signal, i.e., bandwidth $M$ and energy $N$,
since $N=Mn_{S}=M\sinh^{2}\xi$. For this reason, we also use the notation
$T_{epr}=T_{M,N}$. Given this transmitter, we now show how to compute the
error probability which affects the channel discrimination. We consider the
general case of the thermal-loss model, but it is understood that the results
hold for the pure-loss model by setting $\bar{n}=\varepsilon=0$.

Given an arbitrary EPR\ transmitter, the corresponding input state
\begin{equation}
\rho=\left\vert \xi\right\rangle \left\langle \xi\right\vert ^{\otimes M}~,
\label{inputEPRtran}%
\end{equation}
is transformed into the conditional output state%
\begin{equation}
\sigma_{u}=\varphi_{u}^{\otimes M}~,
\end{equation}
where%
\begin{equation}
\varphi_{u}=\mathcal{G}_{u}\left(  \left\vert \xi\right\rangle \left\langle
\xi\right\vert \right)  =\left(  \mathcal{S}_{u}\otimes\mathcal{D}\right)
(\left\vert \xi\right\rangle \left\langle \xi\right\vert )~. \label{Rho_ri}%
\end{equation}
The single-copy output state $\varphi_{u}$\ has zero mean and its CM
$\mathbf{V}_{u}$ can be easily computed from the one of the TMSV state, which
is given in Eq.~(\ref{CM_TMSV}). In fact, if we write the bipartite channel
\begin{equation}
\mathcal{S}_{u}\otimes\mathcal{D}=[\mathcal{N}(\varepsilon)\circ
\mathcal{E}(r_{u},\bar{n})\circ\mathcal{N}(\varepsilon)]\otimes\mathcal{N}%
(2\varepsilon)
\end{equation}
in terms of $(\mathbf{K,N})$\ matrices by using Eqs.~(\ref{KN_1})
and~(\ref{KN_2}), we can exploit the result of Eq.~(\ref{V_out_bip_CH}). Thus,
we get the output CM%
\begin{equation}
\mathbf{V}_{u}=\left(
\begin{array}
[c]{cc}%
\lbrack r_{u}(\mu+\varepsilon)+(1-r_{u})\beta+\varepsilon]\mathbf{I} &
\sqrt{r_{u}(\mu^{2}-1)}\mathbf{Z}\\
\sqrt{r_{u}(\mu^{2}-1)}\mathbf{Z} & (\mu+2\varepsilon)\mathbf{I}%
\end{array}
\right)  ~, \label{outputCM_r}%
\end{equation}
where%
\begin{equation}
\mu:=2n_{S}+1~,~\beta:=2\bar{n}+1~.
\end{equation}
Now, the error probability which affects the channel discrimination
$P_{err}(\mathcal{G}_{0}\neq\mathcal{G}_{1}|T_{M,N})$ is equal to the minimum
error probability in the M-copy discrimination between $\varphi_{0}$ and
$\varphi_{1}$. This quantity is upper-bounded by the quantum Chernoff bound,
i.e.,%
\begin{equation}
P_{err}(\mathcal{G}_{0}\neq\mathcal{G}_{1}|T_{M,N})\leq\mathcal{Q}(M,N)~.
\label{Q_M_Ns}%
\end{equation}
where%
\begin{equation}
\mathcal{Q}(M,N)=\frac{1}{2}\left[  \inf_{t\in(0,1)}\mathrm{Tr}\left(
\varphi_{0}^{t}\varphi_{1}^{1-t}\right)  \right]  ^{M}~.
\end{equation}
To compute $\mathcal{Q}(M,N)$ we perform the normal-mode decomposition of the
Gaussian states $\varphi_{0}$ and $\varphi_{1}$, which corresponds to the
symplectic decomposition of the conditional CM of Eq.~(\ref{outputCM_r})
\cite{MinkoPRA,Alex}. Then, we apply the symplectic formula of
Ref.~\cite{MinkoPRA}. Unfortunately, the result is extremely long to be
presented here (but short analytical expressions are provided afterwards in
the proofs for the pure-loss model). Note that an alternative upper-bound is
the quantum Battacharyya bound, which is given by%
\begin{equation}
\mathcal{B}(M,N)=\frac{1}{2}\left[  \mathrm{Tr}\left(  \varphi_{0}%
^{1/2}\varphi_{1}^{1/2}\right)  \right]  ^{M}\geq\mathcal{Q}(M,N)~.
\label{Batta}%
\end{equation}
This bound is larger but generally easier to compute than the quantum Chernoff
bound (see Ref.~\cite{MinkoPRA} for details).

\section{Quantum-classical comparison\label{APP_comparison}}

In this section we show how to make the comparison between the EPR transmitter
and the classical transmitters in order to reply to the basic questions of
Eqs.~(\ref{Q1}) and~(\ref{Q2}). Our strategy is to derive a sufficient
condition for the superiority of the EPR\ transmitter by comparing the bounds
derived in the previous sections. On the one hand, we know that the error
probability of any classical transmitter $P_{err}^{class} $ is lower-bounded
by $\mathcal{C}$. On the other hand, the error probability of the
EPR\ transmitter $P_{err}^{epr}$ is upper-bound by $\mathcal{Q}$. Thus, a
sufficient condition for $P_{err}^{epr}<P_{err}^{class} $ is provided by the
inequality $\mathcal{Q}<\mathcal{C}$. Depending on the model, our answer can
be analytical or numerical.

\subsection{Pure-loss model}

For the pure-loss model, we have $\mathcal{Q}=\mathcal{Q}(M,N)$ and
$\mathcal{C}=\mathcal{C}(N)$. Since here the classical discrimination bound is
not dependent on the bandwidth, it is sufficient to find a $\bar{M}$ such that
$\mathcal{Q}(\bar{M},N)<\mathcal{C}(N)$. In other words, our basic question of
Eq.~(\ref{Q1}) becomes the following: given two channels $\{\mathcal{E}%
_{0},\mathcal{E}_{1}\}$ and signal-energy $N$, can we find a bandwidth
$\bar{M}$ such that
\begin{equation}
\mathcal{Q}(\bar{M},N)<\mathcal{C}(N)~?
\end{equation}
This clearly implies the existence of an EPR\ transmitter $T_{\bar{M},N}%
$\ such that%
\begin{equation}
P_{err}(\mathcal{E}_{0}\neq\mathcal{E}_{1}|T_{\bar{M},N})<P_{err}%
(\mathcal{E}_{0}\neq\mathcal{E}_{1}|T_{c})~, \label{QQ1}%
\end{equation}
for every $T_{c}$. As stated in our Letter this question has a positive
answer, and this is provided by our \textquotedblleft threshold
energy\textquotedblright\ theorem. According to this theorem, for every pair
of attenuator channels $\{\mathcal{E}_{0},\mathcal{E}_{1}\}$\ and above a
threshold energy $N_{th}$, there is an EPR transmitter (with suitable
bandwidth $\bar{M}$) which outperforms any classical transmitter. Furthermore,
we can prove a stronger result if one of the two channels is just the
identity, e.g., we have $\{\mathcal{E}_{0},\mathcal{I}\}$. In this case we
have $N_{th}=1/2$ and $\bar{M}$ is a minimum bandwidth, as stated by the
\textquotedblleft ideal memory\textquotedblright\ theorem in our Letter. We
give the explicit proofs of these theorems in Sec.~\ref{APP_Pure_Loss}.

In order to quantify numerically the advantage brought by the
EPR\ transmitter, we introduce the minimum information gain
\begin{equation}
G:=1-H(\mathcal{Q})-[1-H(\mathcal{C})]=G(M,N)~, \label{gainG}%
\end{equation}
where%
\begin{equation}
H(x):=-x\log_{2}x-(1-x)\log_{2}(1-x)
\end{equation}
is the binary Shannon entropy. Finding a bandwidth $\bar{M}$\ such that
$G(\bar{M},N)>0$ clearly implies the validity of Eq.~(\ref{QQ1}). In terms of
memory readout, this quantity lowerbounds the number of bits per cell which
are gained by an EPR\ transmitter $T_{\bar{M},N}$\ over any classical
transmitter. By using $G$ one can perform extensive numerical investigations
\cite{NumG}. In particular, one can check that narrowband EPR\ transmitters
(i.e., with low $\bar{M}$) are able to give $G>0$ in the regime of few photons
$N$ and high-reflectivities (i.e., $r_{0}$ or $r_{1}$ sufficiently close to
$1$). These numerical results are shown and discussed in Fig.~2(left) and
Fig.~3 of our Letter.

\subsection{Thermal-loss model}

For the thermal-loss model, we have $\mathcal{Q}=\mathcal{Q}(M,N)$ and
$\mathcal{C}=\mathcal{C}(M,N)$. Here the classical discrimination bound
$\mathcal{C}$ depends on the bandwidth $M$ and tends monotonically to zero for
$M\rightarrow\infty$. This is evident from the general expression of the
fidelity term $F(n_{S})^{M}$ which is present in Eq.~(\ref{Class_Gaussians}).
In fact, from Eq.~(\ref{stronger_F}), we get%
\begin{equation}
F(n_{S})^{M}=\omega^{-M}\exp(-\lambda N)~,
\end{equation}
which decreases to zero for $M\rightarrow\infty$ whenever $\omega>1$ (as is
generally the case for the thermal-loss model). For this reason, let us fix a
maximum bandwidth $M^{\ast}$ and consider the minimum value%
\begin{equation}
\inf_{M\in\lbrack1,M^{\ast}]}\mathcal{C}(M,N)=\mathcal{C}(M^{\ast},N)~.
\end{equation}
It is clear that, given two bipartite channels $\{\mathcal{G}_{0}%
,\mathcal{G}_{1}\}$ and signal-energy $N$, we have%
\begin{equation}
P_{err}(\mathcal{G}_{0}\neq\mathcal{G}_{1}|T_{c})\geq\mathcal{C}(M^{\ast},N)~,
\end{equation}
for every $T_{c}=T(M,L,\rho_{c})$ with signal-bandwidth $M\leq M^{\ast}$. In
other words, using $\mathcal{C}(M^{\ast},N)$ we can bound all the classical
transmitters up to the maximum bandwidth $M^{\ast}$. As a result, our basic
question of Eq.~(\ref{Q2}) becomes the following: given two bipartite channels
$\{\mathcal{G}_{0},\mathcal{G}_{1}\}$ and signal-energy $N $, can we find an
EPR transmitter with suitable bandwidth $\bar{M}$ such that
\begin{equation}
\mathcal{Q}(\bar{M},N)<\mathcal{C}(M^{\ast},N)~?
\end{equation}
This clearly implies the existence of an EPR\ transmitter $T_{\bar{M},N}%
$\ such that%
\begin{equation}
P_{err}(\mathcal{G}_{0}\neq\mathcal{G}_{1}|T_{\bar{M},N})<P_{err}%
(\mathcal{G}_{0}\neq\mathcal{G}_{1}|T_{c})~, \label{QQ2}%
\end{equation}
for every $T_{c}=T(M,L,\rho_{c})$ with $M\leq M^{\ast}$. As explicitly
discussed in the Letter, this question has a positive answer too, and the
value of $M^{\ast}$ can be so high to include all the classical transmitters
which are physically meaningful for the memory model. However, despite the
previous case, here the answer can only be numerical. In fact, the two
bipartite channels $\{\mathcal{G}_{0},\mathcal{G}_{1}\}$ are characterized by
four parameters $\{r_{0},r_{1},\bar{n},\varepsilon\}$ and the comparison
between transmitters involves further three parameters $\{\bar{M},M^{\ast
},N\}$. For this reason, the explicit expression of the minimum information
gain now depends on seven parameters, i.e.,%
\begin{equation}
G=G(\bar{M},M^{\ast},N,r_{0},r_{1},\bar{n},\varepsilon)~.
\end{equation}
By performing numerical investigations, one can analyze the positivity of $G$,
which clearly implies the condition of Eq.~(\ref{QQ2}). According to
Fig.~2(right) of our Letter, we can achieve remarkable positive gains when we
compare narrowband EPR\ transmitters (low $\bar{M}$) with wide sets of
classical transmitters (large $M^{\ast}$) in the regime of few photons (low
$N$) and high reflectivities ($r_{0}$ or $r_{1}$ close to $1$), and assuming
the presence of non-trivial decoherence (e.g., $\varepsilon=\bar{n}=10^{-5}$).
See the Letter for physical discussions.

\section{Analytical results for the pure-loss model\label{APP_Pure_Loss}}

In this section we provide the detailed proofs of the two theorems relative to
the pure-loss model:\ the \textquotedblleft threshold energy\textquotedblright%
\ theorem and the \textquotedblleft ideal memory\textquotedblright\ theorem.
Given two attenuator channels, $\mathcal{E}_{0}=\mathcal{E}(r_{0})$ and
$\mathcal{E}_{1}=\mathcal{E}(r_{1})$, and given the signal-energy $N$, the
basic problem is to show the existence of an EPR transmitter $T_{M,N}$ which
is able to beat the classical discrimination bound $\mathcal{C}(N)$, and,
therefore, all the classical transmitters $T_{c}$. The two theorems provide
sufficient conditions for the existence of this quantum transmitter.

\subsection{\textquotedblleft Threshold energy\textquotedblright\ theorem
\label{APP_Proof1}}

For the sake of completeness we repeat here the statement of the theorem (this
is perfectly equivalent to the statement provided in our Letter).

\begin{theorem}
Let us consider the discrimination of two attenuator channels, $\mathcal{E}%
_{0}=\mathcal{E}(r_{0})$ and $\mathcal{E}_{1}=\mathcal{E}(r_{1})$ with
$r_{0}\neq r_{1}$, by using transmitters with (finite)\ signal energy
\begin{align}
N  &  >N_{th}(r_{0},r_{1}):=\frac{2\ln2}{2-r_{0}-r_{1}-2\sqrt{(1-r_{0}%
)(1-r_{1})}}~.\nonumber\\
&
\end{align}
Then, there exists an EPR\ transmitter $T_{\bar{M},N}$, with suitable
bandwidth $\bar{M}$, such that%
\begin{equation}
P_{err}(\mathcal{E}_{0}\neq\mathcal{E}_{1}|T_{\bar{M},N})<\mathcal{C}(N)~.
\end{equation}

\end{theorem}

\bigskip

\textbf{Proof.~~}For simplicity, in this proof we use the shorthand notation
$P_{err}(T):=P_{err}(\mathcal{E}_{0}\neq\mathcal{E}_{1}|T)$. According to
Corollary~\ref{Theo_FORMS_CORO}, the error probability for an arbitrary
classical transmitter $T_{c}$ is lower-bounded by the classical discrimination
bound, i.e., $P_{err}(T_{c})\geq\mathcal{C}(N)$, where $\mathcal{C}(N)$ is
given in Eq.~(\ref{C_attenuators}). Note that, for every $z\in\lbrack0,1]$, we
have%
\begin{equation}
\frac{1-\sqrt{1-z}}{2}\geq\frac{z}{4}~.
\end{equation}
As a consequence, we can write%
\begin{equation}
\mathcal{C}(N)\geq\frac{e^{-Nx^{2}}}{4}:=\mathcal{\tilde{C}}(N)~,
\label{C_tilde}%
\end{equation}
where%
\begin{equation}
x:=\sqrt{r_{1}}-\sqrt{r_{0}}~. \label{x_exp}%
\end{equation}

Now let us consider an EPR\ transmitter $T_{M,N}$. The corresponding error
probability $P_{err}(T_{M,N})$ can be upper-bounded via the quantum
Battacharyya bound $\mathcal{B}(M,N)$ defined in Eq.~(\ref{Batta}). This bound
can be computed from the conditional CM of Eq.~(\ref{outputCM_r}) by setting
$\bar{n}=\varepsilon=0$\ (see Sec.~\ref{APP_EPR_TRA} and Ref.~\cite{MinkoPRA}%
). In this problem, for fixed (finite) energy $N$, the bound $\mathcal{B}%
(M,N)$ is not monotonic in $M$ (easy to check numerically). However, it is
regular in $M$ and its asymptotic limit $\mathcal{B}_{\infty}(N):=\lim
_{M\rightarrow+\infty}\mathcal{B}(M,N)$ exists. In particular, the asymptotic
Battacharyya bound is equal to%
\begin{equation}
\mathcal{B}_{\infty}(N)=\frac{e^{-Nw}}{2}~, \label{B_infinite}%
\end{equation}
where $w\in\lbrack0,3/2]$ is defined by%
\begin{equation}
w:=\frac{r_{0}+r_{1}+2}{2}-2\sqrt{r_{0}r_{1}}-\sqrt{(1-r_{0})(1-r_{1})}~.
\label{y_exp}%
\end{equation}
Clearly we have%
\begin{equation}
\inf_{M}P_{err}(T_{M,N})\leq\inf_{M}\mathcal{B}(M,N)\leq\mathcal{B}_{\infty
}(N)~,
\end{equation}
which means that $\forall\varepsilon>0$, $\exists\bar{M}\in\mathbb{Z}^{+} $
such that%
\begin{equation}
P_{err}(T_{\bar{M},N})<\mathcal{B}_{\infty}(N)+\varepsilon~. \label{B_eps}%
\end{equation}

In order to prove the result, let us impose the threshold condition%
\begin{equation}
\mathcal{B}_{\infty}(N)<\mathcal{\tilde{C}}(N)~. \label{TH_app}%
\end{equation}
Now, if Eq.~(\ref{TH_app}) is satisfied, then we can always take an
$\varepsilon>0$\ such that%
\begin{equation}
\mathcal{\tilde{C}}(N)=\mathcal{B}_{\infty}(N)+\varepsilon~.
\end{equation}
Then, because of the proposition of Eq.~(\ref{B_eps}), we have that
$\exists\bar{M}\in\mathbb{Z}^{+}$ such that%
\begin{equation}
P_{err}(T_{\bar{M},N})<\mathcal{\tilde{C}}(N)\leq\mathcal{C}(N)~.
\end{equation}

For this reason, the next step is to solve Eq.~(\ref{TH_app}) in order to get
a threshold condition on the energy $N$. It is easy to check that, by using
Eqs.~(\ref{C_tilde})-(\ref{y_exp}), the condition of Eq.~(\ref{TH_app})
becomes%
\begin{equation}
Nf>1~,
\end{equation}
where%
\begin{gather}
f:=\frac{w-x^{2}}{\ln2}\nonumber\\
=\frac{1}{\ln2}\left[  \frac{(1-r_{0})+(1-r_{1})}{2}-\sqrt{(1-r_{0})(1-r_{1}%
)}\right]  ~.
\end{gather}
Since $f$\ is proportional to the difference between an arithmetic mean and a
geometric mean, we have $f\geq0$, and $f=0$ if and only if $r_{0}=r_{1}$.
Thus, if we exclude the singular case $r_{0}=r_{1}$, we can write%
\begin{equation}
N>\frac{1}{f}=N_{th}(r_{0},r_{1})~. \label{eq_N}%
\end{equation}
In conclusion, if the threshold condition of Eq.~(\ref{eq_N}) is satisfied
(where $r_{0}\neq r_{1}$), then there exists a bandwidth $\bar{M}$ such
that\ $P_{err}(T_{\bar{M},N})<\mathcal{C}(N).~\blacksquare$

\subsection{\textquotedblleft Ideal memory\textquotedblright\ theorem}

Here we prove the ideal memory theorem which refers to the case of ideal
memories, i.e., having $r_{0}<r_{1}=1$. This scenario corresponds to
discriminating between an attenuator channel $\mathcal{E}(r_{0})$ and the
identity channel $\mathcal{I}=\mathcal{E}(1)$. For this reason, the quantum
Chernoff bound $\mathcal{Q}(M,N)$ has a simple analytical expression, which
turns out to be decreasing in $M$\ (for fixed energy $N$). Thanks to this
monotony, our result can be proven above a minimum bandwidth $\bar{M}$. In
fact, $\mathcal{Q}(M,N)$\ decreasing in $M$\ implies that $G=1-H[\mathcal{Q}%
(M,N)]-\{1-H[\mathcal{C}(N)]\}$ is increasing in $M$, so that optimal gains
are monotonically reached by broadband EPR transmitters. Note that this was
not possible to prove for the previous theorem (see Sec.~\ref{APP_Proof1})
since, in that case, the quantum Battacharyya bound $\mathcal{B}(M,N)$ turned
out to be non-monotonic in $M$.

For the sake of completeness we repeat here the statement of the
\textquotedblleft ideal memory\textquotedblright\ theorem (this is perfectly
equivalent to the statement provided in our Letter).

\begin{theorem}
Let us consider the discrimination of an attenuator channel $\mathcal{E}%
_{0}=\mathcal{E}(r_{0})$, with $r_{0}<1$, from the identity channel
$\mathcal{I}$, by using transmitters with signal energy
\begin{equation}
N\geq N_{th}:=1/2~.
\end{equation}
Then, there exists a minimum bandwidth $\bar{M}$ such that, for every EPR
transmitter $T_{M,N}$ with $M>\bar{M}$, we have%
\begin{equation}
P_{err}(\mathcal{E}_{0}\neq\mathcal{I}|T_{M,N})<\mathcal{C}(N)~.
\end{equation}

\end{theorem}

\bigskip

\textbf{Proof.}~~Given the discrimination problem $\mathcal{E}_{0}%
\neq\mathcal{I}$ for fixed signal energy $N$, let us consider an
EPR\ transmitter $T_{M,N}$. Its error probability $P_{err}(T_{M,N}%
)=P_{err}(\mathcal{E}_{0}\neq\mathcal{I}|T_{M,N})$ is upper-bounded by the
quantum Chernoff bound $\mathcal{Q}(M,N)$, whose analytical expression is
greatly simplified here. In fact, we have%
\begin{equation}
\mathcal{Q}(M,N)=\frac{1}{2}\left(  1+\frac{N}{M}x\right)  ^{-2M}~,
\end{equation}
where%
\begin{equation}
x=1-\sqrt{r_{0}}\in(0,1]~. \label{x_r0}%
\end{equation}
This expression can be computed using the procedure sketched in
Sec.~\ref{APP_EPR_TRA} by setting $\bar{n}=\varepsilon=0$ and $r_{1}=1$ in the
conditional CM of Eq.~(\ref{outputCM_r}). For fixed energy $N$, it is very
easy to check analytically that $\mathcal{Q}(M,N)$ is decreasing in $M$
(strictly decreasing if we exclude the trivial case $N=0$). Since
$\mathcal{Q}$ is bounded ($\mathcal{Q}\in\lbrack0,1/2]$) and decreasing in
$M$, the broadband limit%
\begin{equation}
\mathcal{Q}_{\infty}(N):=\lim_{M\rightarrow+\infty}\mathcal{Q}(M,N)
\label{Q_limit}%
\end{equation}
exists, and clearly coincides with the infimum, i.e.,%
\begin{equation}
\mathcal{Q}_{\infty}(N)=\inf_{M}\mathcal{Q}(M,N)~. \label{Q_inf}%
\end{equation}
Explicitly, we compute%
\begin{equation}
\mathcal{Q}_{\infty}(N)=\frac{1}{2}\exp(-2Nx)~.
\end{equation}
By definition of limit, Eqs.~(\ref{Q_limit}) and~(\ref{Q_inf}) mean that
$\forall\varepsilon>0$, $\exists\bar{M}\in\mathbb{Z}^{+}$ such that $\forall
M>\bar{M}$%
\begin{equation}
\mathcal{Q}_{\infty}(N)\leq\mathcal{Q}(M,N)<\mathcal{Q}_{\infty}%
(N)+\varepsilon~.
\end{equation}
Now, since we have $P_{err}(T_{M,N})\leq\mathcal{Q}(M,N)$ for every $M$, it is
clear that $\forall\varepsilon>0$, $\exists\bar{M}\in\mathbb{Z}^{+}$ such that
$\forall M>\bar{M}$
\begin{equation}
P_{err}(T_{M,N})<\mathcal{Q}_{\infty}(N)+\varepsilon~. \label{Eq_prop}%
\end{equation}

In order to prove our result, we impose the threshold condition%
\begin{equation}
\mathcal{Q}_{\infty}(N)<\mathcal{C}(N)~, \label{QC_diseq0}%
\end{equation}
where $\mathcal{C}(N)$\ is the classical discrimination bound. According to
Corollary~\ref{Theo_FORMS_CORO}, this is given by%
\begin{equation}
\mathcal{C}(N)=\frac{1-\sqrt{1-z}}{2}~,
\end{equation}
where%
\begin{equation}
z=\exp(-Nx^{2})~.
\end{equation}
Now, if Eq.~(\ref{QC_diseq0}) is satisfied, then we can always take an
$\varepsilon>0$\ such that%
\begin{equation}
\mathcal{C}(N)=\mathcal{Q}_{\infty}(N)+\varepsilon~.
\end{equation}
Then, because of the proposition of Eq.~(\ref{Eq_prop}), we have that
$\exists\bar{M}\in\mathbb{Z}^{+}$ such that $\forall M>\bar{M}$
\begin{equation}
P_{err}(T_{M,N})<\mathcal{C}(N)~.
\end{equation}

\begin{figure}[ptbh]
\vspace{-0.3cm}
\par
\begin{center}
\includegraphics[width=0.47\textwidth] {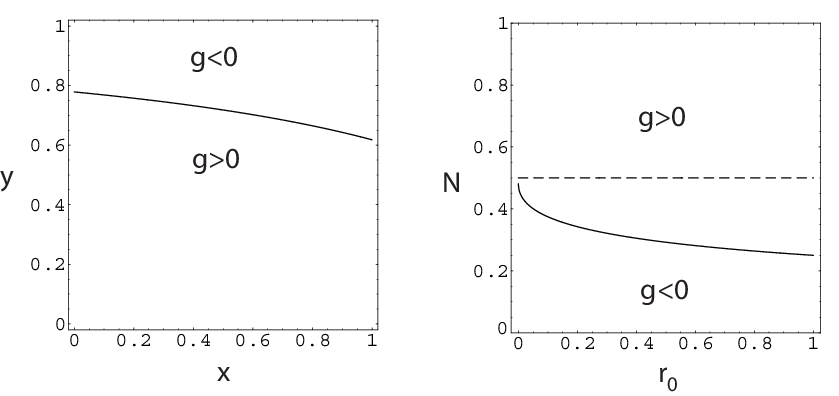}
\end{center}
\par
\vspace{-0.4cm}\caption{\textbf{Left.} Zero-level function $\bar{y}(x)$
plotted in the finite $(x,y)$-plane. Only for $y<\bar{y}(x)$ we have
$g(x,y)>0$. \textbf{Right.} Energy function $\bar{N}(r_{0})$ plotted in the
infinite $(r_{0},N)$-plane (here restricted to the sector $0<N\leq1$). Only
for $N>$ $\bar{N}(r_{0})$ we have $g=g(r_{0},N)>0$. Dashed line corresponds to
the universal energy threshold $N_{th}=1/2$.}%
\label{EneTHR}%
\end{figure}

Clearly, the next step is to solve Eq.~(\ref{QC_diseq0}) and get a threshold
condition on the energy $N$. After simple Algebra, Eq.~(\ref{QC_diseq0}) can
be written as%
\begin{equation}
g(x,y)>0~, \label{g_cond}%
\end{equation}
where%
\begin{equation}
g(x,y):=y^{x^{2}}+y^{4x}-2y^{2x}~,
\end{equation}
and%
\begin{equation}
y:=\exp(-N)\in(0,1)~. \label{y_N_form}%
\end{equation}
Note that, in the definition of $y$, we are excluding the singular points
$N=0,+\infty$. The function $g(x,y)$ can be easily analyzed on the plane
$(0,1]\times(0,1)$. In particular, its zero-level $g(x,y)=0$ is represented by
the continuous function $\bar{y}(x)$ which is plotted in Fig.~\ref{EneTHR}%
(left). The threshold condition of Eq.~(\ref{g_cond}) is equivalent to
\begin{equation}
y<\bar{y}(x)~. \label{thr_y}%
\end{equation}
As evident from Fig.~\ref{EneTHR}(left), the function $\bar{y}(x)$ is
decreasing in $x\in(0,1]$ with extremal values%
\begin{equation}
\lim_{x\rightarrow0^{+}}\bar{y}(x)=\sup_{x\in(0,1]}\bar{y}(x)=e^{-1/4}~,
\end{equation}
and%
\begin{equation}
\lim_{x\rightarrow1^{-}}\bar{y}(x)=\bar{y}(1)=\min_{x\in(0,1]}\bar{y}%
(x)=\frac{\sqrt{5}-1}{2}~.
\end{equation}

Equivalently, by using Eqs.~(\ref{x_r0}) and~(\ref{y_N_form}), we can put the
threshold condition of Eq.~(\ref{thr_y}) in terms of $r_{0}$\ and $N$. In
particular, Eq.~(\ref{thr_y}) takes the form%
\begin{equation}
N>\bar{N}(r_{0})~, \label{EQ_condN}%
\end{equation}
where $\bar{N}(r_{0})$ is the decreasing function of $r_{0}\in\lbrack0,1)$
which is plotted in Fig.~\ref{EneTHR}(right). This function has extremal
values%
\begin{equation}
\bar{N}(0)=\max_{r_{0}\in\lbrack0,1)}\bar{N}(r_{0})=\ln\frac{2}{\sqrt{5}-1}~,
\end{equation}
and%
\begin{equation}
\lim_{r_{0}\rightarrow1^{-}}\bar{N}(r_{0})=\inf_{r_{0}\in\lbrack0,1)}\bar
{N}(r_{0})=\frac{1}{4}~.
\end{equation}
In order to have a criterion which is universal, i.e., $r_{0}$-independent, we
can consider the sufficient condition%
\begin{equation}
N>\max_{r_{0}\in\lbrack0,1)}\bar{N}(r_{0})=\ln\frac{2}{\sqrt{5}-1}\simeq0.48~,
\end{equation}
which clearly implies Eq.~(\ref{EQ_condN}). More easily we can consider the
slightly-larger condition%
\begin{equation}
N\geq N_{th}:=1/2~.
\end{equation}
In conclusion, for every $r_{0}<r_{1}=1$ and (finite) $N\geq N_{th}:=1/2$,
there exists a minimum bandwidth $\bar{M}$ such that $\forall M>\bar{M}$ we
have $P_{err}(T_{M,N})<\mathcal{C}(N)$.$~\blacksquare$

\section{Sub-optimal receiver\label{APP_Bell}}

In our derivations we have assumed that the output receiver is able to perform
an optimal measurement given by the \textquotedblleft Helstrom's
POVM\textquotedblright\ of Eq.~(\ref{optimal_POVM}). Despite this measurement
has an extremely simple formula, it is not straightforward to implement it
using linear optics and photodetection. This problem has been already
considered in Ref.~\cite{Saikat} for the case of quantum illumination, where
an ingenious receiver design has been proven to harness quantum illumination
advantage. Unfortunately this kind of design cannot be applied directly to our
scheme, since quantum reading not only has a different task than quantum
illumination but also works in a completely different regime (high
reflectivities, low thermal-noise and narrowband signals). Note that the
working regime of quantum reading is such that strong correlations survive at
the output of the process. For this reason, we can consider simpler receiver
designs than the ones for quantum illumination. In fact, here we prove that an
output Bell detection followed by a simple classical processing is a
sub-optimal receiver able to provide remarkable advantages, i.e., comparable
to the Helstrom's POVM. Since Bell detection is a standard measurement
(involving linear optics and photodetection) our reading apparatus can be
easily implemented in today's quantum optics labs.

\begin{figure}[ptbh]
\vspace{-0.1cm}
\par
\begin{center}
\includegraphics[width=0.45\textwidth] {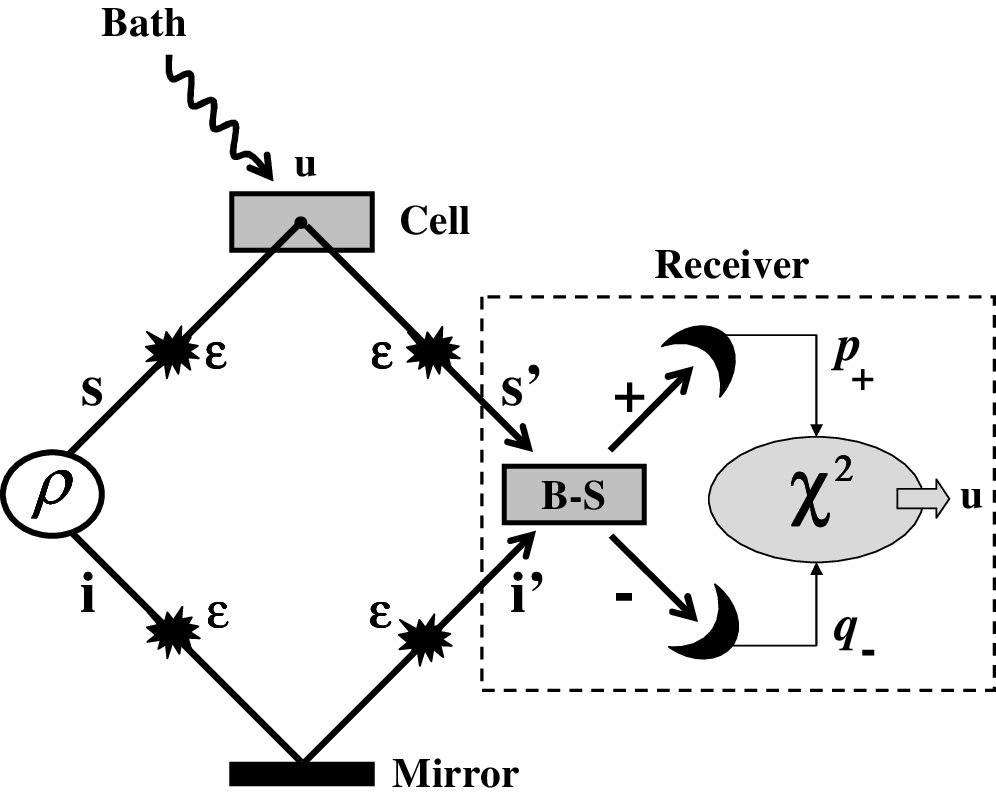}
\end{center}
\par
\vspace{-0.1cm}\caption{Quantum reading scheme with sub-optimal receiver. This
receiver consists of two parts. The first part is a Bell measurement, which
consists in a balanced beam-splitter followed by two homodyne detectors. The
second part is a classical processing of the outcome $\{q_{-},p_{+}\}$ by
means of a $\chi^{2}$-test. See text for more details.}%
\label{suboptimalPIC}%
\end{figure}

The quantum reading scheme with the sub-optimal receiver is depicted in
Fig.~\ref{suboptimalPIC}. The input modes $\{s,i\}$ belong to a TMSV state
$\rho=\left\vert \xi\right\rangle \left\langle \xi\right\vert $. These modes
are processed as usual, i.e., the signal mode $s$ is conditionally transformed
by the cell (reflectivity $r_{u}$\ encoding the bit $u$) while the idler mode
$i$ is sent directly to the receiver. Their quadratures, $\mathbf{\hat{x}}%
_{s}=(\hat{q}_{s},\hat{p}_{s})^{T}$ and $\mathbf{\hat{x}}_{i}=(\hat{q}%
_{i},\hat{p}_{i})^{T}$, are transformed according to the input-output
relations%
\begin{equation}
\mathbf{\hat{x}}_{s}\rightarrow\mathbf{\hat{x}}_{s^{\prime}}=\sqrt{r_{u}%
}(\mathbf{\hat{x}}_{s}+\mathbf{\hat{x}}_{\varepsilon})+\sqrt{1-r_{u}%
}\mathbf{\hat{x}}_{b}+\mathbf{\hat{x}}_{\varepsilon}^{\prime}~,
\end{equation}
and%
\begin{equation}
\mathbf{\hat{x}}_{i}\rightarrow\mathbf{\hat{x}}_{i^{\prime}}=\mathbf{\hat{x}%
}_{i}+2\mathbf{\hat{x}}_{\varepsilon}^{\prime\prime}~,
\end{equation}
where $\mathbf{\hat{x}}_{b}$ are the quadratures of the external thermal bath,
while $\mathbf{\hat{x}}_{\varepsilon}$, $\mathbf{\hat{x}}_{\varepsilon
}^{\prime}$ and $\mathbf{\hat{x}}_{\varepsilon}^{\prime\prime}$ are
quadratures associated with the internal thermal channels. At the receiver the
two modes $\{s^{\prime},i^{\prime}\}$ are combined in a balanced beam-splitter
which outputs the modes $\{-,+\}$ with quadratures%
\begin{equation}
\mathbf{\hat{x}}_{-}:=\left(
\begin{array}
[c]{c}%
\hat{q}_{-}\\
\hat{p}_{-}%
\end{array}
\right)  =\frac{\mathbf{\hat{x}}_{s^{\prime}}-\mathbf{\hat{x}}_{i^{\prime}}%
}{\sqrt{2}}~,
\end{equation}
and%
\begin{equation}
\mathbf{\hat{x}}_{+}:=\left(
\begin{array}
[c]{c}%
\hat{q}_{+}\\
\hat{p}_{+}%
\end{array}
\right)  =\frac{\mathbf{\hat{x}}_{s^{\prime}}+\mathbf{\hat{x}}_{i^{\prime}}%
}{\sqrt{2}}~.
\end{equation}
Here the \textquotedblleft EPR quadratures\textquotedblright\ $\{\hat{q}%
_{-},\hat{p}_{+}\}$ are measured by using two homodyne detectors. The
corresponding outcome $\{q_{-},p_{+}\}$ is classically processed via a
$\chi^{2}$-test, whose output is the value of the bit stored in the memory
cell. In general, for arbitrary bandwidth $M\geq1$, we have $M$ identical
copies $\left\vert \xi\right\rangle \left\langle \xi\right\vert ^{\otimes M}$
at the input and, therefore, $2M$ output variables $\{q_{-},p_{+}\}^{M}$ which
are processed by the $\chi^{2}$-test.

Let us analyze the detection process in detail in order to derive the
error-probability which affects the decoding. First of all we can easily
verify that the two classical outputs, $q_{-}$ and $p_{+}$, are identical
Gaussian variables, so that we can simply use $z$ to denote $q_{-}$ or $p_{+}%
$. Clearly the Gaussian variable $z$ is tested two times for each single-copy
state $\left\vert \xi\right\rangle \left\langle \xi\right\vert $ and,
therefore, $2M$ times during the reading time of the logical bit. This
variable has zero mean and variance $V=V(u)$ which is conditioned to the value
of the stored bit $u=0,1$. Explicitly, we have%
\begin{align}
V(u)  &  =\frac{1}{2}\left[  r_{u}(\mu+\varepsilon)+(1-r_{u})\beta+\right.
\nonumber\\
&  \left.  \mu+3\varepsilon-2\sqrt{r_{u}(\mu^{2}-1)}\right]  ~, \label{V_u}%
\end{align}
where $r_{u}$ is the conditional reflectivity of the cell, $\mu:=2n_{S}+1$ is
the variance associated with the signal-mode, $\beta:=2\bar{n}+1$ is the
noise-variance of the external thermal bath, and $\varepsilon$ is the
noise-variance of the internal thermal channel. It is clear that the decoding
of the logical bit $u$ corresponds to the statistical discrimination between
two variances $V(0)$ and $V(1)$. In other words, the variable $z$ is subject
to the hypothesis test%
\begin{equation}
\left\{
\begin{array}
[c]{c}%
\mathrm{H}_{0}~(u=0)~:~V=V(0)~,\\
\\
\mathrm{H}_{1}~(u=1)~:~V=V(1)~.
\end{array}
\right.  \label{test1}%
\end{equation}
It is easy to show that $V$ of Eq.~(\ref{V_u}) is a decreasing function of the
reflectivity $r_{u}$ as long as
\begin{equation}
r_{u}<\frac{\mu^{2}-1}{(\varepsilon-\beta+\mu)^{2}}~,
\end{equation}
which is always true in the regime considered here (i.e., $\varepsilon$ small
and $\beta$ close to $1$). This means that $r_{0}<r_{1}$ implies $V(0)>V(1)$,
which makes Eq.~(\ref{test1}) a one-tailed test. Equivalently, we can
introduce the normalized variable%
\begin{equation}
z^{\prime}:=\frac{z}{\sqrt{V(1)}}~,
\end{equation}
with zero mean and variance $V^{\prime}(u)=V(u)/V(1)$, and replace
Eq.~(\ref{test1}) with the one-tailed test%
\begin{equation}
\left\{
\begin{array}
[c]{c}%
\mathrm{H}_{0}~:~V^{\prime}=1+\Sigma~,\\
\\
\mathrm{H}_{1}~:~V^{\prime}=1~.~~~~~~
\end{array}
\right.  \label{test2}%
\end{equation}
where%
\begin{equation}
\Sigma:=\frac{V(0)-V(1)}{V(1)}>0~.
\end{equation}
Thus, assuming $r_{0}<r_{1}$, the decoding of the logical bit is equivalent to
the statistical discrimination between the two hypotheses in Eq.~(\ref{test2}%
), i.e., $V^{\prime}>1$ and $V^{\prime}=1$. For arbitrary bandwidth $M$, we
collect $2M$ independent outcomes $\{z_{1},\cdots z_{2M}\}$ and we construct
the $\chi^{2}$-variable%
\begin{equation}
\theta:=\sum_{k=1}^{2M}(z_{k}^{\prime})^{2}~.
\end{equation}
Then, we select one of the two hypotheses according to the following rule
\begin{equation}
\left\{
\begin{array}
[c]{c}%
\text{Accept }\mathrm{H}_{0}~\Leftrightarrow~\theta\geq\mathcal{Q}_{1-\varphi
}^{2M-1}~,\\
\\
\text{Accept }\mathrm{H}_{1}~\Leftrightarrow~\theta<\mathcal{Q}_{1-\varphi
}^{2M-1}~.
\end{array}
\right.
\end{equation}
where $\mathcal{Q}_{1-\varphi}^{2M-1}$ is the $(1-\varphi)^{th}$ quantile of
the $\chi^{2}$ distribution with $2M-1$ degrees of freedom. Here the quantity
$\varphi$ represents the significance level of the test, corresponding to the
asymptotic probability of wrongly rejecting the hypothesis $\mathrm{H}_{1}$,
i.e.,%
\begin{equation}
\varphi=\lim_{M\rightarrow\infty}P(\mathrm{H}_{0}|\mathrm{H}_{1})~.
\end{equation}
This quantity must be fixed and its value characterizes the test. In
particular, for finite $M$, there will be an optimal value of $\varphi$ which
maximizes the performance of the test.

\begin{figure}[ptbh]
\vspace{-0.1cm}
\par
\begin{center}
\includegraphics[width=0.48\textwidth] {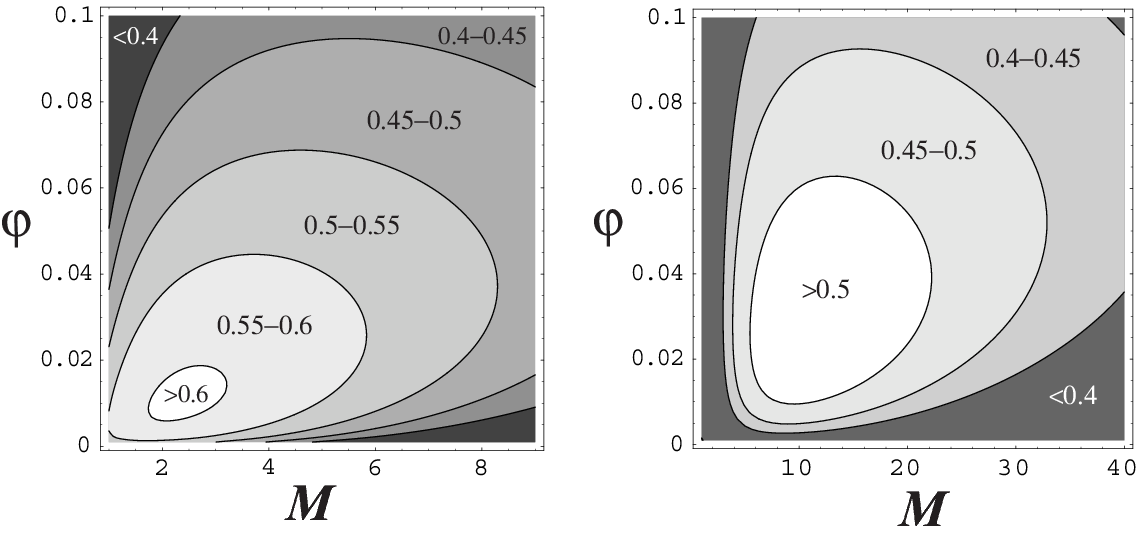}
\end{center}
\par
\vspace{-0.7cm}\caption{(This is a repetition of Fig.~\ref{BellPIC}).
\textbf{Left.} Information gain $G_{test}$ optimized over the bandwidth $M$
(i.e., number of TMSV states) and the significance level of the test $\varphi
$. The gain can be higher than $0.6$ bits per cell. The results are shown for
the pure-loss model ($\varepsilon=\bar{n}=0$) considering $r_{0}=0.85$,
$r_{1}=1$ and $N=35$. \textbf{Right.} Information gain $G_{test}$ optimized
over $M$ and $\varphi$. The results are shown for the thermal-loss model
($\bar{n}=\varepsilon=10^{-5}$) considering $r_{0}=0.85$, $r_{1}=0.95$,
$N=100$ and $M^{\ast}=10^{6}$.}%
\label{Bellpic2}%
\end{figure}

Let us explicitly compute the error probability affecting the classical test.
For arbitrary variance $V^{\prime}$, the $\chi^{2}$ distribution with $2M-1$
degrees of freedom is equal to%
\begin{equation}
P_{M,V^{\prime}}(\theta)=\frac{\theta^{M-1}\exp\left(  -\frac{\theta
}{2V^{\prime}}\right)  }{(2V^{\prime})^{M}\Gamma\left(  M\right)  }~,
\end{equation}
where $\Gamma(x)$ is the Gamma function. The probability of finding $\theta$
bigger than $t$ is given by the integral%
\begin{equation}
I\left(  M,V^{\prime},t\right)  :=\int_{t}^{+\infty}P_{M,V^{\prime}}%
(\theta)d\theta=\frac{\Gamma\left(  M,\frac{t}{2V^{\prime}}\right)  }%
{\Gamma\left(  M\right)  }~.
\end{equation}
where $\Gamma\left(  x,y\right)  $ is the incomplete Gamma function. Thus,
given $\mathrm{H}_{1}$ (i.e., $V^{\prime}=1$) the probability of accepting
$\mathrm{H}_{0}$ (i.e., $\theta\geq\mathcal{Q}_{1-\varphi}^{2M-1}$) is given by%

\begin{equation}
P(\mathrm{H}_{0}|\mathrm{H}_{1})=I\left(  M,1,\mathcal{Q}_{1-\varphi}%
^{2M-1}\right)  ~.\label{ph1}%
\end{equation}
By contrast, given $\mathrm{H}_{0}$ (i.e., $V^{\prime}=1+\Sigma$) the
probability of accepting $\mathrm{H}_{1}$ (i.e., $\theta<\mathcal{Q}%
_{1-\varphi}^{2M-1}$) is given by%
\begin{equation}
P(\mathrm{H}_{1}|\mathrm{H}_{0})=1-I\left(  M,1+\Sigma,\mathcal{Q}_{1-\varphi
}^{2M-1}\right)  ~.\label{ph2}%
\end{equation}
From Eqs.~(\ref{ph1}) and~(\ref{ph2}) we compute the error probability
affecting the test, which is given by%
\begin{gather}
P_{test}:=\frac{1}{2}\left[  P(\mathrm{H}_{0}|\mathrm{H}_{1})+P(\mathrm{H}%
_{1}|\mathrm{H}_{0})\right]  \nonumber\\
=\frac{1}{2}+\frac{1}{2}\left[  I\left(  M,1,\mathcal{Q}_{1-\varphi}%
^{2M-1}\right)  -I\left(  M,1+\Sigma,\mathcal{Q}_{1-\varphi}^{2M-1}\right)
\right]  \nonumber\\
=P_{test}\left(  r_{0},r_{1},N,M,\bar{n},\varepsilon,\varphi\right)  ~.
\end{gather}
Clearly this quantity depends on all the parameters of the model, i.e., memory
reflectivities $\{r_{0},r_{1}\}$, energy and bandwidth of the signal
$\{N,M\}$, levels of noise $\{\bar{n},\varepsilon\}$ and significance level of
the test $\varphi$. This error probability must be compared with the classical
discrimination bound $\mathcal{C}$. For simplicity let us first consider the
pure-loss model ($\bar{n}=\varepsilon=0$). In this case $P_{test}%
=P_{test}\left(  r_{0},r_{1},N,M,\varphi\right)  $ must be compared with
$\mathcal{C}=\mathcal{C}\left(  r_{0},r_{1},N\right)  $\ of
Eq.~(\ref{C_attenuators}). Then, for given memory $\{r_{0},r_{1}\}$ and
signal-energy $N$, we have that quantum reading is superior if we find a
bandwidth $M$ and a significance level $\varphi$ such that $P_{test}%
<\mathcal{C}$. This is equivalent to prove the positivity of the (sub-optimal)
information gain
\begin{equation}
G_{test}=1-H(P_{test})-[1-H(\mathcal{C})]~,\label{gainGtest}%
\end{equation}
which provides a lowerbound to the number of bits per cell which are gained by
quantum reading over any classical strategy. In Fig.~\ref{Bellpic2}(left) we
optimize $G_{test}$ over $M$\ and $\varphi$ for an ideal memory in the
few-photon regime. Remarkably, we can find an area where $G_{test}>0.6$ bits
per cell. 

Then, let us consider the thermal-loss model, with $\bar{n}=\varepsilon
=10^{-5}$. In this case, $P_{test}=P_{test}\left(  r_{0},r_{1},N,M,\bar
{n},\varepsilon,\varphi\right)  $ must be compared with $\mathcal{C}%
=\mathcal{C}\left(  r_{0},r_{1},N,M^{\ast},\bar{n},\varepsilon\right)  $ where
the value of $M^{\ast}$ is high enough to include all the classical
transmitters which are meaningful for the model (here we take $M^{\ast}%
=10^{6}$). In Fig.~\ref{Bellpic2}(right), we consider a memory with
$r_{0}=0.85$ and $r_{1}=0.95$ (high-reflectivity regime), which is illuminated
by a signal with $N=100$ (few-photon regime). The numerical optimization over
$M$\ and $\varphi$ shows an area where $G_{test}>0.5$ bits per cell.

Thus, the remarkable advantages of quantum reading are still evident when the
optimal Helstrom's POVM is replaced by a sub-optimal receiver, consisting of a
Bell measurement followed by a suitable classical post-processing. This
sub-optimal receiver can be easily realized in today's quantum optics labs,
thus making quantum reading a technique within the catch of current technology.

\section{Memory Model with Error Correction\label{SEC_ECcodes}}

In our study, we have considered a theoretical model of memory where each cell
stores exactly one bit of information. Then, by fixing the mean total number
of photons which are irradiated over each memory cell, we have computed the
average information which is retrieved by an input transmitter $T$ and an
optimal output detection. This quantity ranges from zero to one and can be
written as $I(T):=1-H(P_{err})$, were $P_{err}$ is the error-probability
corresponding to $T$. In this scenario, we have compared the information which
is retrieved by a non-classical EPR\ transmitter, lower-bounded by
$I_{epr}:=1-H(Q)$, with the information retrieved by an arbitrary classical
transmitter, upper-bounded by $I_{c}:=1-H(C)$. This comparison has been
quantified by the gain of information $G=I_{epr}-I_{c}$\ introduced in
Eq.~(\ref{gainG}). In doing this comparison, we have also considered the
presence of thermal noise and the practical case of a sub-optimal receiver,
which implies a weaker gain of information as quantified by
Eq.~(\ref{gainGtest}). Remarkably, we have found regimes (few photons and high
reflectivities) where $G$ is strictly positive and can be even close to $1$,
meaning that $I_{epr}\simeq1$ (i.e., an EPR\ transmitter reads all the
information) while $I_{c}\simeq0$ (i.e., no information can be read by any
classical transmitter).

Now it is important to note that our comparison can also be stated in another
equivalent way. Instead of storing one bit per cell and evaluating the
information read by a transmitter $T$, we can store a logical bit in a block
of $m$ cells by using a classical error correcting (EC) code and calculate the
minimum $m$ which is needed for a flawless readout by $T$. Then, we compare
the block sizes, $m_{epr}$ and $m_{c}$, which are allowed by an
EPR\ transmitter and an optimal classical transmitter. In this scenario, the
positivity condition $G>0$ corresponds to $m_{epr}<m_{c}$, meaning that an EPR
transmitter involves less overhead of error correction. In particular, for
$G\rightarrow1$ we have $m_{epr}\rightarrow1$ and $m_{c}\rightarrow\infty$.
This corresponds to having negligible overhead for an EPR\ transmitter versus
infinite\ overhead for any classical transmitter. In this model, the reading
time of a logical bit is clearly proportional to the block size $m$. As a
result, $m\rightarrow\infty$ corresponds to infinite reading time, i.e., the
impossibility to read information. Equivalently, if we fix the total data-size
of the memory, its logical capacity is inversely proportional to the
block-size $m$. Then, $m\rightarrow\infty$ corresponds to zero logical capacity.

Despite theoretically equivalent, this alternative approach is interesting for
practical implementations. In fact, actual digital memories are written using
EC codes. As an example, today's CDs are written using Reed-Solomon codes,
which are responsible for an error-correction overhead that is around the
15-20\% of the total data \cite{DVDbook}. In the remainder of the section, we
discuss in more detail the memory model based on error-correction, by showing
explicit cases where $m_{epr}$ is low while $m_{c}\gg1$. We start by
considering repetition codes since they can easily provide the order of
magnitude of the ratio $m_{c}/m_{epr}$. Then, we refine our derivation by
considering optimal EC codes, for which we exploit both the Hamming
\cite{HammREF} and Gilbert-Varshamov bounds \cite{GilbertREF,VarshaREF}.

\subsection{Repetition Codes}

\begin{figure}[ptbh]
\vspace{-0.4cm}
\par
\begin{center}
\includegraphics[width=0.50\textwidth] {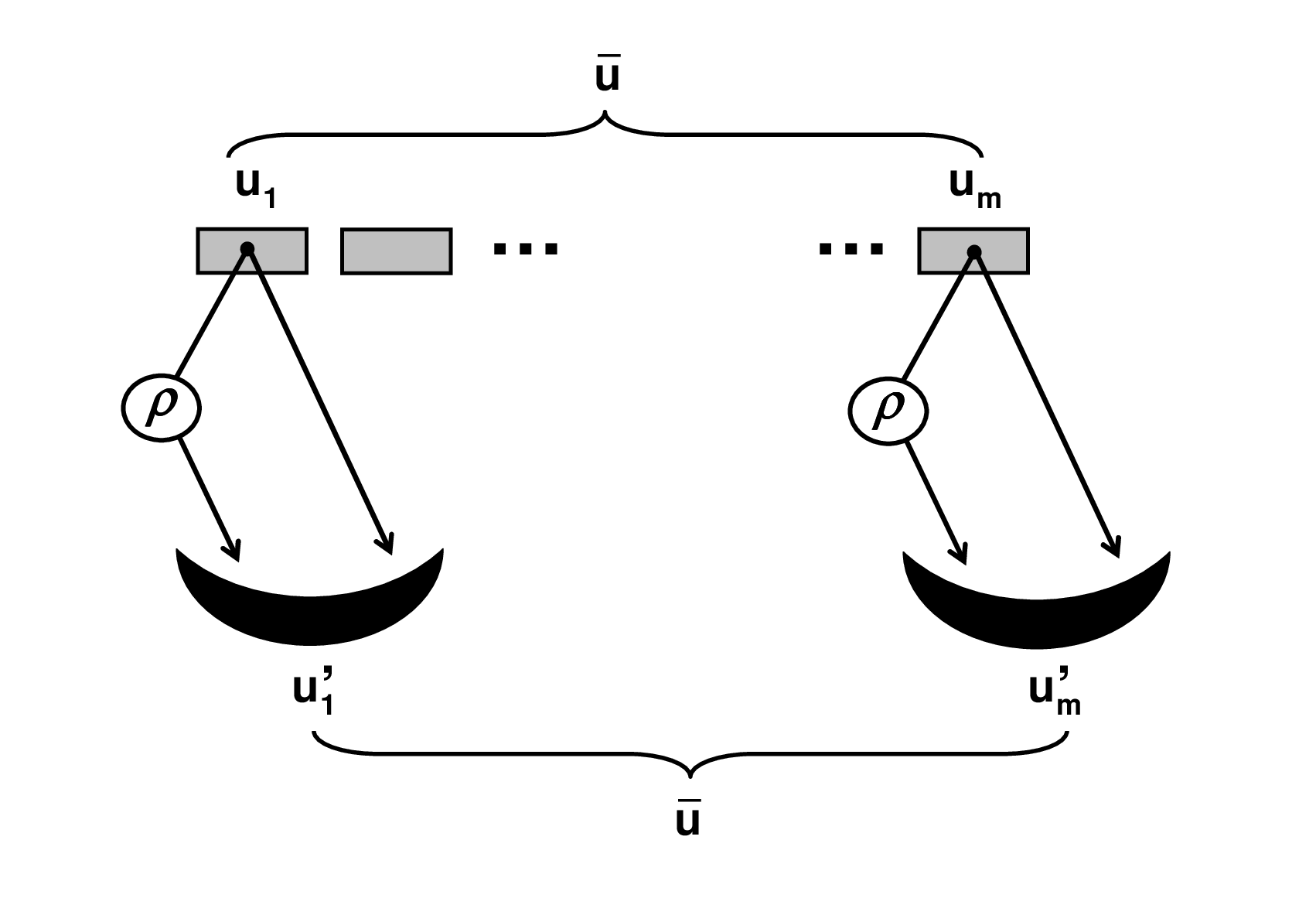}
\end{center}
\par
\vspace{-0.9cm}\caption{Model of memory with classical error correction
(repetition codes). A logical bit $\bar{u}$ is encoded in a block of $m$ cells
by using the codeword $u_{1}u_{2}\cdots u_{m}$ of a repetition code. Each cell
of the block is sequentially read by a transmitter $T(M,L,\rho)$ and an
optimal output detector. After the whole block is read, the output noisy
codeword $u_{1}^{\prime}u_{2}^{\prime}\cdots u_{m}^{\prime}$ is corrected by
majority voting to provide the encoded bit $\bar{u}$ up a logical error
probability $\bar{p}(m,P_{err})$.}%
\label{BlockMODEL}%
\end{figure}

Let us store a logical bit $\bar{u}$ in a block of $m$ cells by using an
$m$-bit repetition code, where $m=2t+1$ is an odd number ($t=0,1,...$) This
means that the logical bit $\bar{u}=\bar{0},\bar{1}$ is encoded in $m$
physical bits $u_{1}u_{2}\cdots u_{m}$ via the codewords%
\begin{equation}
\bar{0}=\underbrace{00\cdots0}_{m},\quad\bar{1}=\underbrace{11\cdots1}_{m}.
\end{equation}
Each physical bit $u_{i}$\ is stored in a corresponding cell (see
Fig.~\ref{BlockMODEL}). Each cell of the block is sequentially read by an
input transmitter $T=T(M,L,\rho)$, signalling $N$\ photons, and an optimal
output detector. The value of each physical bit $u_{i}$ is retrieved up to an
error probability $P_{err}$ which depends on the specific transmitter $T$.
After reading all the cells of the block, the output codeword $u_{1}^{\prime
}u_{2}^{\prime}\cdots u_{m}^{\prime}$ is corrected by majority voting (see
Fig.~\ref{BlockMODEL}).

\begin{figure}[ptbh] \vspace{-0.0cm}
\par
\begin{center}
\includegraphics[width=0.32\textwidth] {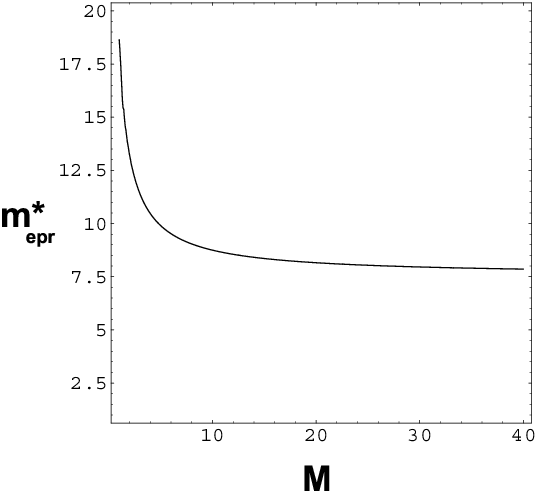}
\end{center}
\par
\vspace{-0.7cm}\caption{Quantum upper-bound $m_{epr}^{\ast}$ versus the
bandwidth $M$ of the EPR\ transmitter. Other parameters are $r_{0}=0.95$,
$r_{1}=1$ and $N=100$. The curve represents an upper-bound for the size of the
encoding block which enables an EPR\ transmitter $T_{M,N}$ to retrieve logical
data without errors (i.e., up to a cut-off $\varepsilon=10^{-9}$). }%
\label{Bsize}%
\end{figure}

Depending on the number of bit-flips, error recovery may fail or not. Up to
$t=(m-1)/2$ bit-flips are correctable, while more than $t$ bit-flips leads to
a logical error. As a result, the input logical bit $\bar{u}$ is retrieved up
to a logical error probability%
\begin{equation}
\bar{p}(m,P_{err})=\sum_{i=\frac{m+1}{2}}^{m}\left(
\begin{array}
[c]{c}%
m\\
i
\end{array}
\right)  (P_{err})^{i}(1-P_{err})^{m-i}~. \label{uncorregible_nbits}%
\end{equation}
By definition, we say that the readout is \textquotedblleft
flawless\textquotedblright\ if $\bar{p}(m,P_{err})\leq\varepsilon$ for some
small cut-off $\varepsilon$. The value of $\varepsilon$ depends on the size of
the memory and the error tolerance that we allow. For instance, for memories
of $1$~Gbit, the value $\varepsilon=10^{-9}$\ corresponds to having less than
$1$ bit of logical data corrupted. Given a transmitter $T$\ (and therefore an
error probability $P_{err}$), the resolution of the equation $\bar
{p}(m,P_{err})=10^{-9}$ identifies a corresponding size $m^{\ast}$\ for the
encoding block. Clearly, the best transmitter is the one who minimizes the
value of $m^{\ast}$. Thus, for fixed signal-energy $N$, we compare the
block-sizes of EPR\ and classical transmitters. Given an EPR\ transmitter
$T_{M,N}$, with bandwidth $M$ and energy $N$, we use the bound $P_{err}%
\leq\mathcal{Q}(M,N)$ of Eq.~(\ref{Q_M_Ns}) to over-estimate the corresponding
block-size $m_{epr}^{\ast}$ (upper-bound). Then, for every classical
transmitter $T_{c}$ irradiating $N$ photons, we use the bound $P_{err}%
\geq\mathcal{C}(N)$ of Eq.~(\ref{C_class_bound}) to under-estimate
$m_{c}^{\ast}$ (lower-bound) \cite{NOTA1}. It is clear that $m_{epr}^{\ast
}<m_{c}^{\ast}$ provides a sufficient condition for the superiority of quantum
reading. In particular, we can easily find configurations where $m_{epr}%
^{\ast}$\ is of the order of units while $m_{c}^{\ast}\gg1$. As a
numerical example, let us consider an ideal memory with
$r_{0}=0.95$ and $r_{1}=1$, where each cell is irradiated by
$N=100$ photons. From Fig.~\ref{Bsize}, we can see that
$m_{epr}^{\ast}$ is decreasing in $M$. In particular, we have
$m_{epr}^{\ast}\lesssim9$ already for small values of $M$, i.e.,
narrowband EPR transmitters. This numerical result shows that
using EPR\ transmitters we can perfectly read data up to a limited
error-correction overhead (as we will show afterwards this
overhead can be reduced by using more efficient EC
codes).

Under the same conditions, the classical lower-bound is extremely high, since
we have $m_{c}^{\ast}\approx560$. In other words, classical transmitters need
a huge overhead of error correction, which is more than $60$ times the one
needed by EPR transmitters. This clearly makes classical transmitters useless
in the present configuration. Similar results can be numerically found with
other choices of parameters in the regime of few-photons and
high-reflectivities. In the following section we study the model by using
optimal EC codes instead of repetition codes. In this more refined derivation,
we will see that the EC overhead for classical transmitters is more than $80$
times the one needed by EPR\ transmitters.

\subsection{Optimal Error Correcting Codes}

The previous derivation based on repetition codes can be generalized to more
powerful EC codes. In fact, a more general model of memory consists of
encoding $k\geq1$ logical bits in a block of $m$ cells by using an arbitrary
EC\ code $[m,k,d]$ with distance. This EC code is able to correct up to
$t=\lfloor\frac{d-1}{2}\rfloor$ bit-flips, where $\lfloor x\rfloor$ represents
the floor function (i.e., the largest integer not greater than $x$). In other
words, it corrects up to $(d-1)/2$ bit-flips for odd $d$, and up to $(d/2)-1$
bit-flips for even $d$. By definition, we call $R=k/m$ the rate of the
code\ and $\delta=d/m$ its relative distance. In our memory we fix the size of
the block to be very large, e.g., $m=2000$, which is comparable to the size of
the data-blocks in current CDs and DVDs. Then, we determine the value of the
relative distance $\delta$ which allows a flawless readout of the memory (up
to a cut-off $\varepsilon$). Given this value, we choose an EC code which
optimizes the rate $R$. The optimal value of $R$ falls in the following range
\begin{equation}
\underline{R}:=1-H(\delta)\leq R\leq1-H(\delta/2):=\overline{R}~,
\label{Hamming}%
\end{equation}
where $H$ is the binary Shannon entropy, i.e.,%
\begin{equation}
H(x):=-x\log_{2}x-(1-x)\log_{2}(1-x)~. \label{binSHA}%
\end{equation}
The inequality of Eq.~(\ref{Hamming}) comes from combining the Hamming
(upper)bound and the Gilbert-Varshamov (lower)bound, that we report in
Sec.~\ref{Hamming_SEC}\ for the sake of completeness.

In our quantum-classical comparison, we fix the signal-energy $N$ irradiated
over each cell. Then, we compare the optimal rate $R_{epr}$ which is
achievable by using an EPR transmitter $T_{M,N}$ with the optimal rate $R_{c}$
achievable by classical transmitters $T_{c}$. More exactly, we compare the
quantum lower-bound $\underline{R}_{epr}$ with the classical upper-bound
$\overline{R}_{c}$, since $\underline{R}_{epr}>\overline{R}_{c}$ implies
$R_{epr}>R_{c}$. In particular, we can show configurations where
$\underline{R}_{epr}$ is close to one while $\overline{R}_{c}$ is close to
zero. This means that EPR\ transmitters need a minimal error correction
overhead for retrieving the data perfectly. By contrast, classical
transmitters need an overhead which is unfeasible for practical
implementations. In the following we discuss the quantum-classical comparison
in more detail.

\begin{figure}[ptbh]
\vspace{-0.4cm}
\par
\begin{center}
\includegraphics[width=0.50\textwidth] {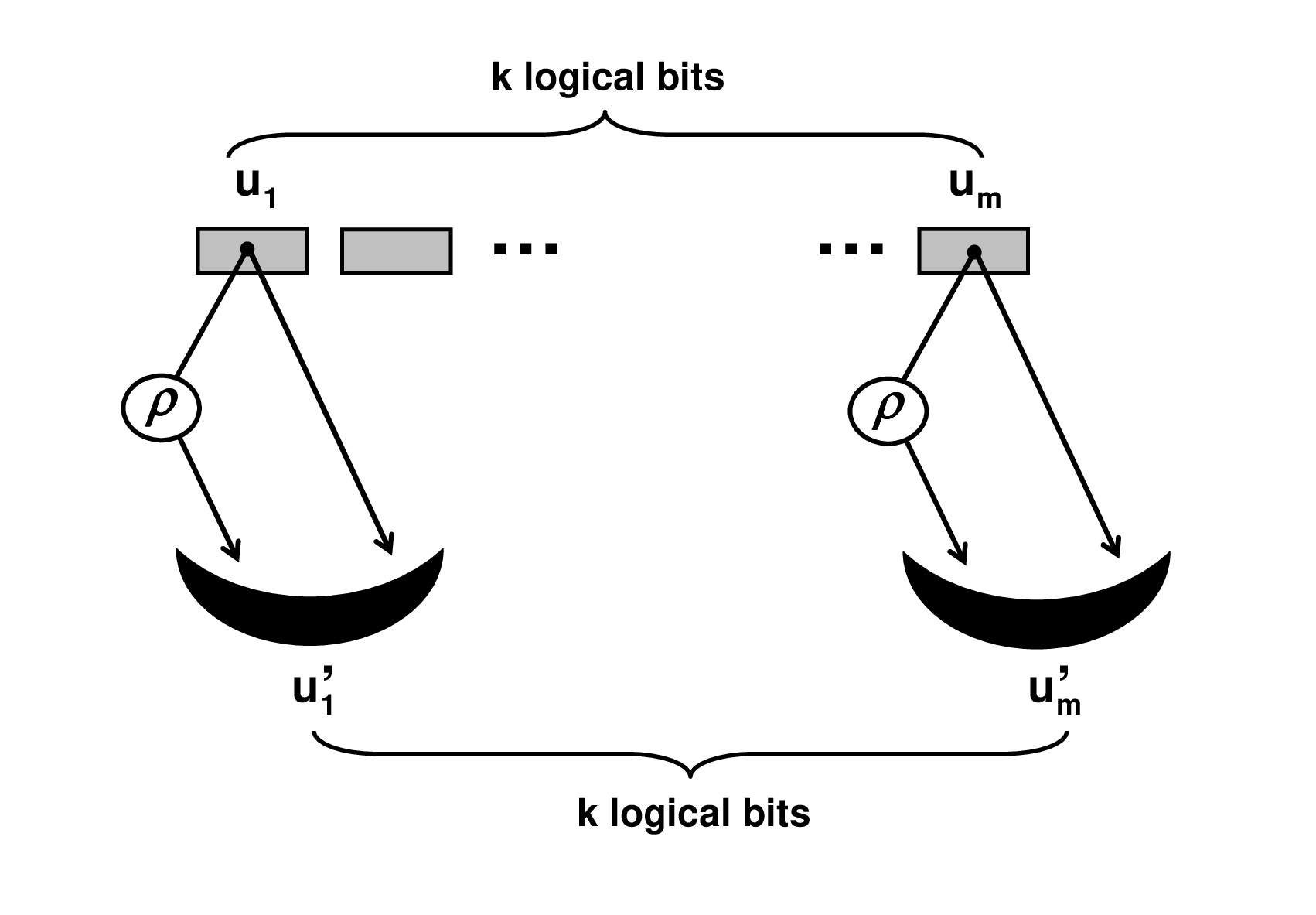}
\end{center}
\par
\vspace{-0.8cm}\caption{Model of memory with classical error correction
(optimal EC\ codes). By using an optimal EC\ code $[m,k,d]$ we encode $k$
logical bits in a large block of $m$ cells. Each cell of the block is
sequentially read by a transmitter $T(M,L,\rho)$ and an optimal output
detector. After the whole block is read, the output noisy codeword
$u_{1}^{\prime}u_{2}^{\prime}\cdots u_{m}^{\prime}$ is corrected by standard
procedures of syndrome detection and error recovery. The final outcome
provides the encoded $k$ logical bits up an error probability $\bar
{p}(d,m,P_{err})$.}%
\label{BlockMODELgen}%
\end{figure}

Let us consider a memory which is subdivided in large blocks, each one
composed of $m=2000$ cells. In each block the information is stored by using
an optimal EC\ code $[m,k,d]$ where $k$ and $d$ are to be determined or,
equivalently, their relative quantities $R=k/m$ and $\delta=d/m$. For a given
a transmitter $T$, we have a corresponding error probability $P_{err}$
affecting the readout of every single cell. After all the cells of one block
are read, the output codeword is processed using standard procedures of
syndrome detection and error recovery (see Fig.~\ref{BlockMODELgen}). The
probability of an uncorrectable error (which causes the lost of $k$\ logical
bits) corresponds to the probability of having more than $t=\lfloor\frac
{d-1}{2}\rfloor$ bit-flips in the block. This quantity is given by
\cite{NOTA2}%
\begin{equation}
\bar{p}(d,m,P_{err})=\sum_{i=\lfloor\frac{d+1}{2}\rfloor}^{m}\left(
\begin{array}
[c]{c}%
m\\
i
\end{array}
\right)  (P_{err})^{i}(1-P_{err})^{m-i}~.
\end{equation}
As usual we say that the readout is \textquotedblleft
flawless\textquotedblright\ if $\bar{p}(d,m,P_{err})\leq\varepsilon$ for some
small cut-off $\varepsilon$, that we set equal to $10^{-9}$\ as before. Now
for fixed block-size $m=2000$ and cut-off $\varepsilon=10^{-9}$, the
resolution of the equation $\bar{p}(d,m,P_{err})=\varepsilon$ provides the
distance $d$\ as a function of $P_{err}$. Thus, given a transmitter $T$, i.e.,
an error probability $P_{err}$, we have a corresponding minimum distance
$d^{\ast}$ for the code. Once that $d^{\ast}$ is determined, we consider the
maximum number of logical bits $k^{\ast}$ which can be stored by an EC code.
It is clear that $k^{\ast}$ is decreasing in $d^{\ast}$, i.e., we can encode
less logical bits if we must correct more errors. In terms of relative
quantities, this means that a transmitter $T\ $is associated with a relative
distance $\delta^{\ast}=d^{\ast}/m$ (which is determined by $P_{err}$)\ and a
corresponding optimal rate $R^{\ast}=k^{\ast}/m$ (which is determined by the
maximization over all the EC\ codes with relative distance $\delta^{\ast}$).

\begin{figure}[ptbh]
\vspace{-0.0cm}
\par
\begin{center}
\includegraphics[width=0.42\textwidth] {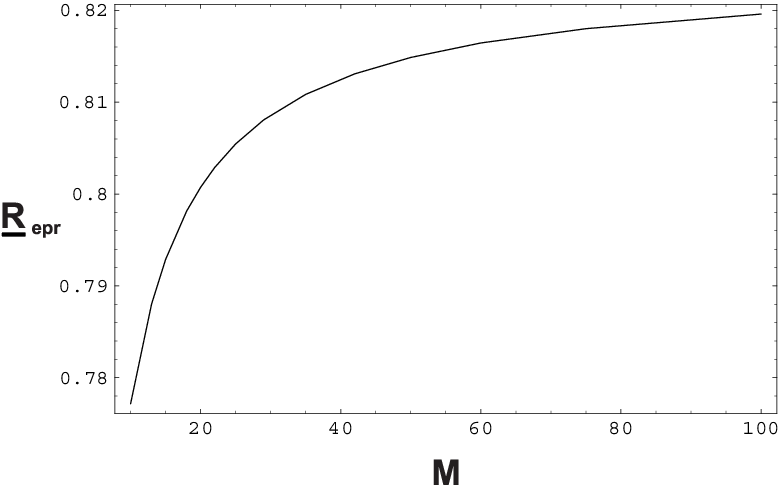}
\end{center}
\par
\vspace{-0.4cm}\caption{Quantum lower-bound $\underline{R}_{epr}$ versus the
bandwidth $M$ of the EPR\ transmitter. Other parameters are $r_{0}=0.95$,
$r_{1}=1,$ $N=100$, $m=2000$ and $\varepsilon=10^{-9}$. The curve represents a
lower-bound for the rate (information bits per cell) which is achievable by an
EPR\ transmitter $T_{M,N}$ combined with an optimal EC\ code.}%
\label{rateECS}%
\end{figure}

As mentioned before we perform our quantum-classical comparison by resorting
to lower and upper bounds. By fixing the energy $N$ irradiated over each cell,
we compare EPR transmitters $T_{M,N}$\ with classical transmitters $T_{c}$.
For an EPR transmitter $T_{M,N}$ we use the bound $P_{err}\leq\mathcal{Q}%
(M,N)$ to over-estimate the relative distance $\delta_{epr}^{\ast}$. This
provide an under-estimation of the optimal rate $R_{epr}^{\ast}$. Then, we use
the bound $\underline{R}_{epr}:=1-H(\delta_{epr}^{\ast})\leq R_{epr}^{\ast}$
of Eq.~(\ref{Hamming}) to provide a further under-estimation of the rate. For
every classical transmitter $T_{c}$ irradiating $N$ photons, we use the bound
$P_{err}\geq\mathcal{C}(N)$ to under-estimate $\delta_{c}^{\ast}$. This
provides an over-estimation of the optimal rate $R_{c}^{\ast}$. Then, we use
the bound $\overline{R}_{c}:=1-H(\delta_{c}^{\ast}/2)\geq R_{c}^{\ast}$ of
Eq.~(\ref{Hamming}) to provide a further over-estimation of this rate. Thus,
we compare the quantum lower-bound $\underline{R}_{epr}$ with the classical
upper-bound $\overline{R}_{c}$, since $\underline{R}_{epr}>\overline{R}_{c}$
provides a sufficient condition for the superiority of quantum reading. In
order to show $\underline{R}_{epr}>\overline{R}_{c}$, let us consider the
previous example of an ideal memory with $r_{0}=0.95$ and $r_{1}=1$, where
each cell is irradiated by $N=100$ photons. This memory is now divided in
large blocks of $m=2000$ cells which are written by using an optimal EC code.
The reading of the memory is flawless up to a cut-off $\varepsilon=10^{-9}$,
corresponding to having less than $2$ kilobits of uncorrectable errors every
$2$ terabits of data. In this configuration, the classical upper-bound is
equal to $\overline{R}_{c}\simeq0.01$. This means that, using classical
transmitters, we can encode less than $20$ logical bits in a block of $2000$
cells. In other words, we need more than $100$ cells to encode a single bit of
information, which is a huge overhead of error correction, making classical
transmitters completely unsuitable for reading data. The situation is
completely different for EPR\ transmitters $T_{M,N}$ as shown in
Fig.~\ref{rateECS}. The quantum lower-bound $\underline{R}_{epr}$\ is
increasing in the bandwidth $M$, reaching its maximum value of $0.824$ for
$M\rightarrow\infty$. Most importantly, we have $\underline{R}_{epr}>0.8$
already for small values of $M$, i.e., narrowband EPR transmitters. This
numerical result shows that, using EPR\ transmitters, we can read data
perfectly up to a limited error-correction overhead (on average, we need less
than $1.25$ cells to store a bit of information). This overhead is more than
$80$ times smaller than the one needed by classical transmitters in the same
physical conditions. Similar results can be found with other choices of
parameters in the regime of few-photons and high-reflectivities.

In conclusion, we have considered alternative models of memories where the
information is stored in EC\ blocks in such a way that the readout is almost
flawless. In the regime of high reflectivities and few photons, we have
checked that narrowband EPR\ transmitters are able to retrieve the logical
data with a small EC overhead. By contrast, using classical transmitters in
the same situation is completely unfeasible since the corresponding
EC\ overhead is huge (more than $100$ cells to store a bit of information).
This would imply reading times $100$ times longer and logical capacities $100$
times smaller. In other words, by reducing the number of photons, the
classical readout becomes so noisy that no EC code is able to recover the
stored information in an efficient way.

\section{General Bounds for Error Correcting Codes\label{Hamming_SEC}}

Here we recall two important bounds for EC codes. These bounds are used in
Eq.~(\ref{Hamming}).

\begin{description}
\item[Hamming Bound~\cite{HammREF}.~] For large $m$, an EC code $[m,k,d]$ has
rate $R:=k/m$ and relative distance $\delta:=d/m$ such that%
\begin{equation}
R\leq1-H(\delta/2)+O(1/m)~, \label{Hmm_bound}%
\end{equation}
where $H$ is the binary Shannon entropy [defined in Eq.~(\ref{binSHA})].

\item[Gilbert-Varshamov Bound \cite{GilbertREF,VarshaREF}.~] For large $m$ and
$0\leq\delta\leq1/2$, there exists an EC code with rate%
\begin{equation}
R\geq1-H(\delta)+O(1/m)~, \label{GV_bound}%
\end{equation}
where $H$ is the binary Shannon entropy [defined in Eq.~(\ref{binSHA})].
\end{description}

\section{Conclusive Discussions\label{APP_discussion}}

\subsection{Implications of the Few-Photon Regime\label{SEC_implications}}

The advantages of quantum reading are related with the regime of few photons,
roughly given by $N\simeq1\div10^{2}$ photons per cell. Note that this is very
far from the energy which is used in today's classical readers, roughly given
by $N\simeq10^{10}$ photons per cell \cite{Trates,DVDbook}. In order to
understand the advantages connected with the few-photon regime, let us fix the
mean signal power $P_{S}$ which is irradiated over the cell during the reading
time. This quantity is approximately
\begin{equation}
P_{S}=(h\nu)\frac{N}{t}~, \label{FormulaDISC}%
\end{equation}
where $h$ is the Planck's constant, $\nu$ is the carrier frequency of the
light, $N$ is the mean number of photons in the signal, and $t$ is the reading
time of the cell. According to Eq.~(\ref{FormulaDISC}), for fixed power
$P_{S}$, we can decrease $N$ together with $t$. In other words, the regime of
few photons can be identified with the regime of short reading times, i.e.,
high data-transfer rates. Thus our results indicate the existence of
quantum\ transmitters which allow reliable \textit{fast }readout of digital memories.

Another implication of the few-photon regime is the increase of the storage
capacity. As long as the carrier frequency $\nu$ of the reading light is able
to resolve each single cell, we can increase the storage capacity just as a
consequence of the increased data-transfer rates. In other words, because of
the shorter reading time of the cell, we can increase the number of cells per
area unit (density) while keeping the total reading time of the disk as
constant. This is possible until the linear size of each cell is much larger
than the wavelength of the light. For higher densities, we have to increase
the frequency of the light in order to avoid diffraction and still use our
model. It is clear that using frequencies above the visible range will involve
the development of appropriate quantum sources. Note that the increase in the
memory density can also be explained as a consequence of
Eq.~(\ref{FormulaDISC}). In fact, for fixed power $P_{S}$, we can decrease $N$
while increasing $\nu$. This means that the regime of few photons can also be
identified with the regime of high frequencies, i.e., high densities. Thus
there exist quantum\ transmitters which can read, reliably, \textit{dense}
digital memories.

The previous physical discussion (done for fixed signal power $P_{S}$) can be
generalized to include the \textquotedblleft price to pay\textquotedblright%
\ for generating the light source. Given a global initial power $P$, we can
write $P_{S}=\kappa P$, where the conversion factor $\kappa$ depends on the
source to be generated: typically $\kappa\simeq1$ for classical light, while
$\kappa\ll1$ for non-classical light. Then, for fixed $P$, we have reading
times $t=O(\phi)$ where $\phi:=N/\kappa$. Here the ratio $\phi$ can still be
advantageous for quantum reading thanks to its superiority in the few-photon
regime. For instance, today's CDs are classically read using $N\simeq10^{10}$
photons per cell, so that $\phi_{\text{\textit{class}}}\simeq10^{10}$
\cite{Trates,DVDbook}. Using spontaneous parametric down conversion in
periodically poled KTP\ waveguides \cite{Fiorentino} we can generate EPR
correlations around $810$~nm with $\kappa\simeq10^{-9}$. Exploiting this
quantum source in the few-photon regime, e.g., $N\simeq1$, we can get
$\phi_{\text{\textit{quant}}}\simeq10^{9}$, thus realizing shorter reading
times. Clearly, this is a very rough estimate. However, this improvement will
become more and more evident as technology provides cheaper ways to create
non-classical light (see, e.g., the recent achievements of \cite{TPE}).

\subsection{Towards a Pilot Experiment}

It is important to note that the discussions of Sec.~\ref{SEC_implications}
represent general theoretical predictions, mainly based on the simple formula
of Eq.~(\ref{FormulaDISC}). More detailed evaluations are needed for an
experimental implementation of the scheme, where all the technicalities must
be taken into account. From this point of view it is interesting to attempt an
evaluation of the current technological facilities in order to realize a first
pilot experiment able to show the potentialities of quantum reading. In the
following we provide a semi-quantitative estimate of the data-transfer rates
that we could achieve by exploiting the current facilities in quantum technology.

First of all, note that a possible way to realize quantum reading is in the
time domain, where different modes correspond to different laser pulses. Thus,
we can consider an EPR transmitter $T_{M,N}$ emitting $M$ couples of entangled
pulses which irradiate a total of $N$\ photons over the cell. Let us assume
that each pulse has duration $\tau=w^{-1}$, where $w$ is the spectral
bandwidth (around a carrier frequency $\nu\gg w$). Then the minimum reading
time of the cell is $t=M\tau$. Its inverse $R=t^{-1}$ gives the maximum
data-transfer rate (in terms of bits per second). This value of the rate is
however a rough estimate. In the experimental practice, the pulses do not
satisfy the minimum time-bandwidth product relationship ($\tau w=1$), and the
data-transfer rate is better given by $R=R^{\prime}/M$, where $R^{\prime}$ is
the experimental pulse repetition rate (pulses per second), and $M$ is the
number of pulses per cell. Today it is possible to generate femtosecond pulses
of strongly entangled photons with high repetition rates. For instance, in
Refs.~\cite{Kri,Pan}, commercial mode-locked Ti:sapphire lasers are used to
generate 100-fs pulses at 780-810~nm with repetition rates of 76-81~MHz. Using
non-linear crystals, each of these pulses is first frequency-doubled and then
down-converted (via SPDC) in two entangled pulses. The overall process is
clearly inefficient. However, if we use strong input powers the output
entangled pulses are populated with about one photon per pulse \cite{Pan}.
This means that we can generate TMSV states with sufficient squeezing at the
optical-infrared frequencies with high repetition rates (e.g., $R^{\prime}%
=80$~MHz). Using $M$ of these states, we can construct an EPR transmitter
emitting $M$ pulses and signal-energy $N=zM$ where $z\simeq1$ (since we can
have roughly one photon per pulse impinging on the cell \cite{Pan}). Thus, for
the quantum reading scheme, we can consider $R=zR^{\prime}/N$ where $z\simeq1$
photons per pulse, $R^{\prime}$ is around 80~MHz (pulses per second), and $N$
is the mean total number of photons per cell. This estimate must be further
corrected by considering other technical issues. First, we have to introduce a
factor $y<1$ accounting for non-ideal quantum efficiencies of the output
detectors. Thus, we have to consider the lower rate $R=yzR^{\prime}/N$. In
order to evaluate $y$, the fundamental measurement is the time-domain homodyne
detection. This is in fact the measurement at the basis of our sub-optimal
receiver design of Sec.~\ref{APP_Bell} (when it is supposed to work in the
time domain.) Time-resolved homodyne detections have been recently studied by
several experimental groups \cite{Han,Wen,Wenb,Had,Zav,Zavb,Zavc}. According
to Refs.~\cite{Had,Zav,Zavb,Zavc}, it is possible to realize time-resolved
homodyne detections at the optical-infrared frequencies (786-800~nm) with high
repetition rates (54-82~MHz) and high signal-to-noise ratios (12dB). In
particular, at repetition rates around 80~MHz, we can still have acceptable
quantum efficiencies, e.g., $y\simeq0.6$ according to Ref.~\cite{Zavc}.
Finally, there is also the effect of classical error correction. On average,
the readout of a single cell corresponds to decoding $x<1$ logical bits, which
leads to the further correction $R=xyzR^{\prime}/N$. As discussed in the
previous Sec.~\ref{SEC_ECcodes}, we can decode $x\simeq0.8$ information bits
per cell by employing good error correcting codes (see Fig.~\ref{rateECS}\ and
previous Sec.~\ref{SEC_ECcodes}\ for details). Thus, for the quantum reading
of high reflectivity memories, we can consider $R=xyzR^{\prime}/N$ where
$x\simeq0.8$, $y\simeq0.6$, $z\simeq1$ and $R^{\prime}\simeq80~$MHz. The
resulting data-transfer rate $R\simeq38/N$ (Mbit/s) is inversely proportional
to the number of photons $N$. By assuming $N$ in the range of 1-70 photons, we
have $R$ ranging between 0.5~Mbit/s and 38~Mbit/s. This is already comparable
with the data-transfer rates of current optical memories. For instance, in
today's CDs, we have $R=1.23$~Mbit/s at the basic speed of 1X, and
$R=70$~Mbit/s at the speed of 56X (see, e.g., Wikipedia). Thus, quantum
reading can already provide good data-transfer rates by exploiting present
facilities in ultra-fast quantum technology.

It is clear that this technique has the potential to become much faster as
quantum technology provides better quantum sources and faster quantum
detectors. As a matter of fact new efficient ways for producing entanglement
have been recently developed. A very promising method is the
two-photon-emission from semiconductors \cite{TPE}. According to
Ref.~\cite{TPE}, the pair-generation rate in GaAs/AlGaAs quantum well
structure is estimated to be 3 orders of magnitude higher than for traditional
broadband parametric down-conversion sources. By exploiting new sources of
this kind, quantum reading has the long-term potential to go far beyond any
data-transfer rate which is achievable by classical devices.

\subsection{Application to Photodegradable Media}

Besides the potentialities previously described, there is another interesting
application of quantum reading in the few-photon regime: the safe readout of
photodegradable memories. These are memories where faint quantum light can
retrieve the data safely, while classical light could only be destructive.
Memories of this kind could be constructed on purpose. For instance, we could
construct extremely photo-sensitive organic microfilms which melt under very
few photons of visible light. An agency could use these cryptographic devices
to store confidential information. Their security would rely on the
technological complexity of the corresponding readers (e.g., extremely precise
quantum readers held by the agency only.)

Other examples of photodegradable memories are dye-based optical disks. Since
we can read information using few photons, we could construct and make use of
optical disks which are composed of extremely photo-sensitive dyes or other
similar organic materials. The problem of dye-degradation is very important
and involves current optical disks too. In today's recordable optical media
(CD-R and DVD-R), the data is recorded in internal layers of organic dye
(e.g., Azo, Cyanine, or Phthalocyanine) by means of a writing process called
\textquotedblleft dye-sublimation\textquotedblright\ \cite{DVDbook,NOTA3}. The
long-term use of these memories seems to be restricted to visible frequencies,
since their organic layers undergo a rapid UV-degradation at higher
frequencies \cite{DVDbook,UV}. In other words, despite organic disks of higher
densities could be written, their first readout could be completely
destructive as a result of the UV-degradation. In this scenario the use of
faint quantum light could provide a safe and reliable readout, thus enabling
these memories to be developed to higher densities. Clearly this is possible
after the development of appropriate non-classical sources in the UV range.
Thus, from this perspective, our work shows new possible directions in the
technology of organic memories.

More generally, the results of quantum reading can be applied to the study of
photodegradable absorbing materials. Whenever two absorbing media can be
modeled by two attenuator channels with different losses, their discrimination
is equivalent to the readout of an information bit from a memory cell
(according to our basic model of memory). If these media are furthermore
photodegradable, then the use of faint quantum light could represent the only
method to solve their discrimination without destroying the sample.


\begin{thebibliography}{99}
\bibitem {CLASSstate}For any $K$-mode bosonic state $\rho$, we can write the
$\mathcal{P}$-representation $\rho=\int d^{2K}\boldsymbol{\alpha~}%
\mathcal{P}(\boldsymbol{\alpha})\left\vert \boldsymbol{\alpha}\right\rangle
\left\langle \boldsymbol{\alpha}\right\vert $, where $\mathcal{P}%
(\boldsymbol{\alpha})$\ is normalized to $1$ and $\left\vert
\boldsymbol{\alpha}\right\rangle \left\langle \boldsymbol{\alpha}\right\vert $
is a multimode coherent state. Then, $\rho$ is called \textquotedblleft
classical\textquotedblright\ (\textquotedblleft
non-classical\textquotedblright) if $\mathcal{P}(\boldsymbol{\alpha})$ is
positive (non-positive) \cite{Prepres}. For $\rho$ classical, $\mathcal{P}%
(\boldsymbol{\alpha})$ is a probability distribution (implying separability).
\bibitem {EPRcorrelations}For two bosonic modes, $A$ and $B$, with quadratures
$\hat{q}_{A}$, $\hat{p}_{A}$, $\hat{q}_{B}$ and $\hat{p}_{B}$, one can define
the two operators $\hat{q}_{-}:=(\hat{q}_{A}-\hat{q}_{B})/\sqrt{2}$ (relative
position)\ and $\hat{p}_{+}:=(\hat{p}_{A}+\hat{p}_{B})/\sqrt{2}$ (total
momentum). Then, the system has EPR correlations (in these operators) if
$V(\hat{q}_{-})+V(\hat{p}_{+})<2\nu_{0}$, where $V(\cdot)$ is the variance,
and $\nu_{0}$ is the standard quantum limit ($\nu_{0}=1$ in this paper.) A
bosonic system with EPR correlations is entangled, but the contrary is not
necessarily true. In today's quantum optics labs, the EPR\ correlations
represent the standard type of continuous variable entanglement. These
correlations are usually generated via spontaneous parametric down conversion.
\bibitem {G}$G>0$ is a sufficient condition for the superiority of quantum
reading. $G=1$ corresponds to the singular case where only quantum light can
retrieve information.
\bibitem {notationCOMM}With the notation $[\mathbf{\hat{x}},\mathbf{\hat{x}%
}^{T}]$ we mean a matrix with entries $[\mathbf{\hat{x}}_{i},\mathbf{\hat{x}%
}_{j}]$, where $i,j=1,\cdots,2n$. Analogously, $\{\mathbf{\hat{x}%
},\mathbf{\hat{x}}^{T}\}$ is a matrix with entries $\{\mathbf{\hat{x}}%
_{i},\mathbf{\hat{x}}_{j}\}$.
\bibitem {Noteshort}Note that in our Letter we use the short-hand notation
$P_{err}(T):=P_{err}(\mathcal{E}_{0}\neq\mathcal{E}_{1}|T)$.
\bibitem {Jensen}Jensen's inequality states that, for every random variable
$\mathbf{x}$, with values in $\mathbb{R}^{M}$ and expectation value
$\left\langle \mathbf{x}\right\rangle $, and for any concave function
$f:\mathbb{R}^{M}\rightarrow\mathbb{R}$, we have $\left\langle f(\mathbf{x}%
)\right\rangle \leq f\left(  \left\langle \mathbf{x}\right\rangle \right)  $
(the inequality is reversed for convex functions). More in general, the
inequality holds for random variables and concave functions which are defined
in measurable subsets $\Omega\subset\mathbb{R}^{M}$.
\bibitem {NumG}Numerical investigations can also be performed by using the
simpler quantity%
\[
G^{\prime}:=1-H(\mathcal{B})-[1-H(\mathcal{C})]\leq G~,
\]
which is based on the quantum Battacharyya bound.
\bibitem {Trates}A single-layer CD with speed 1X (i.e., 500~rpm) has a
data-transfer rate of 1.23~Mbit/s. This data is read by a laser diode at
780~nm and power $\approx5$~mW, therefore irradiating about $5$~nJ per bit. A
higher-speed CD needs a laser with proportionally more power, in such a way
that at least $1$~nJ per bit is irradiated, corresponding to $N\simeq10^{10}$
photons per bit. By neglecting the error correction overhead, this corresponds
to $N\simeq10^{10}$ photons per cell.
\bibitem {NOTA1}For the sake of simplicity, we consider here the pure-loss
model. As before, the derivation can be extended to the thermal-loss model and
the sub-optimal receiver.
\bibitem {NOTA2}Note that this formula provides the expression of
Eq.~(\ref{uncorregible_nbits}) when we set $d$ odd and $d=m$ (properties of
the repetition codes).
\bibitem {NOTA3}During the writing process, a high-power laser heats the dye
polymer which therefore darkens, with the result of causing less reflection
from the metallic layer underneath.
\end{thebibliography}
\end{document}